\documentclass[prd,aps,floatfix]{revtex4}
\usepackage{amssymb,graphicx}

\newcommand{\Real}{\mathbb{R}}
\newcommand{\Complex}{\mathbb{C}}
\newcommand{\re}{\mbox{Re}}

\newcommand{\bbA}{\mbox{\boldmath $A$}}
\newcommand{\bbB}{\mbox{\boldmath $B$}}

\begin{document}

\title{Boundary conditions for Einstein's field equations: Analytical and numerical analysis}

\author{Olivier Sarbach$^{1,2,3}$, and Manuel Tiglio$^{2,4,5}$.}

\affiliation{$1$ Department of Mathematics, Louisiana State University, 
Lockett Hall, Baton Rouge, Louisiana 70803-4918\\
$2$ Department of Physics and Astronomy, Louisiana State
University, 202 Nicholson Hall, Baton Rouge, LA 70803-4001\\
$3$ Theoretical Astrophysics 130-33
California Institute of Technology, Pasadena, CA 91125\\
$4$ Center for Computation and Technology, 302 Johnston Hall, Louisiana State
University, Baton Rouge, LA 70803-4001\\
$5$ Center for Radiophysics and Space Research, Cornell University, Ithaca, NY 14853}

%%%%%%%%%%%%%%%%%%%%%%%%%%%%%%%%%%
\begin{abstract}
%%%%%%%%%%%%%%%%%%%%%%%%%%%%%%%%%%

Outer boundary conditions for strongly and symmetric hyperbolic
formulations of $3$D Einstein's field equations with a live gauge
condition are discussed. The boundary conditions have the property
that they ensure constraint propagation and control in a sense made
precise in this article the physical degrees of freedom at the
boundary. We use Fourier-Laplace transformation techniques to find
necessary conditions for the well posedness of the resulting
initial-boundary value problem and integrate the resulting
three-dimensional nonlinear equations using a finite-differencing
code. We obtain a set of constraint-preserving boundary conditions
which pass a robust numerical stability test. We explicitly compare
these new boundary conditions to standard, maximally dissipative ones
through Brill wave evolutions. Our numerical results explicitly show
that in the latter case the constraint variables, describing the
violation of the constraints, do not converge to zero when resolution
is increased while for the new boundary conditions, the constraint
variables do decrease as resolution is increased. As an application,
we inject pulses of ``gravitational radiation'' through the boundaries
of an initially flat spacetime domain, with enough amplitude to
generate strong fields and induce large curvature scalars, showing
that our boundary conditions are robust enough to handle nonlinear
dynamics.

We expect our boundary conditions to be useful for improving the
accuracy and stability of current binary black hole and binary neutron
star simulations, for a successful implementation of characteristic or
perturbative matching techniques, and other applications. We also
discuss limitations of our approach and possible future directions.
\end{abstract}


\maketitle

%%%%%%%%%%%%%%%%%%%%%%%%%%%%%%%%%%%%%%%%%%%%%%%%%%%%%%%%%%%%%%%%%%%%
\section{Introduction}
%%%%%%%%%%%%%%%%%%%%%%%%%%%%%%%%%%%%%%%%%%%%%%%%%%%%%%%%%%%%%%%%%%%%

In many numerical simulations in General Relativity one integrates Einstein's
field equations on a spatially compact domain with artificial timelike
boundaries, effectively truncating the computational domain. This raises the
question of how to specify boundary conditions. In this article we address
this question within the context of the Cauchy formulation, in which the field
equations split into evolution and constraint equations. We adopt a free
evolution approach, in which the constraints are solved on the initial time
slice only and the future of the initial slice is computed by integrating the
evolution equations. Boundary conditions within this approach should ideally
satisfy the following three requirements: (i) be compatible with the
constraints in the sense of guaranteeing that initial data which solves the
constraint equations yields constraint-satisfying solutions, (ii) permit to
control, in some sense, the gravitational degrees of freedom at the boundary,
and (iii) be stable in the sense of yielding a well posed initial-boundary
value problem (IBVP).

There are several motivations for the construction of boundary conditions
satisfying the above properties. First, recent detailed analysis of binary
neutron star evolutions \cite{miller} showed that the presence of artificial
boundaries in current state of the art evolutions, which use rather ad hoc
boundary conditions, dramatically affects the dynamics in the strong field
region, near the stars. While it can be argued that this effect should
disappear when placing the boundaries further and further away from the strong
field region, until ideally, the region of interest in the computational
domain is causally disconnected from the boundaries, in practice, this would
require huge computer resources, especially in the three-dimensional case.
Even though some kind of mesh refinement should help in placing the boundaries
far away, there is in any case a minimum resolution needed in the far region
in order to reasonably represent wave propagation, thus constraining the size
of the computational domain for a given amount of memory. Next, an
understanding of boundary conditions is important if the evolution equations
include elliptic equations, since in this case effects from the boundaries can
propagate with infinite speed, and so have an immediate effect on the fields
being evolved. Examples of cases in which elliptic equations arise include
elliptic gauge conditions (such as maximal slicing \cite{max_slicing} or
minimal distortion conditions \cite{min_distortion}) and constraint projection
\cite{cp_AM, cp_S, cp_Caltech} methods. Finally, isolating the physical
degrees of freedom at the boundaries should be important in view of
Cauchy-characteristic (see \cite{ccm} for a review) or Cauchy-perturbative
\cite{cpm1,cpm2,cpm3} matching techniques, where a Cauchy code is coupled to a
characteristic or perturbative ``outer module'' which is well adapted to carry
out the evolution in the far zone. In these cases, it is important to
communicate only the physical degrees of freedom at the boundary between the
two codes, since the Cauchy code and the outer module might be based on
completely different formulations.

Boundary conditions satisfying requirement (i) can be constructed by analyzing
the constraint propagation system, which constitutes an evolution system for
the constraint variables and is a consequence of Bianchi's identities and the
evolution equations. If it can be cast into first order symmetric hyperbolic
form, the imposition of homogeneous maximally dissipative boundary conditions
\cite{LaxPh, Friedrichs} for the constraint propagation system guarantees that
a smooth enough solution of the evolution system (if it exists) which
satisfies the constraints initially automatically satisfies the constraints at
later times. Homogeneous maximally dissipative boundary conditions consist in
a linear relation between the in- and outgoing characteristic fields of the
system, which is chosen such that an energy estimate can be derived. This
energy estimate implies that the unique solution of the constraint propagation
system with zero initial data is zero, i.e. that the constraints are preserved
during evolution. Maximally dissipative boundary conditions for the constraint
variables usually translate into differential conditions for the fields
satisfying the main evolution equations, since the constraint variables depend
on derivatives of the main variables. This means that the resulting boundary
conditions for the main system usually are not of maximally dissipative type
and, as discussed below, analyzing well posedness of the corresponding IBVP
becomes more difficult.

Requirement (ii), controlling the physical degrees of freedom, is a difficult
one since there are no known {\em local} expressions for the energy or the
energy flux density in General Relativity.  Nevertheless, one should be able
to control the physical degrees of freedom in some approximate sense, as for
example in the weak field regime approximation, in which one linearizes the
equations around flat spacetime. In this approximation, it might be a good
idea to specify conditions through the Weyl tensor, since it is invariant with
respect to infinitesimal coordinate transformations of Minkowski spacetime.
More precisely, since there are two gravitational degrees of freedom, we
should specify two linearly independent combinations of the components of the
Weyl tensor. Below we will discuss boundary conditions which involve the
Newman-Penrose complex scalars $\Psi_0$ and $\Psi_4$ (see, for instance,
\cite{Stewart-Book}) with respect to a null tetrad adapted to the
time-evolution vector field and the normal to the boundary. If the boundary is
at null infinity these scalars represent the in- and outgoing gravitational
radiation, respectively. Furthermore, it turns out that these scalars are
invariant with respect to infinitesimal coordinate transformations and tetrad
rotations for linear fluctuations of any Petrov type D solution represented in
an adapted background tetrad \cite{Teukolsky}. This class of solutions not only
comprises flat spacetime but also the family of Kerr solutions describing
stationary rotating black holes.  Since in many physical situations one is
interested in modeling asymptotically flat spacetimes, such that if the outer
boundary is placed sufficiently far away from the strong field region
spacetime can be described by a perturbed Kerr black hole, we expect the
boundary condition $\Psi_0 = 0$ to be a good approximation for a
``non-reflecting'' wave condition. These boundary conditions are actually part
of the family of conditions imposed in the formulation of Ref. \cite{FN}, to
date the only known well posed initial-boundary value formulation of the
vacuum Einstein equations, and were also considered in \cite{BB}.

Requirement (iii), the well posedness of the resulting IBVP, turns out to be a
difficult problem as well: For quasilinear symmetric hyperbolic systems with
maximally dissipative boundary conditions there are well-known well posedness
theorems \cite{Rauch, Secchi1, Secchi2} which state that a (local in time)
solution exists in some appropriate Hilbert space, that the solution is
unique, and that it depends continuously on the initial and boundary data. The
proof of Ref. \cite{FN} is based on these techniques. There, using a
formulation based on tetrad fields, the authors manage to obtain a symmetric
hyperbolic system by adding suitable combinations of the constraints to the
evolution equations in such a way that the constraints propagate tangentially
to the boundary.  In this way, the issue of preserving the constraints
becomes, in some sense, trivial. (See \cite{IR} for a treatment in spherical
symmetry in which the constraints propagate tangentially to the boundary as
well.)  However, for the more commonly used metric formulations it seems
difficult to achieve tangential propagation for the constraints with a
symmetric or strongly hyperbolic system, and therefore, one has to deal with
either constraint propagation across the boundary or systems that are not
strongly or symmetric hyperbolic. Here we choose to deal with constraint
propagation across the boundary and strongly or symmetric hyperbolic systems.
There has been a lot of effort in understanding such systems, both at the
analytical \cite{Stewart,SSW,SW,CPRST,CS,Gioel, Frittelli,GMG1,GMG2,RS} and
numerical \cite{BB, CLT, SSW,SW,Gioel,LS-FatMax,cp_Caltech,CPBC-Bona} level.
Although partial proofs of well posedness have been obtained using symmetric
hyperbolic systems with maximally dissipative boundary conditions
\cite{SSW,SW,CPRST,GMG2}, it seems that these kind of boundary conditions are
not flexible enough since constraint-preserving boundary conditions usually
yield differential conditions.

In this article we construct constraint-preserving boundary conditions
(CPBC) for a family of first order strongly and symmetric hyperbolic
evolution systems for Einstein's equations. For definiteness, we focus
on the formulation presented in Ref. \cite{ST}, which is a
generalization of the Einstein-Christoffel type formulations
\cite{EC-FR, EC-AY, EC-Hern, EC-KST} with a Bona-Mass{\'o} type of
gauge condition \cite{BM-Live1, BM-Live2} for the lapse. However, our
approach is quite general and should also be applicable to other
hyperbolic formulations of Einstein's equations. In Section
\ref{Sect:SHLG} we briefly review the family of formulations
considered here and recall under which conditions the main evolution
equations are strongly or symmetric hyperbolic. In Section
\ref{Sect:ESCV} we discuss the constraint propagation system and
analyze under what circumstances it is symmetric hyperbolic. Having
cast this system in symmetric hyperbolic form we impose CPBC via
homogeneous maximally dissipative boundary conditions for the
constraint variables. These conditions are differential boundary
conditions when expressed in terms of the fields satisfying the main
system. We complete these boundary conditions in Section \ref{Sect:BC}
by imposing extra conditions which control the physical and gauge
degrees of freedom. There are several possibilities for doing so, of
which we analyze the following two: 1) we first specify algebraic
conditions in the form of a coupling between the outgoing and ingoing
characteristic fields (referred to as CPBC without Weyl control later
in this article); 2) we specify boundary conditions via the Weyl
scalars $\Psi_0$ and $\Psi_4$, as discussed above (referred to as CPBC
with Weyl control later in this article).

Section \ref{Sect:FL} is devoted to an analysis of the well posedness of the
resulting IBVP. As mentioned above, this is a difficult problem since our
boundary conditions are not in algebraic form. I particular, they are not in
maximally dissipative form, so we cannot apply the standard theorems for
symmetric systems with maximally dissipative boundary conditions. Therefore,
our goal here is more modest: We analyze the IBVP in the ``high frequency
limit'' by considering high-frequency perturbations of smooth solutions. In
this regime the equations become linear with constant coefficients, and the
domain can be taken to be a half plane. In this case solutions can be
constructed explicitly by performing a Laplace transformation in time and a
Fourier transformation in the spatial directions that are tangential to the
boundary. This leads to the verification that a certain determinant is
non-zero. If this determinant condition is violated, the system admits ill
posed modes growing exponentially in time with an arbitrarily small growth
time.  Therefore, the determinant condition yields {\em necessary} conditions
for well posedness of the resulting IBVP and allows us to discard several
cases which would lead to an ill posed formulation. {\em These ill posed
  formulations appear even in cases in which both the main and the constraint
  propagation system are symmetric hyperbolic}. We also stress that the
determinant condition verified in this article is a weaker version of the
celebrated Kreiss condition \cite{Kreiss, KL-Book, GKO-Book}, which yields
well posed IBVP for hyperbolic problems with {\em algebraic} boundary
conditions \cite{Kreiss, MO}. However, the Kreiss condition guarantees nothing
for the present case of {\em differential} boundary conditions. See \cite{RS}
for an example of an IBVP with differential boundary conditions which
satisfies the Kreiss condition but fails to be well posed in $L^2$.

In section \ref{Sect:NR} we discretize the IBVP by the method of lines. The
spatial derivatives are discretized using finite difference operators that
satisfy the summation by parts property and we explain in detail how we
numerically implement our CPBC. Then, in sections \ref{Sect:Sim1} and
\ref{Sect:Sim} we perform the following $3$D numerical tests: First, we evolve
IBVPs which fail to fulfill the determinant condition and are therefore ill
posed. The goal of evolving these systems is to confirm the expected lack of
numerical stability. We do confirm such instability, finding an obvious lack
of convergence: the results exhibit exponentially in time growing modes, where
the exponential factor gets larger as resolution is increased, as predicted by
our analytical analysis. Next, we focus on evolutions of two systems that do
satisfy the determinant condition, differing only on whether they control
components of the Weyl tensor at the boundary or not. Finally, we compare
stable evolutions of systems with CPBC with evolutions where maximally
dissipative boundary conditions are given for the {\em main} evolution system,
and are therefore expected to violate the constraints.

We first analyze evolutions of these four systems through a {\em robust
  stability test} \cite{robust1, robust2}. In this test, random initial and
boundary data is specified at different resolutions, and the growth rate in
the time evolved fields is observed. A growth rate that becomes {\em larger}
as resolution is increased is a strong indication of a numerical instability,
while for numerical stability the growth rate should be {\em bounded} by a
constant that is independent of resolution. We find that systems that violate
the determinant condition fail to pass the robust stability test, as expected.
However, we {\em also} find that at least some systems with CPBC with Weyl
control which satisfy the determinant condition are numerically unstable as
well, although in a somehow weaker sense (explained in the text) that reminds
the numerical evolution of weakly hyperbolic systems \cite{CPST-Convergence}.
In contrast to this, the systems with CPBC without Weyl control which satisfy
the determinant condition that we have evolved pass the robust stability test
for the length of our simulations (usually between $100$ and $1,000$ crossing
times).

Next, we concentrate on evolutions of Brill waves \cite{BrillWaves}, and
confirm the expectations drawn from the robust stability test in what concerns
numerical stability. Using these waves we further concentrate on a detailed
comparison between the results using maximally dissipative boundary conditions
for the main evolution system and stable CPBC. Our convergence tests strongly
suggest that in the former case the constraint variables {\em do not} converge
to zero in the limit of infinite resolution, implying that one {\em does not}
obtain a solution to Einstein's field equations, while in the latter case for
the same resolutions the constraint variables do converge to zero.

Next we concentrate on the stable CPBC case and evolve pure gauge solutions,
using high order accurate finite difference operators which satisfy the
summation by parts property. The operators used are eighth order accurate in
the interior points and fourth order accurate at and near the boundary points.

As a final numerical experiment, we also concentrate on the stable
CPBC case and inject pulses of gravitational radiation through the
boundaries of an initially flat spacetime. We inject pulses of large
enough amplitude to create very large curvature in the interior (as
measured by curvature invariants), showing that our CPBC are strong
enough to handle very non-linear dynamics. The order of magnitude
achieved by the curvature invariant measured corresponds to being at
roughly $r=0.7$ from the singularity, in a Schwarzschild spacetime of
mass one. In these simulations this curvature is produced {\em solely}
by the injected pulses.

A summary of the results and conclusions are presented in Sect.
\ref{Sect:Conc}. Technical details, like the derivation of the constraint
propagation system and of the characteristic fields, and a special family of
solutions to the linearized IBVP with Weyl control are found in
\ref{App:MECPS}, \ref{App:CharFields&SH} and \ref{App:SFS}.

%%%%%%%%%%%%%%%%%%%%%%%%%%%%%%%%%%%%%%%%%%%%%%%%%%%%%%%%%%%%%%%%%%%%
\section{Strongly hyperbolic formulations with a live gauge}
\label{Sect:SHLG}
%%%%%%%%%%%%%%%%%%%%%%%%%%%%%%%%%%%%%%%%%%%%%%%%%%%%%%%%%%%%%%%%%%%%

In this section we review the family of hyperbolic formulations of Einstein's
field equations constructed in \cite{ST}, which is an extension of the
Einstein-Christoffel type \cite{EC-FR, EC-AY} of formulations which
incorporates a generalization of the Bona-Mass{\'o} \cite{BM-Live1, BM-Live2}
slicing conditions. It consists of $34$ evolution equations for the variables
$\{N, g_{ij}, K_{ij}, d_{kij}, A_i\}$, where $N$ is the lapse function,
$g_{ij}$ the three-metric, $K_{ij}$ the extrinsic curvature, and where the
extra variables $d_{kij}$ and $A_i$ represent the first order spatial
derivatives of the three-metric and of the logarithm of the lapse,
respectively.

The evolution equations are obtained from the ADM evolution equations in
vacuum by adding suitable constraints to the right-hand side of the equations
(see \ref{App:MECPS} for more details). Following the notation of Ref.
\cite{ST}, we have
\begin{eqnarray}
\partial_0 N &=& - F(N,K,x^{\mu} ),
\label{Eq:Ndot}\\
%%%%%%%%%%%%%%%%%%%%%%%
\partial_0 g_{ij} &=&  -2K_{ij} \; ,
\label{Eq:gdot}\\
%%%%%%%%%%%%%%%%%%%%%%%
\partial_0 K_{ij} &=& R_{ij} - \partial_{(i} A_{j)} + \Gamma^k_{\;
ij} A_k - A_i A_j - 2 K_i^{\;\; l} K_{lj} + K K_{ij}
\nonumber\\
 &+& \gamma\, g_{ij}C + \zeta\, g^{kl}C_{k(ij)l} \; ,
\label{Eq:Kdot}\\
%%%%%%%%%%%%%%%%%%%%%%%
\partial_0 d_{kij} &=& -2\partial_k K_{ij} - 2 A_k K_{ij}
 + \eta\, g_{k(i}C_{j)} + \chi\, g_{ij}C_k 
 + \frac{2}{N} g_{l(i} \partial_{j)}\partial_k\beta^l ,
\label{Eq:ddot}\\
%%%%%%%%%%%%%%%%%%%%%%%
\partial_0 A_i &=& -\frac{\partial F}{\partial N} A_i 
- \frac{1}{N}\frac{\partial F}{\partial K} \left( g^{kl}\partial_i K_{kl} - d_{ikl} K^{kl} \right) 
- \frac{1}{N}\frac{\partial F}{\partial x^i} + \xi \,  C_i\, .
\label{Eq:Adot}
\end{eqnarray}
Here, $\partial_0 = (\partial_t - \pounds_\beta)/N$, and $F(N,K,x^{\mu} )$ is
a function that is smooth in all its arguments and that satisfies $\sigma =
(\partial_K F)/(2N) > 0$. For the simulations below we shall choose $F = N
\cdot K$ which corresponds to time-harmonic slicing, but for our analytical
results we shall leave $F$ unspecified for generality. We assume that the
shift vector $\beta^i$ is a fixed, a priori specified vector field. The
parameters \footnote{These parameters can be constants or a priori specified
  functions on spacetime. See Ref. \cite{DynControl, TLN-BlackHole} where
  time-dependent parameters are used in order to control growth of constraint
  violations.} $\gamma$, $\zeta$, $\eta$, $\chi$, $\xi$ control the dynamics
off the constraint hypersurface, defined by the vanishing of the following
expressions
\begin{eqnarray}
C &\equiv& \frac{1}{2}(g^{kl} R_{kl} - K^{kl} K_{kl} + K^2),
\label{Eq:C}\\
C_i &\equiv& g^{kl}\left( \partial_k K_{li} - \partial_i K_{kl} \right) 
  + \frac{1}{2}(d^k - 2b^k)K_{ki} + \frac{1}{2} d_i^{\;\; kl} K_{kl}\; ,
\label{Eq:Ci}\\
C_{kij} &\equiv& d_{kij} - \partial_k g_{ij}\; ,
\label{Eq:Ckij}\\
C_{lkij} &\equiv& \partial_{[l}d_{k]ij}\; ,
\label{Eq:Clkij}\\
C^{(A)}_i &\equiv& A_i - \partial_i N/N\; ,
\label{Eq:CAi}\\
C^{(A)}_{ij} &\equiv& \frac{1}{N} \partial_{[i} (N A_{j]})\; .
\label{Eq:CAij}
\end{eqnarray}
Here, $C=0$ is the Hamiltonian constraint, $C_i = 0$ the momentum one, and
$C_{kij}=0$, $C_{lkij}=0$, $C^{(A)}_i = 0$ and $C^{(A)}_{ij} = 0$ are
artificial constraints that arise as a consequence of the introduction of the
extra variables $d_{kij}$, $A_i$ \footnote{Notice that the constraint
  variables $C_{lkij}$ and $C^{(A)}_{ij}$ are automatically zero if the
  constraints $C_{kij}=0$, $C^{(A)}_i = 0$ are satisfied, and so they are
  redundant. However, we will need these variables in the next section in
  order to cast the constraint propagation system into first order form.}. The
Ricci tensor $R_{ij}$ belonging to the three-metric is written as
\begin{eqnarray}
R_{ij} &=& \frac{1}{2} g^{kl} \left(
 -\partial_k d_{lij} + \partial_k d_{(ij)l} + \partial_{(i} d_{|kl|j)} - \partial_{(i} d_{j)kl} \right)
\nonumber\\
 &+& \frac{1}{2} d_i^{\;\; kl} d_{jkl} + \frac{1}{2}(d_k - 2b_k)\Gamma^k_{\; ij}
  - \Gamma^k_{\; lj} \Gamma^l_{\; ik}\; ,
\label{Eq:Rij}
\end{eqnarray}
where $b_j \equiv d_{kij} g^{ki}$, $d_k \equiv d_{kij} g^{ij}$ and
\begin{displaymath}
\Gamma^k_{\; ij} = \frac{1}{2} g^{kl} \left( 2d_{(ij)l} - d_{lij} \right).
\end{displaymath}

The evolution equations
(\ref{Eq:Ndot},\ref{Eq:gdot},\ref{Eq:Kdot},\ref{Eq:ddot},\ref{Eq:Adot}) have
the form of a quasilinear first order system,
\begin{equation}
\partial_0 u = {\cal A}(u)^j\partial_j u + {\cal F}(u),
\end{equation}
where $u = (N, g_{ij}, K_{ij}, d_{kij}, A_i)$ and the matrix-valued
functions ${\cal A}^j$, $j=1,2,3$, and the vector-valued function
${\cal F}$ are smooth.  We are looking for solutions with given
initial data on some three-dimensional manifold which satisfies the
constraints equations. In order to guarantee the existence of
solutions we restrict the freedom in choosing the parameters $\sigma$,
$\gamma$, $\zeta$, $\eta$, $\chi$, $\xi$ by demanding that the
corresponding initial-value formulation is well posed. This means that
given smooth initial data, a (local in time) solution should exist in
some appropriate Hilbert space, be unique, and depend continuously on
the initial data. Although in this article we are interested in the
numerical evolution of spacetimes on a spatially compact region with
boundaries, well posedness of the problem in the absence of boundaries
is a necessary condition for obtaining a numerical stable and
consistent evolution inside the domain of dependence of the initial
slice. An easy and intuitive way of finding necessary conditions for
well posedness is to look at high frequency perturbations of smooth
solutions: Let $p = (t,x^j)$ be a fixed point and $u_0$ a smooth
solution in a neighborhood of $p$. Perturb $u_0$ according to
$u(t,x^j) = u_0(t,x^j) + \varepsilon\tilde{u}(\omega t)\exp(i\omega
n_j x^j)$, with $\varepsilon$, $\omega$ real, and $n_i$ a constant
one-form on $\Real^3$ which is normalized such that $g(p)^{ij} n_i n_j
= 1$. If we evaluate the evolution equations at a point near $p$,
divide by $\omega\varepsilon$, and take first the limit
$\omega\rightarrow \infty$ and then the limit $\varepsilon\rightarrow
0$, the evolution equations reduce to
\begin{equation}
\partial_t\tilde{u} = \left[ N(p) {\cal A}(p,n) + \beta(p)^j n_j \right] \tilde{u},
\label{Eq:Frozen}
\end{equation}
where the matrix ${\cal A}(p,n)$ is given by
\begin{eqnarray}
{\cal A}(p,n)u = \left( \begin{array}{c} 0 \\ 0 \\ 
\frac{1}{2}\left[  - d_{nij} + (1+\zeta ) d_{(ij)n} - \zeta\, n_{(i} b_{j)} 
  + n_{(i}\Delta_{j)} - 2n_{(i} A_{j)} + \gamma\, g_{ij} \Delta_n \right] \\
 -2n_k K_{ij} + \eta\,g_{k(i}\left(K_{j)n} - n_{j)} K \right) 
              + \chi\,g_{ij}\left( K_{kn} - n_k K \right) \\
 -2\sigma n_i K + \xi\,\left( K_{in} - n_i K \right)
\end{array} \right),
\nonumber\\
\label{Eq:PrincipalSymb}
\end{eqnarray}
where we have set $\Delta_j = b_j - d_j$ and where the index $n$ refers to the
contraction with $n^i$. Notice that by rescaling and rotating the coordinates
$x^i$ one can always achieve that $g(p)_{ij} = \delta_{ij}$, and by rescaling
the coordinate $t$ one can achieve that $N(p) = 1$. For this reason, we drop
the entry $p$ in the following. We call the system (\ref{Eq:Frozen}) the
associated frozen coefficient problem. A necessary condition for the well
posedness of the initial-value formulation defined by the (non-linear) Eqs.
(\ref{Eq:Ndot},\ref{Eq:gdot},\ref{Eq:Kdot},\ref{Eq:ddot},\ref{Eq:Adot}) is the
well posedness of the associated frozen coefficient problem. If some extra
smoothness properties are satisfied (see \ref{App:CharFields&SH}) this
condition is also a sufficient one.  The frozen coefficient problem is well
posed if the matrix $N {\cal A}(n) + \beta^i n_i$ is diagonalizable and has
only real eigenvalues for each $n$. Clearly, this is true if and only if the
matrix ${\cal A}(n)$ is diagonalizable and has only real eigenvalues. As we
have shown in \cite{ST} this can be easily analyzed by taking advantage of the
block structure of ${\cal A}(n)$: Suppose $u$ is an eigenvector of ${\cal
  A}(n)$ with eigenvalue $\mu$. Then
\begin{displaymath}
\mu\left( \begin{array}{c} u^{(0)} \\ u^{(1)} \\ u^{(2)} \end{array} \right)
 = \left( \begin{array}{ccc} 0 & 0 & 0 \\ 0 & 0 & \bbA \\ 0 & \bbB & 0 \end{array} \right)
   \left( \begin{array}{c} u^{(0)} \\ u^{(1)} \\ u^{(2)} \end{array} \right),
\end{displaymath}
where $u^{(0)} = (N, g_{ij})$, $u^{(1)} = (K_{ij})$, $u^{(2)} = (d_{kij},
A_i)$, and where the matrices $\bbA$ and $\bbB$ are read off from Eq.
(\ref{Eq:PrincipalSymb}). A sufficient condition for ${\cal A}(n)$ to be
diagonalizable and posses only real eigenvalues can be obtained by considering
the equation
\begin{equation}
\mu^2 K_{ij} = \bbA\bbB K_{ij}\, .
\label{Eq-WaveK}
\end{equation}
Explicitly, we have
\begin{displaymath}
\mu^2 K_{ij} = K_{ij} + A n_{(i} K_{j)n}
 + B n_i n_j K + C g_{ij} \left( K_{nn} - K\right),
\end{displaymath}
where the coefficients $A$, $B$ and $C$ are
\begin{eqnarray}
A &=& -2 - \frac{\chi}{2} - \frac{1}{4}(3\zeta - 1)\eta - \xi\, ,
\label{Eq:CoeffA}\\
B &=& \frac{\chi}{2} + \frac{1}{4}(3\zeta-1)\eta + \xi + 1 + 2\sigma,
\label{Eq:CoeffB}\\
-2C &=& \chi - \frac{1}{2}(\zeta + 1)\eta + \gamma(2 - \eta + 2\chi).
\label{Eq:CoeffC}
\end{eqnarray}
We now demand that $\bbA\bbB$ is diagonalizable and has only positive
eigenvalues. As we have shown in \cite{ST} this guarantees that ${\cal A}(n)$
is diagonalizable and has only real eigenvalues. Representing $\bbA\bbB$ in an
orthonormal basis $e^1_i$, $e^2_i$, $e^3_i$ such that $e^1_i = n_i$, we find
\begin{eqnarray}
\mu^2 K_{nn} &=& (1+A+B) K_{nn} + (B-C)K^A_A\; ,\\
\mu^2 K_{nA} &=& (1 + A/2) K_{nA}\; ,\\
\mu^2 K^A_A &=& (1 -2C) K^A_A\; ,\\
\mu^2\hat{K}_{AB} &=& \hat{K}_{AB}\; ,
\end{eqnarray}
where $A = 2,3$ and $\hat{K}_{AB} = K_{AB} - \delta_{AB}\delta^{CD} K_{CD}/2$.
From this we immediately see that $\bbA\bbB$ is diagonalizable with only
positive eigenvalues if and only if
\begin{eqnarray}
\lambda_1 &=& 2\sigma ,
\label{Eq:lam1}\\
\lambda_2 &=& \frac{1}{2} (2\gamma + 1)(2 + 2\chi - \eta) - \frac{1}{2}\,\zeta\eta\, ,
\label{Eq:lam2}\\
\lambda_3 &=& -\frac{1}{8} (2\chi-\eta + 4\xi) - \frac{3}{8}\,\zeta\eta\, ,
\label{Eq:lam3}
\end{eqnarray}
are positive and if $B - C = \lambda_1 + \frac{1}{2}(\lambda_2 + 1) -
2\lambda_3 = 0$ whenever $\lambda_1=\lambda_2$. In \ref{App:CharFields&SH} we
derive the characteristic fields, which are given by the projections of $u$
onto the eigenspaces of ${\cal A}(n)$.  These fields play an important role in
the construction of boundary conditions. Using these fields, we also derive in
\ref{App:CharFields&SH} sufficient conditions for the non-linear evolution
system to be strongly hyperbolic and thus yield a well posed initial value
formulation.

%%%%%%%%%%%%%%%%%%%%%%%%%%%%%%%%%%%%%%%%%%%%%%%%%%%%%%%%%%%%%%%%%%%%
\section{Constraint propagation system}
\label{Sect:ESCV}
%%%%%%%%%%%%%%%%%%%%%%%%%%%%%%%%%%%%%%%%%%%%%%%%%%%%%%%%%%%%%%%%%%%%

In order to obtain a solution to Einstein's equations, not only do we have to
solve the evolution equations but also the constraints. We want to follow a
free evolution scheme, in which the constraints are solved initially only. For
this scheme to be valid, we have to guarantee that any solution for such
initial data automatically satisfies the constraints in the computational
domain everywhere and at every time. At the numerical level, since the
constraints are already violated initially due to truncation or roundoff
errors, we have to guarantee that the numerical solution to the evolution
equations converge to a constraint-satisfying solution to the continuum
equations in the limit of infinite resolution. In order to show this, the
constraint propagation system, which gives the change of the constraint
variables under the flux of the main evolution system, plays an important
role. In this section we show that for a suitable range of the parameters
$\gamma$, $\zeta$, $\eta$, $\chi$, $\xi$ this system can be cast into first
order symmetric hyperbolic form. We then specify boundary conditions that
guarantee that zero initial data for this system leads to zero constraint
variables at later times.

The constraint propagation system is derived in \ref{App:MECPS}, up to lower
order terms which are linear algebraic expressions in the constraint variables
and whose precise form are not needed for the purpose of this article. In
order to analyze under which conditions the system is symmetric hyperbolic, it
is convenient to decompose $C_{lkij}$ into its trace and trace-less parts,
\begin{displaymath}
C_{lkij} = E_{lkij} + \frac{1}{2} \left( g_{l(i} B_{j)k} - g_{k(i} B_{j)l} \right) 
 + \frac{1}{3}\, g_{ij} W_{lk}\, ,
\end{displaymath}
where $E_{lkij} = E_{[lk](ij)}$ is trace-less with respect to all pair of
indices and where $B_{ij}$ and $W_{ij}$ are determined by the traces $S_{ki}
\equiv C^s_{\; (ki)s}$, $A_{ki} \equiv C^s_{\; [ki]s}$ and $V_{lk} \equiv
C_{lks}^{\;\;\;\; s}$,
\begin{displaymath}
B_{ki} = \frac{4}{3}\, S_{ki} - \frac{12}{5}\, A_{ki} - \frac{4}{5}\, V_{ki}\, ,\qquad
W_{lk} = \frac{9}{5}\, V_{lk} + \frac{12}{5}\, A_{lk}\, .
\end{displaymath}
(Notice that $S_{ij}$ is symmetric trace-less while $A_{ij}$ and $V_{ij}$ are
antisymmetric.) In terms of the variables $U =
(C^{(A)}_i,C_{kij},E_{lkij},C,C_i,S_{ij} \hat{A}_{ij},V_{ij},C^{(A)}_{ij})$,
where $\hat{A}_{ij} = A_{ij} + V_{ij}/2 + C^{(A)}_{ij}$, the non-linear
constraint propagation system has the form
\begin{equation}
\partial_0 U = {\cal A}_C(u)^j\partial_j U + {\cal B}(u) U,
\label{Eq:EvolConstr}
\end{equation}
where the principal symbol ${\cal A}_C(u,n) = {\cal A}_C(u)^j n_j$ is given by
\begin{displaymath}
{\cal A}_C(u,n) U = \left( \begin{array}{c} 0 \\ 0 \\ 0 \\
 -\frac{1}{2}(2 + 2\chi - \eta) C_n \\
 -(2\gamma+1)n_i C + \zeta S_{ni} - \hat{A}_{ni} \\
 -\frac{3\eta}{4}\left[ n_{(i} C_{j)} - \frac{1}{3} g_{ij} C_n \right] \\
 \left[\frac{1}{4}(2\chi-\eta) + \xi \right]\, n_{[i} C_{j]} \\
 (\eta + 3\chi)\, n_{[i} C_{j]} \\
 \xi\; n_{[i} C_{j]}
\end{array} \right),
\end{displaymath}
and where the matrix ${\cal B}(u)$ depends on the main fields $u$ and their
spatial derivatives, but not on $U$. The system Eq. (\ref{Eq:EvolConstr}) is
called {\em symmetric hyperbolic} if there exists a symmetric positive
definite matrix ${\bf H}$ which may depend on $u$ but not on $n$ such that
${\bf H}{\cal A}_C(u,n)$ is symmetric for all $u$ and $n$. From the above
representation of the principal symbol it is not difficult to see that the
system is symmetric hyperbolic if the following conditions hold
\begin{eqnarray}
&& \left(2\gamma + 1 \right)(2 + 2\chi - \eta) > 0,
\label{Eq:ConstrSym1}\\
&& \zeta \eta < 0,
\label{Eq:ConstrSym2}\\
&& 2\chi - \eta + 4\xi < 0.
\label{Eq:ConstrSym3}
\end{eqnarray}
Notice that these conditions automatically imply that $\lambda_2 > 0$ and
$\lambda_3 > 0$, which are necessary conditions for the main evolution system
to be strongly hyperbolic. If the conditions
(\ref{Eq:ConstrSym1},\ref{Eq:ConstrSym2},\ref{Eq:ConstrSym3}) are satisfied, a
symmetrizer is given by the quadratic form
\begin{eqnarray}
U^T {\bf H}\, U &=& g^{ij} C^{(A)}_i C^{(A)}_j 
  + C^{kij} C_{kij} + E^{lkij} E_{lkij}
  + \hat{V}^{ij}\hat{V}_{ij} + \hat{C}^{ij}\hat{C}_{ij}
\nonumber\\
 &+& \frac{2(2\gamma+1)}{2+2\chi-\eta}\, C^2 + C^i C_i 
  + \frac{4}{3}\left| \frac{\zeta}{\eta} \right| S^{ij} S_{ij}
  + \frac{4}{|2\chi-\eta + 4\xi|}\, \hat{A}^{ij}\hat{A}_{ij}\; ,
\label{Eq:Sym}
\end{eqnarray}
where
\begin{eqnarray*}
\hat{A}_{ij} &=& A_{ij} + \frac{1}{2}\, V_{ij} + C^{(A)}_{ij}\; ,\\
\hat{V}_{ij} &=& V_{ij} - \frac{4(\eta+3\chi)}{(2\chi-\eta)+4\xi}\, \hat{A}_{ij}\; ,\\
\hat{C}_{ij} &=& C^{(A)}_{ij} - \frac{4\xi}{(2\chi-\eta)+4\xi}\, \hat{A}_{ij}\; .
\end{eqnarray*}

The symmetrizer allows us to obtain an energy estimate for solutions to Eq.
(\ref{Eq:EvolConstr}) on a domain $O$ of $\Real^3$ with smooth boundary
$\partial O$ and suitable boundary conditions on $\partial O$. Defining the
energy norm
\begin{equation}
E_{constraints} = \int_{O} U^T {\bf H}\, U\; d^3 x,
\end{equation}
differentiating with respect to $t$ and using the constraint propagation
system, Eq. (\ref{Eq:EvolConstr}), we obtain
\begin{eqnarray}
\frac{d}{dt} E_{constraints} 
 &=& 2\int_{O} U^T {\bf H} \left[ N({\cal A}_C^i\partial_i U + {\cal B}\, U) 
  + \beta^j\partial_i U \right]\; d^3 x
\nonumber\\
 &=& \int_{\partial O} U^T {\bf H} N{\cal A}_C(n) U \; d^2 x
\nonumber\\
 &+& \int_{O} U^T \left[ N {\bf H}{\cal B} + N {\cal B}^T {\bf H} 
  - \partial_i(N{\bf H}{\cal A}_C^i + {\bf H}\beta^i) \right] U\; d^3 x,
\end{eqnarray}
where here $n_i$ denotes the unit outward one-form to the boundary. In
the last step we have used the fact that ${\cal A}_C(n)$ is symmetric
with respect to the scalar product defined by (\ref{Eq:Sym}) and
assumed that the shift is tangential to the boundary at $\partial
O$. As one can easily verify,
\begin{equation}
U^T {\bf H}  {\cal A}_C(n) U
 = \frac{1}{2} g^{ij}\left( V^{(+)}_i V^{(+)}_j - V^{(-)}_i V^{(-)}_j \right),
\end{equation}
where $V^{(\pm)}_i = -C_i \pm (\hat{A}_{ni} - \zeta S_{ni}) \pm (2\gamma +
1)n_i C$ are the in $(+)$ and out $(-)$ going characteristic constraint
fields. Therefore, if we impose the boundary conditions
\begin{equation}
V^{(+)}_i = S_i^{\; j} V^{(-)}_j \; ,
\label{Eq:CPBC}
\end{equation}
where the matrix $(S_i^{\; j})$ satisfies $g^{ij} S_i^{\; k} S_j^{\; l} v_l
v_k \leq g^{kl} v_k v_l$ for all one-forms $v_k$, we obtain the energy
estimate
\begin{equation}
\frac{d}{dt} E_{constraints} \leq const\cdot E_{constraints}\; ,
\end{equation}
where the constant only depends on bounds for $N{\cal B}$ and ${\bf
H}^{-1}\partial_i(N{\bf H}{\cal A}_C^i + {\bf H}\beta^i)$. Since
$E_{constraints} \geq 0$ this proves that $E_{constraints}(t) = 0$ for
all $t > 0$ provided that $E_{constraints}(t=0) = 0$. For this reason,
we call the three conditions (\ref{Eq:CPBC}) {\it
constraint-preserving boundary conditions}.

%%%%%%%%%%%%%%%%%%%%%%%%%%%%%%%%%%%%%%%%%%%%%%%%%%%%%%%%%%%%%%%%%%%%
\section{Boundary conditions for the main evolution system}
\label{Sect:BC}
%%%%%%%%%%%%%%%%%%%%%%%%%%%%%%%%%%%%%%%%%%%%%%%%%%%%%%%%%%%%%%%%%%%%

In this section we consider the main evolution system, Eqs.
(\ref{Eq:Ndot},\ref{Eq:gdot},\ref{Eq:Kdot},\ref{Eq:ddot},\ref{Eq:Adot}),
on a open domain $O$ of $\Real^3$ with smooth boundary $\partial
O$. We also assume that the shift vector is chosen such that it is
tangential to $\partial O$ at the boundary. This means that at the
boundary we have six ingoing characteristic fields, denoted by
$v^{(+)}_{nn}$, $\hat{v}^{(+)}_{AB}$, $v^{(+)}_{AA}$, $v^{(+)}_{nB}$
(see \ref{App:CharFields&SH} for their definition; here and in the
following, $A,B = 2,3$, and quantities with a hat denote trace-free
two by two matrices), and thus we have to provide six independent
boundary conditions. Following the classification scheme of
Ref. \cite{CPRST} one can show that the first field is a gauge field,
the second ones are physical fields, and the last are
constraint-violating fields. We stress that this classification scheme
does only make precise sense in the linearized regime for plane waves
propagating in the normal direction to the boundary (see \cite{CPRST}
for a more detailed discussion about this).

If we forgot about the constraints, we could give data to the six
ingoing fields. The simplest possibility would be to freeze the
ingoing fields to their values given by the initial data. Provided the
evolution system is symmetric hyperbolic this would yield a well posed
IBVP. However, in the presence of constraints, the boundary conditions
have to ensure that no constraint-violating modes enter the boundary,
i.e. the boundary conditions have to be compatible with the
constraints. In fact, the numerical results of section \ref{Sect:Sim}
show explicitly that freezing of the ingoing fields to their initial
values does not, in general, provide a solution to Einstein's
equations: The constraints are violated.

Instead of the freezing non-constraint preserving boundary conditions
just mentioned, we impose the three constraint-preserving boundary
conditions (\ref{Eq:CPBC}). This fixes three of the six conditions we
are allowed to specify at $\partial O$. We complete these three
conditions in the following two ways:
\begin{enumerate}
\item {\bf CPBC without Weyl control}\\
  Here we adopt the simplest possibility and impose algebraic boundary
  conditions on the ``gauge'' and ``physical'' fields,
\begin{eqnarray}
v^{(+)}_{nn} &=& a\, v^{(-)}_{nn} + h,
\label{Eq:Gauge}\\
\hat{v}^{(+)}_{AB} &=& b\,\hat{v}^{(-)}_{AB} + \hat{h}_{AB}\; ,
\label{Eq:MaxDissModes}
\end{eqnarray}
where $a$ and $b$ are constants satisfying $|a| \leq 1$, $|b|\leq 1$,
and $h$ and $\hat{h}_{AB}$ are functions on $\partial O$ describing
the boundary data. The justification for the bounds on $a$ and $b$
will become clear in the next section. Notice that the choice $a=b=-1$
results in Dirichlet conditions in the sense that data is imposed on
some components of the extrinsic curvature while the choice $a=b=1$
yields boundary conditions that impose data on combinations of spatial
derivatives of the three metric (see the definitions of $v^{(+)}_{nn}$
and $\hat{v}^{(+)}_{AB}$ in \ref{App:CharFields&SH}). In our
simulations below, we choose $a=b=0$ which yields Sommerfeld-like
conditions.

\item {\bf CPBC with Weyl control}\\
  Here we replace Eq. (\ref{Eq:MaxDissModes}) by a similar condition for the
  Weyl tensor. We impose the boundary conditions
\begin{eqnarray}
v^{(+)}_{nn} &=& a\, v^{(-)}_{nn} + s,
\label{Eq:GaugeBis}\\
\hat{w}^{(+)}_{AB} &=& c\,\hat{w}^{(-)}_{AB} + \hat{H}_{AB}\; ,
\label{Eq:MaxDissWeyl}
\end{eqnarray}
where $|c|\leq 1$ and where $\hat{w}^{(\pm)}_{AB}$ are defined in terms of the
electric ($E_{ij}$) and magnetic ($B_{ij}$) parts of the Weyl tensor in the
following way: Let $e_1$, $e_2$, $e_3$ be an orthonormal triad with respect to
the three metric $g_{ij}$ such that $e_1$ coincides with the unit outward
normal $n^i$ to the boundary.  Then, $\hat{w}^{(\pm)}_{AB} = \left( e_A^{(i}
  e_B^{j)} - \delta_{AB} \delta^{CD} e_C^i e_D^j/2 \right) \left( E_{ij} \pm
  n^k\varepsilon_{kil} B_j^{\;\; l} \right)$, where $\varepsilon_{kij}$ is the
volume element associated to $g_{ij}$. If the vacuum equations hold, the
electric and magnetic components can be determined by
\begin{eqnarray}
E_{ij} &=& \partial_0 K_{ij} + K_{ia} K^a_{\;\; j} + \nabla_{(i} A_{j)} + A_i A_j\; ,\\
B_{ij} &=& \varepsilon_{rs(i}\nabla^r K^s_{\;\; j)}\; ,
\end{eqnarray}
where one uses the evolution equation (\ref{Eq:Kdot}) in order to
reexpress $\partial_0 K_{ij}$ in terms of spatial derivatives of the
main variables. The boundary conditions (\ref{Eq:MaxDissWeyl})
correspond to the conditions imposed by Friedrich and Nagy
\cite{FN}. In the symmetric hyperbolic system considered in \cite{FN},
where the components of the Weyl tensor are evolved as independent
fields, these boundary conditions arise naturally when analyzing the
structure of the equations since they give rise to maximally
dissipative boundary conditions. In particular, a well posed
initial-boundary value problem incorporating the condition
(\ref{Eq:MaxDissWeyl}) is derived in \cite{FN}. In contrast to this,
the boundary conditions (\ref{Eq:MaxDissWeyl}) are not maximally
dissipative for our symmetric hyperbolic system since they are not
even algebraic.

The conditions (\ref{Eq:MaxDissWeyl}) can also be expressed in terms of the
Newman-Penrose scalars $\Psi_0$ and $\Psi_4$ with respect to a null tetrad
which is adapted to the boundary in the following sense: Let $e_0$ denote the
future-directed unit normal to the $t=const$ slices. Together with the above
vectors $e_1$, $e_2$, $e_3$ it forms a tetrad. From this, we construct the
following Newman-Penrose null tetrad:
\begin{displaymath}
l^\mu = \frac{1}{\sqrt{2}} \left( e_0^\mu + e_1^\mu \right),\qquad
k^\mu = \frac{1}{\sqrt{2}} \left( e_0^\mu - e_1^\mu \right),\qquad
m^\mu = \frac{1}{\sqrt{2}} \left( e_2^\mu + i e_3^\mu \right).
\end{displaymath}
Then, we find
\begin{displaymath}
\Psi_0 = \hat{w}_{22}^{(+)} - i\hat{w}_{23}^{(+)}, \qquad
\Psi_4 = \hat{w}_{22}^{(-)} + i\hat{w}_{23}^{(+)},
\end{displaymath}
and the boundary condition (\ref{Eq:MaxDissWeyl}) is simply
\begin{equation}
\Psi_0 = c \Psi_4^*,
\label{Eq:MaxDissWeylBis}
\end{equation}
where the star denotes complex conjugation (we could generalize this boundary
condition by allowing for complex values of $c$). Notice that $\Psi_0$ and
$\Psi_4$ are not uniquely determined by the unit normals $e_0$ and $n$: with
respect to a rotation of $e_2$, $e_3$ about the angle $\phi$, these quantities
transform through $\Psi_0 \mapsto e^{2i\phi}\Psi_0$, $\Psi_4 \mapsto
e^{-2i\phi}\Psi_4$, the factor $2$ reflecting the spin of the graviton.
However, the boundary condition (\ref{Eq:MaxDissWeylBis}) is indeed invariant
with respect to such transformations. In our simulations below, we shall
choose $c=0$ corresponding to an outgoing radiation condition.

Finally, as discussed in the introduction, $\Psi_0$ and $\Psi_4$ represent,
respectively, in- and outgoing radiation when evaluated at null infinity and
are gauge-invariant quantities for linearizations about a Kerr background.
\end{enumerate}

%%%%%%%%%%%%%%%%%%%%%%%%%%%%%%%%%%%%%%%%%%%%%%%%%%%%%%%%%%%%%%%%%%%%
\section{Fourier-Laplace analysis}
\label{Sect:FL}
%%%%%%%%%%%%%%%%%%%%%%%%%%%%%%%%%%%%%%%%%%%%%%%%%%%%%%%%%%%%%%%%%%%%

We now analyze the well posedness of the IBVP defined by the evolution
equations
(\ref{Eq:Ndot},\ref{Eq:gdot},\ref{Eq:Kdot},\ref{Eq:ddot},\ref{Eq:Adot}),
the CPBC (\ref{Eq:CPBC}) and the boundary conditions (\ref{Eq:Gauge})
and (\ref{Eq:MaxDissModes}) or (\ref{Eq:Gauge}) and
(\ref{Eq:MaxDissWeyl}). We derive necessary conditions for well
posedness by verifying a certain determinant condition in the high
frequency limit. We assume that the parameters are such that the
evolution equations are strongly hyperbolic since otherwise the
problem is ill posed even in the absence of boundaries, and such that
the constraint propagation system is symmetric hyperbolic.

Let $p\in\partial O$ be a point on the boundary. By taking the high frequency
limit we obtain the associated frozen coefficient problem at $p$. After
rescaling and rotating the coordinates if necessary, we can achieve that $N(p)
= 1$, $g(p)_{ij} = \delta_{ij}$ and that the domain of integration is $x^1 >
0$. In this way, we obtain a constant coefficient problem on the half space.
Introducing the operator $\hat{\partial}_0 = \partial_t - \beta^i\partial_i$,
it is given by
\begin{eqnarray}
\hat{\partial}_0 K_{ij} &=& \frac{1}{2}\left( -\partial^k d_{kij} 
 + (1+\zeta )\partial^k d_{(ij)k} + (1-\zeta)\partial_{(i} b_{j)} 
 - \partial_{(i} d_{j)} -2\partial_{(i} A_{j)} \right)
\nonumber\\
 &+& \frac{\gamma}{2}\,\delta_{ij}\partial^k\left( b_k - d_k \right)
\label{Eq:LinKij}\\
%%%%%%%%%%%%%%%%%%%%%%%%%%%%%%%
\hat{\partial}_0 d_{kij} &=& -2\partial_k K_{ij} 
 + \eta\,g_{k(i}\left(\partial^l K_{j)l} - \partial_{j)} K \right) 
 + \chi\,g_{ij}\left( \partial^l K_{kl} - \partial_k K \right),
\label{Eq:Lindkij}\\
%%%%%%%%%%%%%%%%%%%%%%%%%%%%%%%
\hat{\partial}_0 A_i &=& -2\sigma\partial_i K + \xi\,\left( \partial^j K_{ij} - \partial_i K \right).
\label{Eq:LinAi}
\end{eqnarray}
Notice that this system is equivalent to the one that one would obtain by
linearizing the evolution equations around flat spacetime in a slicing with
respect to which the three metric is flat, the lapse is one and the shift is
constant and tangential to the boundary, but not necessarily zero.

Since we have a linear constant coefficient problem on the half plane,
we can solve these equations by means of a Laplace transformation in
time and a Fourier transformation in the $x^2$ and $x^3$ directions
\cite{GKO-Book,KL-Book}. That is, we write the solution as a
superposition of solutions of the form $u(t,x^1,x^2,x^3) =
\tilde{u}(\omega x^1) \exp(\omega(z t + i\hat{\omega}_A( x^A + \beta^A
t)))$, where $z\in\Complex$ with $\re(z) > 0$,
$\delta^{AB}\hat{\omega}_A\hat{\omega}_B = 1$, $A=2,3$ and
$\tilde{u}\in L^2(\Real_+)$. (Notice that for such solutions,
$\hat{\partial}_0 u = \omega z u$.)  Substituting this into
Eqs. (\ref{Eq:LinKij},\ref{Eq:Lindkij},\ref{Eq:LinAi}) one obtains a
system of ordinary differential equations coupled to algebraic
conditions.  Since there are six in- and six outgoing modes, there are
twelve independent differential equations. The remaining equations
which are algebraic can be used in order to eliminate the
characteristic fields which have zero speeds, and one ends up with a
closed system of twelve linear ordinary differential equations.
Because the system is strongly hyperbolic we expect \cite{Kreiss, MO}
exactly six linearly independent solutions that decay as
$x^1\rightarrow \infty$, and six solutions that blow up as
$x^1\rightarrow \infty$. Since we require the solution to lie in $L^2$
we only consider the six decaying solutions. The determinant condition
consists in verifying that the boundary conditions with homogeneous
boundary data annihilate these six solutions. If the determinant
condition is violated, the problem admits solutions of the form
$u(t,x^1,x^2,x^3) = \tilde{u}(\omega x^1) \exp(\omega(z t +
i\hat{\omega}_A( x^A + \beta^A t)))$ for some $\re(z) > 0$, where
$\omega$ can be arbitrarily large, and the system is ill posed since
$|u(t,x^1,x^2,x^3)|/|u(0,x^1,x^2,x^3)| = \exp(\omega\re(z) t)
\rightarrow \infty$ as $\omega\rightarrow\infty$ for each fixed
$t$. Thus, the determinant condition is necessary for the well
posedness of the IBVP and, as we will see, will yield nontrivial
conditions.

A convenient way for finding the six decaying solutions is to look at the
second order equation for $K_{ij}$, which is a consequence of Eqs.
(\ref{Eq:LinKij},\ref{Eq:Lindkij},\ref{Eq:LinAi}),
\begin{equation}
\hat{\partial}_0^2 K_{ij} = \partial^k\partial_k K_{ij} 
 + A\,\partial_{(i}\partial^k K_{j)k}
 + B\,\partial_i\partial_j K 
 + C\,\delta_{ij}\left(\partial^k\partial^l K_{kl} - \partial^l\partial_l K \right),
\label{Eq:SecondKij}
\end{equation}
where the coefficients $A$, $B$ and $C$ are given in Eqs.
(\ref{Eq:CoeffA},\ref{Eq:CoeffB},\ref{Eq:CoeffC}). We show below that there
are exactly six solutions to this second order system which have the form
$K_{ij}(t,x^1,x^2,x^3) = \tilde{K}_{ij}(\omega x^1) \exp(\omega(z t +
i\hat{\omega}_A( x^A + \beta^A t)))$, with $\re(z) > 0$ and such that
$\tilde{K}_{ij}$ decays as $x^1 \rightarrow \infty$.  Since in Fourier-Laplace
space the operator $\hat{\partial}_0$ is just multiplication with the nonzero
factor $\omega z$, corresponding solutions to the original system can be
obtained by determining $d_{kij}$ and $A_i$ from Eqs. (\ref{Eq:Lindkij}) and
(\ref{Eq:LinAi}), respectively.

In order to find the decaying solutions of Eq. (\ref{Eq:SecondKij}) we make
the ansatz $\tilde{K}_{ij} = k_{ij}\exp(\omega\nu x)$, where here and in the
following we set $x \equiv x^1$ for simplicity, and where $\nu$ is a complex
number with negative real part to be determined.  It is also convenient to
introduce a unit two-vector $\hat{\eta}^A$ which is orthogonal to
$\hat{\omega}_A$, and to decompose $k_{ij}$ in the components $k_{xx}$,
$k_{x\omega} = k_{xB}\hat{\omega}^B$, $k_{x\eta} = k_{xB}\hat{\eta}^B$, $k^A_A
= \delta^{AB} k_{AB}$, $k_{\omega\eta} = k_{AB}\hat{\omega}^A\hat{\eta}^B$ and
$\hat{k}_{\omega\omega} = (\hat{\omega}^A\hat{\omega}^B -
\hat{\eta}^A\hat{\eta}^B)k_{AB}/2$. Using this ansatz, Eq.
(\ref{Eq:SecondKij}) splits into the following two decoupled systems,
\begin{equation}
z^2\left( \begin{array}{c} k_{x\eta} \\ k_{\omega\eta} \end{array} \right)
 = M_1(\nu)\left( \begin{array}{c} k_{x\eta} \\ k_{\omega\eta} \end{array} \right),
\qquad
z^2\left( \begin{array}{c} k_{xx} \\ k^A_A \\ k_{x\omega} \\ \hat{k}_{\omega\omega} \end{array} \right)
 = M_2(\nu)\left( \begin{array}{c} k_{xx} \\ k^A_A \\ k_{x\omega} \\ \hat{k}_{\omega\omega} \end{array} \right),
\end{equation}
where the matrices $M_1(\nu)$ and $M_2(\nu)$ are given by
\begin{eqnarray}
&& M_1(\nu) = \left( \begin{array}{cc}
  \lambda_3\nu^2-1 & \frac{i}{2}\, A\nu \\
  \frac{i}{2}\, A\nu & \nu^2-1-\frac{A}{2}
\end{array} \right),
\nonumber\\
&& M_2(\nu) = \left( \begin{array}{cccc} 
   \lambda_1\nu^2 + C-1 & (B-C)\nu^2+\frac{C}{2} & i(A+2C)\nu & -C \\
   2C-B & \lambda_2\nu^2 - \frac{A}{2}-B+C-1 & i(A+4C)\nu & -(A+2C) \\
   \frac{i}{2}(A+2B)\nu & \frac{i}{4}(A+4B)\nu & \lambda_3\nu^2 - 1-\frac{A}{2} & \frac{i}{2}\, A\nu \\
   -\frac{B}{2} & -\frac{1}{4}(A+2B) & \frac{i}{2}A\nu & \nu^2-1-\frac{A}{2}
\end{array} \right),
\nonumber
\end{eqnarray}
and $\lambda_1$, $\lambda_2$, $\lambda_3$ are defined in Eqs.
(\ref{Eq:lam1},\ref{Eq:lam2},\ref{Eq:lam3}).

The matrix $M_1(\nu)$ has the eigenvalue-eigenvector pair
\begin{eqnarray}
z^2 = \nu^2-1,  && 
\left( \begin{array}{c} k_{x\eta} \\ k_{\omega\eta} \end{array} \right)
 = \left( \begin{array}{c} 1 \\ i\nu \end{array} \right),
\nonumber\\
z^2 = \lambda_3(\nu^2-1),  && 
\left( \begin{array}{c} k_{x\eta} \\ k_{\omega\eta} \end{array} \right)
 = \left( \begin{array}{c} -i\nu \\ 1 \end{array} \right).
\nonumber
\end{eqnarray}
The two vectors are always linearly independent from each other since $\re(z)
> 0$. This yields the solution
\begin{equation}
\left( \begin{array}{c} \tilde{K}_{x\eta} \\ \tilde{K}_{\omega\eta} \end{array} \right)
 = \sigma_1\left( \begin{array}{c} 1 \\ -i\nu_4 \end{array} \right)  e^{-\omega\nu_4 x}
 + \sigma_2\left( \begin{array}{c} i\nu_3 \\ 1 \end{array} \right) e^{-\omega\nu_3 x},
\label{Eq:K1}
\end{equation}
where $\sigma_1$, $\sigma_2$ are two constants and where here and in
the following $\nu_l = \sqrt{\lambda_l^{-1}z^2+1}$, $l=1,2,3,4$
($\lambda_4=1$) where the branch of the square root for which
$\re(\nu_l) > 0$ for $\re(z) > 0$ is chosen. Similarly, after
obtaining the eigenvalues and eigenvectors of $M_2(\nu)$ one obtains
the solution
\begin{equation}
\left( \begin{array}{c} \tilde{K}_{xx} \\ \tilde{K}^A_A \\ 
   \tilde{K}_{x\omega} \\ \hat{\tilde{K}}_{\omega\omega} \end{array} \right)
 = \sigma_3 v_3\, e^{-\omega\nu_4 x} + \sigma_4 v_4\, e^{-\omega\nu_3 x}
 + \sigma_5 v_5\, e^{-\omega\nu_2 x} + \sigma_6 v_6\, e^{-\omega\nu_1 x},
\label{Eq:K2}
\end{equation}
where
\begin{eqnarray}
&& v_3 = \left( \begin{array}{c} 1 \\ -1 \\ -i\nu_4 \\ \frac{1}{2} - \nu_4^2 \end{array} \right), \qquad
   v_4 = \left( \begin{array}{c} -i\nu_3 \\ i\nu_3 \\ -\frac{1}{2}(\nu_3^2+1) \\ \frac{i}{2}\nu_3 \end{array} \right),
\qquad 
   v_6 = \left( \begin{array}{c} \nu_1^2 \\ -1 \\ -i\nu_1 \\ -\frac{1}{2} \end{array} \right),
\nonumber\\
&& v_5 = \left( \begin{array}{c} 2(B-C)\nu_2^2 + A+B+2C \\ -2(A+B+2C)\nu_2^2+A-B+4C \\ -i\nu_2(A+3B) \\ -\frac{1}{2}(A+3B) \end{array} \right)
\quad\hbox{if $\lambda_1 \neq \lambda_2$,  }
   v_5 = \left( \begin{array}{c} 1 \\ 2 \\ 0 \\ 0 \end{array} \right) \quad\hbox{if $\lambda_1 = \lambda_2$.}
\nonumber
\end{eqnarray}
Therefore, we have obtained six linearly independent solutions which
decay exponentially as $x\rightarrow\infty$ and thus lie in $L^2(O)$.
They are parameterized by the constants (which depend on $\omega$ and
$z$) $\sigma_1$,...,$\sigma_6$. A necessary condition for the IBVP to
be well posed is that these constants are uniquely determined by the
boundary data. Before checking this condition, it is instructive to
have a closer look at the six-parameter family of solutions given by
Eqs. (\ref{Eq:K1}) and (\ref{Eq:K2}).  Let us first compute the
momentum constraint variable $C_i = \partial^j K_{ij} - \partial_i K$:
It has the form $C_i(t,x^1,x^2,x^3) = \tilde{C}_i(\omega x^1)
\exp(\omega(z t + i\hat{\omega}_A( x^A + \beta^A t)))$ where
\begin{eqnarray}
\tilde{C}_\eta &=& i\omega(1-\nu_3^2)\sigma_2\, e^{-\omega\nu_3 x},
\label{Eq:Ceta}\\
\tilde{C}_x &=& -\frac{i}{2}\omega(1-\nu_3^2)\sigma_4\, e^{-\omega\nu_3 x}
 + 2\omega\nu_2\tilde{\sigma_5}\, e^{-\omega\nu_2 x}, 
\label{Eq:Cx}\\
\tilde{C}_\omega &=& -\frac{1}{2}\omega\nu_3(1-\nu_3^2)\sigma_4\, e^{-\omega\nu_3 x}
 -2i\omega\tilde{\sigma_5}\, e^{-\omega\nu_2 x},
\label{Eq:Comega}
\end{eqnarray}
where $\tilde{\sigma}_5 = (1-\nu_2^2)(\lambda_1-\lambda_2)\sigma_5$ if
$\lambda_1 \neq \lambda_2$ and $\tilde{\sigma}_5 = \sigma_5$ otherwise. Thus,
$C_i = 0$ for this family of solutions if and only if $\sigma_2 = \sigma_4 =
\sigma_5 = 0$. In other words, the three-parameter subfamily of solutions
parametrized by $\sigma_2$, $\sigma_4$ and $\sigma_5$ are {\em
  constraint-violating} modes.  Next, consider an infinitesimal coordinate
transformation parametrized by a vector field $(X^\mu) = (f,X^i)$, and assume
zero shift for simplicity. With respect to such a transformation, the
linearized lapse and extrinsic curvature change according to
\begin{eqnarray}
&& N \mapsto N + \partial_t f,\\
&& K_{ij} \mapsto K_{ij} - \partial_i\partial_j f.
\end{eqnarray}
On the other hand, the linearization of Eq.(\ref{Eq:Ndot}) around a
Minkowski background\footnote{Here we also assume that
$F(N,K=0,x^\mu)=0$ for all $N$, which is satisfied by the
time-harmonic slicing condition adopted in our simulations.} yields
\begin{equation}
\partial_t N = -2\sigma K.
\label{Eq:NdotLin}
\end{equation}
We see that the choice $f(t,x^1,x^2,x^3) = \omega^{-2} \exp(\omega(z t
-\nu_1 x + i\hat{\omega}_A x^A))$ leaves Eq. (\ref{Eq:NdotLin})
invariant and induces the transformation
\begin{equation}
\left( \begin{array}{c} \tilde{K}_{xx} \\ \tilde{K}^A_A \\ \tilde{K}_{x\omega} \\ \hat{\tilde{K}}_{\omega\omega} \end{array} \right)
\mapsto
\left( \begin{array}{c} \tilde{K}_{xx} \\ \tilde{K}^A_A \\ \tilde{K}_{x\omega} \\ \hat{\tilde{K}}_{\omega\omega} \end{array} \right)
 - \left( \begin{array}{c} \nu_1^2 \\ -1 \\ -i\nu_1 \\ -\frac{1}{2} \end{array} \right) e^{-\omega\nu_1 x},
\end{equation}
while $\tilde{K}_{x\eta}$ and $\tilde{K}_{\omega\eta}$ remain invariant.
Therefore, it is possible to gauge away the solution parametrized by
$\sigma_6$ and we call this solution a {\em gauge} mode from hereon. The
remaining family of solutions parametrized by $\sigma_1$ and $\sigma_3$ are
{\em physical} modes: They satisfy the constraints and $\sigma_1$ and
$\sigma_3$ are gauge-invariant.

Next, we verify that the integration constants $\sigma_1$,...,$\sigma_6$ are
uniquely determined by the boundary conditions. First, we notice that the
expressions (\ref{Eq:Ceta},\ref{Eq:Cx},\ref{Eq:Comega}) for the
Fourier-Laplace transformation of the momentum constraint yield a
three-parameter family of solutions for the constraint propagation system, Eq.
(\ref{Eq:EvolConstr}). Since this system is symmetric hyperbolic and since we
specify homogeneous maximally dissipative boundary conditions for it (see Eq.
(\ref{Eq:CPBC})), the corresponding IBVP is well posed. In particular, zero is
the only solution with trivial initial data. This implies that $\sigma_2 =
\sigma_4 = \sigma_5 = 0 $\footnote{This can also be verified directly by
  introducing the expressions (\ref{Eq:Ceta},\ref{Eq:Cx},\ref{Eq:Comega}) into
  the Fourier-Laplace transformed of the CPBC (\ref{Eq:CPBC}).}. We stress
that such a conclusion cannot be drawn if the constraint propagation is
strongly but not symmetric hyperbolic, see Ref. \cite{CS} for a
counterexample.

Next, using Eqs. (\ref{Eq:Lindkij}) and (\ref{Eq:LinAi}), we find the
following expressions for the relevant characteristic fields at the boundary
\begin{eqnarray}
v^{(\pm)}_{xx} &=& K_{xx} + \Omega K^A_A \mp \frac{1}{\sqrt{\lambda_1}\,\omega z}\left[ 
 \lambda_1\partial_x K + \zeta(1-\Omega)\partial^A K_{Ax} + \lambda_1(1-\Omega)C_x \right],\\
\hat{v}^{(\pm)}_{AB} &=& \hat{K}_{AB} \mp \frac{1}{\omega z}\left[ 
  \partial_x\hat{K}_{AB} - (1+\zeta)\partial_{(A} K_{B)x} \right]^{TF},\\
\hat{w}^{(\pm)}_{AB} &=& \omega z\hat{K}_{AB} \mp \left[ \partial_x\hat{K}_{AB} 
 - \partial_{(A} K_{B)x} \right]^{TF} 
 + \frac{1}{\omega z}\left[ -2\sigma\partial_A\partial_B K + \xi\partial_{(A} C_{B)} \right]^{TF},\\
\end{eqnarray}
where
\begin{equation}
\Omega = \frac{1 + 2\lambda_1 + \lambda_2 - 4\lambda_3}{2(\lambda_1-\lambda_2)}\, ,
\quad \hbox{if $\lambda_1\neq\lambda_2$ and $\Omega$ arbitrary otherwise,}
\label{Eq:Omega}
\end{equation}
and where $[...]^{TF}$ denotes the trace-free part with respect to the metric
$\delta_{AB}$. Plugging into this the six-parameter family of solutions given
in Eqs. (\ref{Eq:K1}) and (\ref{Eq:K2}), taking into account the vanishing of
the constraint-violating modes, $\sigma_2=\sigma_4=\sigma_5=0$, and evaluating
at $x=0$, we obtain
\begin{eqnarray}
\tilde{v}^{(\pm)}_{xx} &=& \frac{z}{\lambda_1}\left[ z \pm \sqrt{z^2 + \lambda_1} \right] \sigma_6
 + (1-\Omega)\left[ \sigma_3 + \sigma_6 \mp \frac{\zeta}{\sqrt{\lambda_1}\, z}\left( \nu_4\sigma_3 + \nu_1\sigma_6 \right) \right],
\label{Eq:vxx}\\
\tilde{\hat{v}}^{(\pm)}_{\omega\omega} &=& -\left[ \left(z^2 + \frac{1}{2}\right) \pm
  \left( z - \frac{\zeta}{2z} \right)\sqrt{z^2+1} \right]\sigma_3
 - \frac{1}{2z}\left[ z \mp \nu_1\zeta \right]\sigma_6\;  ,
\\
\tilde{\hat{w}}^{(\pm)}_{\omega\omega} &=& -\omega z\left[ \left(z^2+\frac{1}{2}\right) \pm z\sqrt{z^2+1} \right]\sigma_3\; ,
\label{Eq:hatwww}\\
\tilde{v}^{(\pm)}_{\omega\eta} &=& -\frac{i}{z}\left[ z\sqrt{z^2+1} \pm 
  \left( z^2 + \frac{1}{2}(1-\zeta) \right) \right]\sigma_1\; ,
\\
\tilde{w}^{(\pm)}_{\omega\eta} &=& -i\omega\left[ z\sqrt{z^2+1} \pm 
  \left( z^2 + \frac{1}{2} \right) \right]\sigma_1\; .
\label{Eq:hatwwe}
\end{eqnarray}
In particular, the fields $\tilde{\hat{w}}^{(\pm)}_{\omega\omega}$ and
$\tilde{w}^{(\pm)}_{\omega\eta}$ are gauge-invariant. This is to be expected
since these fields are linear combinations of the components of the Weyl
tensor, which is a gauge-invariant quantity in the linearized regime. The
fields $\tilde{v}^{(\pm)}_{\omega\eta}$ are gauge-invariant too, but the
remaining fields depend on the constant $\sigma_6$ which represents a pure
gauge degree of freedom. In particular, if the parameters of the formulation
are such that one can choose $\Omega=1$, $\tilde{v}^{(\pm)}_{xx}$ are pure
gauge fields.  Now it is not difficult to check the determinant condition for
the different boundary conditions proposed in the previous section.

\subsection{CPBC with Weyl control}

Here we analyze the family of boundary conditions
(\ref{Eq:Gauge},\ref{Eq:MaxDissWeyl}), where we set the boundary data $h$ and
$\hat{H}_{AB}$ to zero. We first analyze the boundary conditions
$\tilde{\hat{w}}^{(+)}_{AB} = c\,\tilde{\hat{w}}^{(-)}_{AB}$, where $|c|\leq
1$: Plugging into this the expressions from Eqs.
(\ref{Eq:hatwww},\ref{Eq:hatwwe}), we obtain
\begin{eqnarray}
&& \left[ (z + \sqrt{z^2+1})^2 + c (z - \sqrt{z^2+1})^2 \right]\sigma_1 = 0,
\label{Eq:sigma1}\\
&& \left[ (z + \sqrt{z^2+1})^2 - c (z - \sqrt{z^2+1})^2 \right]\sigma_3 = 0.
\label{Eq:sigma3}
\end{eqnarray}
Since the function $z\mapsto (z + \sqrt{z^2+1})^2$ maps $\re(z) > 0$ onto the
{\em outside} of the unit disk minus the negative real axis, while the
function $z\mapsto (z - \sqrt{z^2+1})^2$ maps $\re(z) > 0$ onto the {\em
  inside} of the unit disk minus the negative real axis, and since $|c|\leq
1$, it follows that the gauge-invariant constants $\sigma_1$ and $\sigma_3$
vanish. Now it also becomes clear why we restricted the range of the parameter
$c$: If $|c| > 1$, there are nontrivial exponential growing solutions with
either $\sigma_1\neq 0$ or $\sigma_3 \neq 0$, and the system is ill posed.

In order to analyze the consequences of the boundary condition
$\tilde{v}^{(+)}_{xx} = a\,\tilde{v}^{(-)}_{xx}$, we assume that either
$\Omega=1$ (in which case $\tilde{v}^{(\pm)}_{xx}$ are pure gauge fields) or
$\zeta=-\lambda_1$. In both cases we obtain, using $\sigma_3=0$ and Eq.
(\ref{Eq:vxx}),
\begin{displaymath}
\left[ z^2 + \lambda_1(1-\Omega) \right]\left[ (1-a)z + (1+a)\sqrt{z^2+\lambda_1} \right]\sigma_6 = 0.
\end{displaymath}
The term in the second bracket is non-vanishing for $\re(z) > 0$ if and only
if $|a|\leq 1$ (see Lemma 1 of Ref. \cite{RS} for a proof). The term in the
first bracket never vanishes provided that $\Omega \leq 1$.

Summarizing, the CPBC with Weyl control and a choice of parameters
that allows for $\Omega=1$ or such that $\zeta = -\lambda_1$ and
$\Omega\leq 1$ does always yield an initial-boundary value formulation
that satisfies the determinant condition, as long as the main
evolution system is strongly hyperbolic and the constraint propagation
system is symmetric hyperbolic (see sections \ref{Sect:SHLG} and
\ref{Sect:ESCV}) .

\subsection{CPBC without the Weyl control}

Next, we analyze the family of boundary conditions
(\ref{Eq:Gauge},\ref{Eq:MaxDissModes}) where we set the boundary data $h$ and
$\hat{h}_{AB}$ to zero. In this case, the result is less robust and depends
more strongly on the choice of the parameters. For example, assume again that
$\Omega=1$ which implies that $\sigma_6=0$.  Now Eq. (\ref{Eq:sigma3}) has to
be replaced by
\begin{displaymath}
\left[ (1+b)(2z^2 - \zeta)\sqrt{z^2+1} + (1-b)z(2z^2+1) \right]\sigma_3 = 0.
\end{displaymath}
By taking $z$ real and positive and by taking the limits $z \rightarrow 0$ and
$z \rightarrow\infty$ one sees that the expression inside the square bracket
changes its sign if $-1 < b \leq 1$ and $\zeta > 0$. In this case it does not
follow that $\sigma_3$ is zero and the system suffers from the presence of ill
posed modes.

For simplicity, let us assume that $\lambda_1=\lambda_2=\lambda_3=1$ and that
$\zeta=-1$. All of the simulations below will satisfy these conditions. In
those cases, we obtain
\begin{eqnarray}
f_b(z)\sqrt{z^2+1}\;\sigma_1 = 0,\\
f_b(z)\left[ (2z^2+1)\sigma_3 + \sigma_6 \right] = 0,\\
f_a(z)\left[ (1-\Omega)(\sigma_3 + \sigma_6) + z^2\sigma_6 \right] = 0,
\end{eqnarray}
where $f_a(z) = (1-a)z + (1+a)\sqrt{z^2+1}$. Since $|a|\leq 1$, $|b|\leq 1$,
$f_a$ and $f_b$ never vanish for $\re(z) > 0$ and the only condition we obtain
is $\Omega \leq 3/2$.

%%%%%%%%%%%%%%%%%%%%%%%%%%%%%%%%%%%%%%%%%%%%%%%%%%%%%%%%%%%%%%%%%%%%
\section{Numerical implementation}
\label{Sect:NR}
%%%%%%%%%%%%%%%%%%%%%%%%%%%%%%%%%%%%%%%%%%%%%%%%%%%%%%%%%%%%%%%%%%%%

In this section we discuss how to numerically implement the IBVP with or
without Weyl control. For quasilinear first order symmetric hyperbolic systems
with maximally dissipative boundary conditions there are well known methods
for discretizing the problem such that numerical stability is guaranteed at
the linearized level (see
\cite{numerics-Let,numerics1,numerics2,TLN-BlackHole} for a recent application
in the context of numerical relativity, and references therein for the
original work). Unfortunately, in our case, the boundary conditions are more
complicated since the CPBC (\ref{Eq:CPBC}) and the conditions
(\ref{Eq:MaxDissWeyl}) which control the Weyl tensor are not even algebraic,
and so we cannot apply this methods for discretizing the boundary conditions.
So we first describe our method for implementing the constraint-preserving
boundary conditions, and then explain how the differential equations are
discretized.

For simplicity, we focus on the case where the coupling coefficients
$S_i^j$, $a$, $b$ and $c$ vanish, which corresponds to Sommerfeld-like
conditions, as discussed in Section \ref{Sect:BC}. The boundary
conditions (\ref{Eq:CPBC}) and (\ref{Eq:MaxDissWeyl}) are implemented
in the following way: For all points that lie on the boundary we add
to the right-hand side of the evolution equations terms that are
linear in the quantities $V_i^{(+)}$ and $\hat{w}_{AB}^{(+)} -
\hat{H}_{AB}$ (which are zero if the boundary conditions are
satisfied). More precisely, we replace the right-hand side of
equations (\ref{Eq:Kdot},\ref{Eq:ddot},\ref{Eq:Adot}) by
\begin{eqnarray}
\partial_0 K_{ij} &=& (\hbox{as before}) + p_{ij}^a V_a^{(+)} + q_{ij}^{AB}(\hat{w}_{AB}^{(+)} - \hat{H}_{AB}),
\label{Eq:Kdotmod}\\
\partial_0 d_{kij} &=& (\hbox{as before}) + p_{kij}^a V_a^{(+)} + q_{kij}^{AB}(\hat{w}_{AB}^{(+)} - \hat{H}_{AB}),
\label{Eq:ddotmod}\\
\partial_0 A_{i} &=& (\hbox{as before}) + p_{i}^a V_a^{(+)} + q_{i}^{AB}(\hat{w}_{AB}^{(+)} - \hat{H}_{AB}),
\label{Eq:Adotmod}
\end{eqnarray}
where the matrix coefficients $p_{ij}^a$, $q_{ij}^{AB}$, etc. are
allowed to depend on the three metric and the unit outward normal
one-form $n_i$ to the boundary. (The $q's$ are set to zero for the
case without Weyl control.)  It is clear that these extra terms change
the principal part of the equations.  The idea is to choose the $p$'s
and $q's$ in such a way that with respect to the unit outward normal
$n_i$ to the boundary the ingoing characteristic fields
$v_{AA}^{(+)}$, $v_{nA}^{(+)}$ and $\hat{v}_{AB}^{(+)}$ become zero
speed fields while the speed of the other fields remains
unchanged. This has the effect of eliminating the normal derivatives
in the evolution equations for $v_{AA}^{(+)}$, $v_{nA}^{(+)}$ and
$\hat{v}_{AB}^{(+)}$, and hence these equations are {\em intrinsic} to
the boundary. We will see that this requirement uniquely determines
the $p$'s and $q$'s. In order to see this we first notice that the
ingoing characteristic constraint fields have the form
\begin{eqnarray}
V_n^{(+)} &=& \frac{1}{2}\frac{\partial}{\partial n} 
  \left[ (1+\sqrt{\lambda_2}) v_{AA}^{(+)} + (1-\sqrt{\lambda_2}) v_{AA}^{(-)} \right]
 + {\cal Q}( \partial_{||} u, u), \\
V_A^{(+)} &=& -\frac{1}{2}\frac{\partial}{\partial n}
  \left[ (1+\sqrt{\lambda_3}) v_{nA}^{(+)} + (1-\sqrt{\lambda_3}) v_{nA}^{(-)} \right]
 + {\cal Q}_A( \partial_{||} u, u), \\
\hat{w}_{AB}^{(+)} &=& \frac{\partial}{\partial n}\, \hat{v}_{AB}^{(+)}
 + {\cal Q}_{AB}( \partial_{||} u, u),
\end{eqnarray}
where $\partial/\partial n$ denotes the normal derivative to the boundary and
where the expressions ${\cal Q}$, ${\cal Q}_A$ and ${\cal Q}_{AB}$ only depend
on the variables $u$ and their derivatives $\partial_{||} u$ tangential to the
boundary. Therefore, the principal symbol corresponding to the modified system
(\ref{Eq:Kdotmod},\ref{Eq:ddotmod},\ref{Eq:Adotmod}) and the direction of the
unit outward normal $n_i$ is given by
\begin{eqnarray}
\mu\, v_{nn}^{(\pm)} &=& \pm\sqrt{\lambda_1}\, v_{nn}^{(\pm)}
 + p_{nn}^{(\pm)a} \tilde{V}_a^{(+)} + q_{nn}^{(\pm)CD}\hat{v}_{CD}^{(\pm)}, \\
\mu\, v_{AA}^{(\pm)} &=& \pm\sqrt{\lambda_2}\, v_{AA}^{(\pm)}
 + p_{AA}^{(\pm)a} \tilde{V}_a^{(+)} + q_{AA}^{(\pm)CD}\hat{v}_{CD}^{(\pm)}, \\
\mu\, v_{nA}^{(\pm)} &=& \pm\sqrt{\lambda_3}\, v_{nA}^{(\pm)}
 + p_{nA}^{(\pm)a} \tilde{V}_a^{(+)} + q_{nA}^{(\pm)CD}\hat{v}_{CD}^{(\pm)}, \\
\mu\,\hat{v}_{AB}^{(\pm)} &=& \pm \hat{v}_{AB}^{(\pm)}
 + \hat{p}_{AB}^{(\pm)a} \tilde{V}_a^{(+)} + \hat{q}_{AB}^{(\pm)CD}\hat{v}_{CD}^{(\pm)}, \\
\mu\, v_{Ann}^{(0)} &=& 0
 + p_{Ann}^{(0)a} \tilde{V}_a^{(+)} + q_{Ann}^{(0)CD}\hat{v}_{CD}^{(\pm)}, \\
...
\end{eqnarray}
where $p_{nn}^{(\pm)}$, ... $p_{Ann}^{(0)}$ and $q_{nn}^{(\pm)}$, ...
$q_{Ann}^{(0)}$ are defined in terms of $(p_{ij}^a,p_{kij}^a,p_i^a)$ and
$(q_{ij}^a,q_{kij}^a,q_i^a)$ in the same way as $v_{nn}^{(\pm)}$, ...
$v_{Ann}^{(0)}$ in terms of $(K_{ij}, d_{kij}, A_i)$, and where
\begin{eqnarray}
\tilde{V}_n^{(+)} &=& \frac{1}{2}\left[ (1+\sqrt{\lambda_2}) v_{AA}^{(+)} + (1-\sqrt{\lambda_2}) v_{AA}^{(-)} \right],\\
\tilde{V}_A^{(+)} &=& -\frac{1}{2}\left[ (1+\sqrt{\lambda_3}) v_{nA}^{(+)} + (1-\sqrt{\lambda_3}) v_{nA}^{(-)} \right].
\end{eqnarray}
Assume first that $\lambda_2 = \lambda_3 = 1$. In this case, we see
immediately that $v_{nn}^{(\pm)}$, ... $v_{Ann}^{(0)}$ can only remain
characteristic fields (i.e. fields with respect to which the principal symbol
is diagonal) if all $p$'s and $q$'s are zero except for $p_{AA}^{(+)n}$,
$p_{nA}^{(+)B}$ and $\hat{q}_{AB}^{(+)CD}$. Then, the choice
\begin{eqnarray}
p_{AA}^{(+)n} V_n^{(+)} &=& -V_n^{(+)},\\
p_{nA}^{(+)B} V_B^{(+)} &=& V_A^{(+)},\\
\hat{q}_{AB}^{(+)CD}\hat{w}_{CD}^{(+)} &=& -\hat{w}_{AB}^{(+)}
\end{eqnarray}
yields zero speeds for the variables $v_{AA}^{(+)}$, $v_{nA}^{(+)}$ and
$\hat{v}_{AB}^{(+)}$. The matrix coefficients $p_{ij}^a$, ... $q_i^{AB}$ are
easily obtained from this by applying the inverse transformation to the one
that defines the characteristic fields in terms of the main variables (see
\ref{App:CharFields&SH}).  If $\lambda_2 \neq 1$ or $\lambda_3 \neq 1$ it is
not possible to retain the fields $v_{AA}^{(+)}$, $v_{nA}^{(+)}$ as
characteristic fields. In this case, we replace $v_{AA}^{(+)}$, $v_{nA}^{(+)}$
by $\bar{v}_{AA}^{(+)} \equiv \lambda_2 K^A_A + D^A_A$ and $\bar{v}_{nA}^{(+)}
\equiv \lambda_3 K_{nA} + D_{nA}$, respectively, where $D^A_A$ and $D_{nA}$
are given in \ref{App:CharFields&SH}, and set
\begin{eqnarray}
p_{AA}^{(+)n} V_n^{(+)} &=& \frac{-2\sqrt{\lambda}_2}{\sqrt{\lambda}_2 + 1}\, V_n^{(+)},\\
p_{nA}^{(+)B} V_B^{(+)} &=& \frac{2\sqrt{\lambda}_3}{\sqrt{\lambda}_3 + 1}\, V_A^{(+)},\\
\hat{q}_{AB}^{(+)CD}\hat{w}_{CD}^{(+)} &=& -\hat{w}_{AB}^{(+)},
\end{eqnarray}
and all other $p$'s and $q$'s to zero. This implies $\mu\,\bar{v}_{AA}^{(+)} =
0$, $\mu\,\bar{v}_{nA}^{(+)} = 0$, $\mu \hat{w}_{AB}^{(+)} = 0$, so
$\bar{v}_{AA}^{(+)}$, $\bar{v}_{nA}^{(+)}$ and $\hat{w}_{AB}^{(+)}$ are zero
speed fields while the speeds of the remaining characteristic fields are
unchanged.

Next, we discretize the evolution system
(\ref{Eq:Ndot},\ref{Eq:gdot},\ref{Eq:Kdotmod},\ref{Eq:ddotmod},\ref{Eq:Adotmod})
using the method of lines. The matrices $p$ and $q$ are set to zero at
interior points and are chosen as described above at boundary points.
The spatial discretization uses difference operators satisfying
summation by parts (SBP) (see, for example, \cite{GKO-Book}). In this
paper we use two of these difference operators, which we call D2-1 and
D8-4. These operators satisfy SBP with respect to diagonal norms; it
can be seen \cite{kreiss_scherer} that the use of these kind of norms
implies that the order of the operator at and near boundary points is
half that one in the interior.
 
D2-1 is a simple operator which is second order accurate in the interior and
first order accurate at boundaries. If the gridpoints range from $0$ to $N$
and the gridspacing is denoted by $\Delta x$, this operator is given by
\begin{eqnarray}
Du_0 &=& \frac{1}{\Delta x}\left( u_1 - u_0 \right),
\nonumber\\
Du_i &=& \frac{1}{2 \Delta x}\left( u_{i+1} - u_{i-1} \right)  \; \mbox{ for } i=1,\ldots,N-1,
\nonumber \\
Du_N &=& \frac{1}{\Delta x}\left( u_N - u_{N-1} \right).
\nonumber
\end{eqnarray}
We typically add Kreiss-Oliger like dissipation \cite{kreiss_oliger} modified
at and near boundaries such that the resulting operator is negative definite
with respect to the norm for which SBP holds (much in the way Kreiss-Oliger
dissipation is negative definite in the absence of boundaries). This operator,
which we call $Q_d$, is added to the right-hand side of the evolution
equations with a free (nonnegative) multiplicative parameter $\sigma_{diss}$
(see \cite{numerics1} for more details):
\begin{eqnarray} 
Q_d u_0 &=& -2\sigma_{diss} \Delta x D_+^2 u_0,\nonumber\\ 
Q_d u_1 &=& -\sigma_{diss} \Delta x (D_+^2 -2D_+D_-)u_1,\nonumber\\ 
Q_d u_i &=& -\sigma_{diss} (\Delta x)^3(D_+D_-)^2 u_i,
           \; \mbox{ for } i=2,\ldots,N-2, \label{eq:KOdiss21}\\
Q_d u_{N-1} &=& -\sigma_{diss} \Delta x (D_-^2 -2D_+D_-)u_{N-1}, \nonumber\\ 
Q_d u_N &=& -2\sigma_{diss} \Delta x D_-^2 u_N. \nonumber
\end{eqnarray} 
D8-4 is one of Strand's three-parametric difference operators of order
eight in the interior and four at and close to boundaries, chosen in
Ref. \cite{multipatch} to minimize its spectral radius and therefore
its associated Courant limit. We have not been able to extend the
associated Kreiss-Oliger dissipation for this operator in a
straightforward way such as to include the presence of boundaries (see
\cite{multipatch} for more details). Therefore, here, we simply set the
dissipative operator to zero near boundaries. However, as we will
discuss later, the resulting dissipation operator does not seem to
yield a stable numerical scheme in the presence of CPBC (as opposed to
maximally dissipative boundary conditions) and a better dissipation
operator is needed.
%is not enough to stabilize the scheme in the presence of CPBC (as opposed to
%maximally dissipative) conditions when we inject pulses through the
%boundaries, and a better dissipation operator (which is not zero near the
%boundary) would be needed. 

For the time discretization we use a third order Runge-Kutta
integrator. The boundary conditions (\ref{Eq:Gauge}) and (for the case
without Weyl control) (\ref{Eq:MaxDissModes}) are imposed in the
following way: We first compute the vector field corresponding to the
discretized right-hand side of the system
(\ref{Eq:Ndot},\ref{Eq:gdot},\ref{Eq:Kdotmod},\ref{Eq:ddotmod},\ref{Eq:Adotmod})
and represent it in the characteristic basis. Then, the fields
corresponding to $v_{nn}^{(+)}$ are overwritten by the time derivative
of $h$ and (for the case without Weyl control) the field corresponding
to $\hat{v}_{AB}^{(+)}$ are overwritten with the time derivative of
$\hat{h}_{AB}$. Finally, the resulting vector field is transformed
back to the original basis. In all the simulations presented below we
use the same gridspacing in all directions, $\Delta x = \Delta y =
\Delta z$ and a Courant factor $\lambda$ of either $\lambda = 0.25$ or
$\lambda = 0.5$. The code we use is based on the one described in Ref.
\cite{TLN-BlackHole}.

%%%%%%%%%%%%%%%%%%%%%%%%%%%%%%%%%%%%%%%%%%%%%%%%%%%%%%%%%%%%%%%%%%%%
\section{Simulations: Spacetimes evolved and formulations of the IBVP used}
\label{Sect:Sim1}
%%%%%%%%%%%%%%%%%%%%%%%%%%%%%%%%%%%%%%%%%%%%%%%%%%%%%%%%%%%%%%%%%%%%

\subsection{Spacetimes evolved}

The following initial data is used in the $3$D evolutions of the next section.
In each case, the spatial domain is the cube $\{ (x,y,z) \in [-1,1]^3 \}$ of
side length $2$.

\subsubsection{Random data or robust stability test}

Here we consider initial data corresponding to flat space and add a random
perturbation to it. Therefore, initially, the fields are chosen to be
\begin{eqnarray}
g_{ij} &=& \delta_{ij} + \epsilon {\cal R}^g_{ij} \\
K_{ij} &=& \epsilon {\cal R}^k_{ij} \\
d_{kij} &=& \epsilon {\cal R}^d_{kij} \\
N &=& 1 + \epsilon {\cal R}^n \\
A_i &=& \epsilon {\cal R}^a_i
\end{eqnarray}
where the different $\cal R$ quantities are random numbers which are uniformly
distributed in $[-1,1]$. Similarly, at each timestep the boundary data $h$,
$\hat{h}_{AB}$ and $\hat{H}_{AB}$ (see Eqs. (\ref{Eq:Gauge}),
(\ref{Eq:MaxDissModes}) and (\ref{Eq:MaxDissWeyl}), respectively) for the
``gauge'' and ``physical'' degrees of freedom is set to a random number of the
same order, $\epsilon \cal {R}$. In the simulations shown below we choose
$\epsilon=10^{-5}$.

This {\it robust stability test} \cite{robust1, robust2} is designed to test
the numerical stability of the scheme by finding out whether there is, at
fixed time, unbounded growth as resolution is increased. One of the advantages
of the test is that it excites all frequency modes allowed by a given
resolution and it is therefore useful in spotting instabilities (if present)
that could otherwise remain hidden in some convergence tests. However, this
test is {\em not} a convergence test, since the constraints are not satisfied
and since different random data is used when resolution is changed.

\subsubsection{Brill waves}

Axisymmetric Brill waves \cite{BrillWaves} are evolved in the next section.
The corresponding initial data for the three-metric is given in Cartesian
Coordinates $x,y,z$ in the form
\begin{equation}
ds^2 = g_{ij} dx^i dx^j 
     = \Psi^4 \left[ e^{2q} \left( \frac{x dx + y dy}{\rho} \right)^2 + e^{2q} dz^2
+ \left( \frac{x dy - y dx}{\rho} \right)^2 \right],
\label{eqn:brillmetric}
\end{equation}
where $\rho=\sqrt{x^2+y^2}$, the function $q$ has the form
\begin{equation}
q = A \rho^2\exp\left(-\frac{x^2+y^2+z^2}{\sigma_r^2} \right),
\end{equation}
and the conformal factor $\Psi$ is obtained by solving the Hamiltonian
constraint for time-symmetric initial data. In order to do so, we use
the numerical elliptic solver BAM and the IDBrill Thorn, both publicly
available from the CACTUS distribution \cite{cactus}. The initial data
for $K_{ij}$, $N$, $A_i$ and $d_{kij}$ is given by
\begin{eqnarray}
K_{ij} &=& 0,
\label{ki}\\
N &=& 1,
\label{ni}\\
A_i &=& 0,
\label{ai}\\
d_{kij} &=& D_k g_{ij}\; ,
\label{di}
\end{eqnarray}
where $D_k$ is the D2-1 finite difference operator in the $k$ direction. For
these Brill wave simulations we choose the D2-1 operator not only for
computing the initial data for $d_{kij}$ but also for the discretization of
the right-hand side of the evolution equations. There is no advantage in using
a higher order accurate difference operator because the elliptic solver is
only second order accurate. In the evolutions shown below, $\sigma_r=1/5$. For
weak enough waves the solution bounces at the origin and disperses to
infinity, while for strong waves an apparent horizon is typically found
\cite{critical}.  In the evolutions below we use very weak waves,
corresponding to an amplitude $A=10^{-2}$ (while the critical solution is
believed to correspond to $A \approx 4.8$ \cite{critical}). As we shall see,
even in these very weak field evolutions there is a clear violations of the
constraints when non constraint-preserving boundary conditions are used.  Even
though these waves are axisymmetric, we have evolved them in full $3$D.

\subsubsection{Gauge solutions}

Initial data for a static solution corresponding to a gauge transformation of
flat spacetime can be obtained by setting $K_{ij}$, $N$ and $A_i$ to the same
expressions as above [Eqs.(\ref{ki},\ref{ni},\ref{ai})], while the
three-metric is obtained by starting from the flat metric in spherical
coordinates,
\begin{displaymath}
ds^2 = dr^2 + r^2(d\vartheta^2 + \sin^2{\vartheta}\, d\varphi^2),
\end{displaymath}
performing a coordinates transformation
\begin{displaymath}
r = \bar{r}\left( 1 - a\, e^{-\bar{r}^2/\sigma_0^2} \right)
\end{displaymath}
and transforming the resulting metric to the Cartesian coordinates $(x,y,z) =
(\bar{r} \sin\vartheta\cos\phi, \bar{r} \sin\vartheta\sin\phi, \bar{r}
\cos\vartheta)$. Initial data for $d_{kij}$ is obtained by computing
analytically the gradient of the resulting Cartesian components of the three
metric, $d_{kij} = \partial_k g_{ij}$. This gauge solution has been used to
compare the stability properties of different formulations of Einstein's
vacuum equations \cite{bssn}. In our simulations, we choose $a=0.1$ and
$\sigma_0=0.2$.

\subsubsection{Injecting pulses of gravitational radiation through the boundaries}

Finally, we consider an example where we start with flat initial data, i.e. $N
= 1$, $g_{ij} = \delta_{ij}$, $K_{ij} = 0$, $d_{kij} = 0$, $A_k = 0$, but
inject a pulse of gravitational radiation by choosing as boundary data
\begin{eqnarray}
&& h = 0, \\
&& \hat{h}_{22} = -\hat{h}_{33} = \hat{h}_{23} = -2\alpha(t-t_0)\sigma_t^{-2} e^{\left[
    -(t-t_0)^2/\sigma_t^2 - \rho^2/\sigma_r^2\right]}
\end{eqnarray}
where $\rho$ is a ``polar coordinate at each face''. For example, at
the $x= \pm 1$ faces $\rho = \sqrt{y^2 + z^2}$ and similarly for the
other faces. In the simulations shown below we choose $t_0=1.2$,
$\sigma_t = \sigma_r = 0.2$ and amplitudes $\alpha = 0.01$ and $2$.

%%%%%%%%%%%%%%%%%%%%%%%%%%%%%%%%%%%%%%%%%%%%%%%
\subsection{Formulations of the IBVP used}
\label{formulations}
%%%%%%%%%%%%%%%%%%%%%%%%%%%%%%%%%%%%%%%%%%%%%%%

In the simulations below, we choose $F = N K$ corresponding to time harmonic
slicing, set the shift to zero everywhere and at all times and restrict
ourselves to the case in which the parameter $\zeta$ in Eq. (\ref{Eq:Kdot}) is
set to $\zeta=-1$. Furthermore, we only consider choices of parameters for
which $\lambda_2=1$ and $\lambda_3=1$, and set the parameter $\Omega$ defined
in Eq. (\ref{Eq:Omega}) to zero.  This implies that the characteristic
directions lie along the light cone or the normal to the $t=const$
hypersurfaces.

There are two subsets of parameter space which fulfill these requirements:
\begin{enumerate}
\item Mono-parametric family (parametrized by the nonzero value of $\chi$):
\begin{equation}
\gamma = -\frac{1}{2}\; ,\qquad
\zeta = -1, \qquad
\eta = 2, \qquad
\xi = -\frac{\chi}{2}\; ,\qquad
\chi \ne 0.
\label{monoparametric}
\end{equation}

\item Bi-parametric family (parametrized by the parameters $\eta$ and $\gamma
  \neq -1/2$):
\begin{equation}
\zeta = -1, \qquad
\chi = -\frac{\gamma (2-\eta)}{1+2\gamma}\; ,\qquad
\xi = -\frac{\chi}{2} + \eta - 2, \qquad
\gamma \ne -\frac{1}{2}\; , \qquad
\eta. 
\label{biparametric}
\end{equation}
\end{enumerate}
One can show that for these families the evolution system
(\ref{Eq:Ndot},\ref{Eq:gdot},\ref{Eq:Kdot},\ref{Eq:ddot},\ref{Eq:Adot}) is
symmetric hyperbolic. However, notice that the mono-parametric family violates
the condition (\ref{Eq:ConstrSym1}) which means that the constraint
propagation system is not symmetric hyperbolic. The bi-parametric family
satisfies the conditions
(\ref{Eq:ConstrSym1},\ref{Eq:ConstrSym2},\ref{Eq:ConstrSym3}) if and only if
$0 < \eta < 2$. According to the analysis in the previous section the
determinant condition is satisfied for this subfamily. In this article, we
consider the following four cases:
\begin{itemize}

\item{\it CPBC without Weyl control: a completely ill posed case}\\
  The bi-parametric family is used in this case, with parameters
  $(\eta,\gamma) = (3,0)$, and CPBC without Weyl control. As mentioned above,
  the resulting parameters {\em violate} the condition for the constraint
  propagation system to be symmetric hyperbolic. Therefore, there is no
  guarantee that the IBVP is well posed. As a matter of fact, it turns out
  that the determinant condition is violated in this case and thus the problem
  admits ill posed modes. All the simulations performed below confirm this
  fact.

\item{\it CBPC with Weyl control}\\
  Here we choose the bi-parametric family and CPBC with Weyl control. The
  parameters $\eta$ and $\gamma$ are chosen to be either $(\eta,\gamma) =
  (1,0)$ or $(\eta,\gamma) = (7/4,-2/3)$; both choices satisfy the determinant
  condition. Although the growth rate is smaller for the second choice, in
  both cases we find that the system is numerically unstable. However, we also
  notice that the instability seems to be milder than in the completely ill
  posed case: In the weak Brill wave runs, for instance, one needs to run for
  several crossing times or use very high resolution before noticing the lack
  of convergence. A similar situation arises in the numerical evolution of
  weakly hyperbolic systems \cite{CPST-Convergence} with periodic boundary
  conditions.

\item{\it CPBC without Weyl control}\\
  Again, we choose the bi-parametric family, but now we consider CPBC without
  Weyl control, and choose $(\eta,\gamma) = (1,0)$. The numerical simulations
  presented below suggest that the system is numerically stable.

\item{\it Maximally dissipative}\\
  The mono-parametric family with $\chi=-1$ is used with maximally dissipative
  boundary conditions. This system should be stable, since the evolution
  equations are symmetric hyperbolic and so the IBVP is well posed. The
  simulations below confirm this. But more importantly, they provide an
  explicit demonstration that evolutions of a system with
  constraint-preserving boundary conditions, when numerically stable, are more
  accurate than standard, maximally dissipative ones, as in the latter case
  the boundary conditions introduce constraint violations that do not converge
  to zero.

\end{itemize}

%%%%%%%%%%%%%%%%%%%%%%%%%%%%%%%%%%%%%%%%%%%%%%%%%%%%%%%%%%%%%%%%%%%%
\section{Simulations}
\label{Sect:Sim}
%%%%%%%%%%%%%%%%%%%%%%%%%%%%%%%%%%%%%%%%%%%%%%%%%%%%%%%%%%%%%%%%%%%%

For each of the four IBVP described above we run the robust stability test and
evolve weak Brill waves. For the case of CPBC without Weyl control, we also
evolve the gauge solution and inject pulses of gravitational radiation from
the boundary of an initially flat spacetime.

%%%%%%%%%%%%%%%%%%%%%%%%%%%%%%%%%%%%%%%%%%%%%%%%%%%%%%%%%%%%%%%%%%%%
\subsection{Random, or robust stability test}
%%%%%%%%%%%%%%%%%%%%%%%%%%%%%%%%%%%%%%%%%%%%%%%%%%%%%%%%%%%%%%%%%%%%

\begin{figure}[ht] 
\begin{center} 
  \includegraphics*[height=6cm]{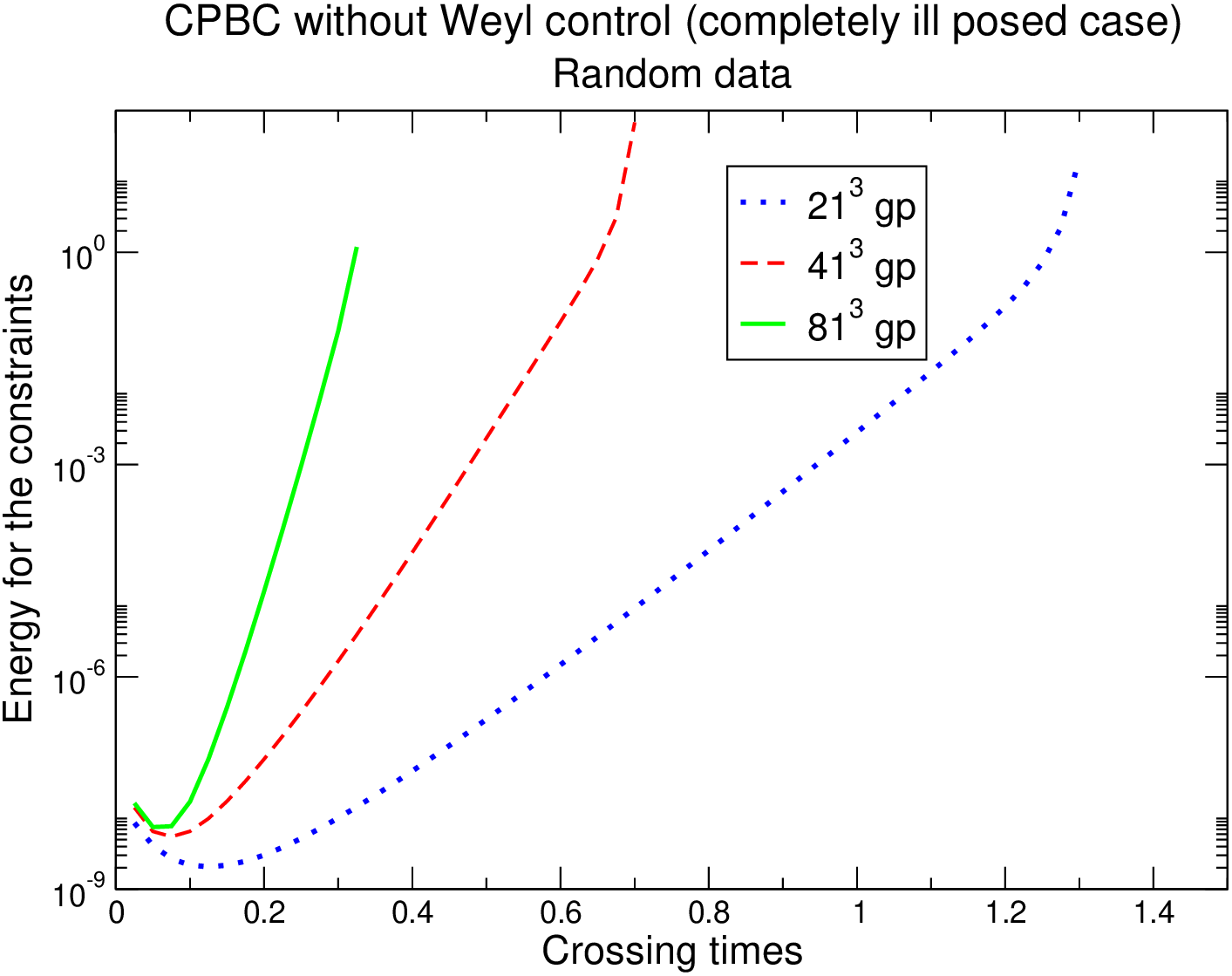}
  \includegraphics*[height=6cm]{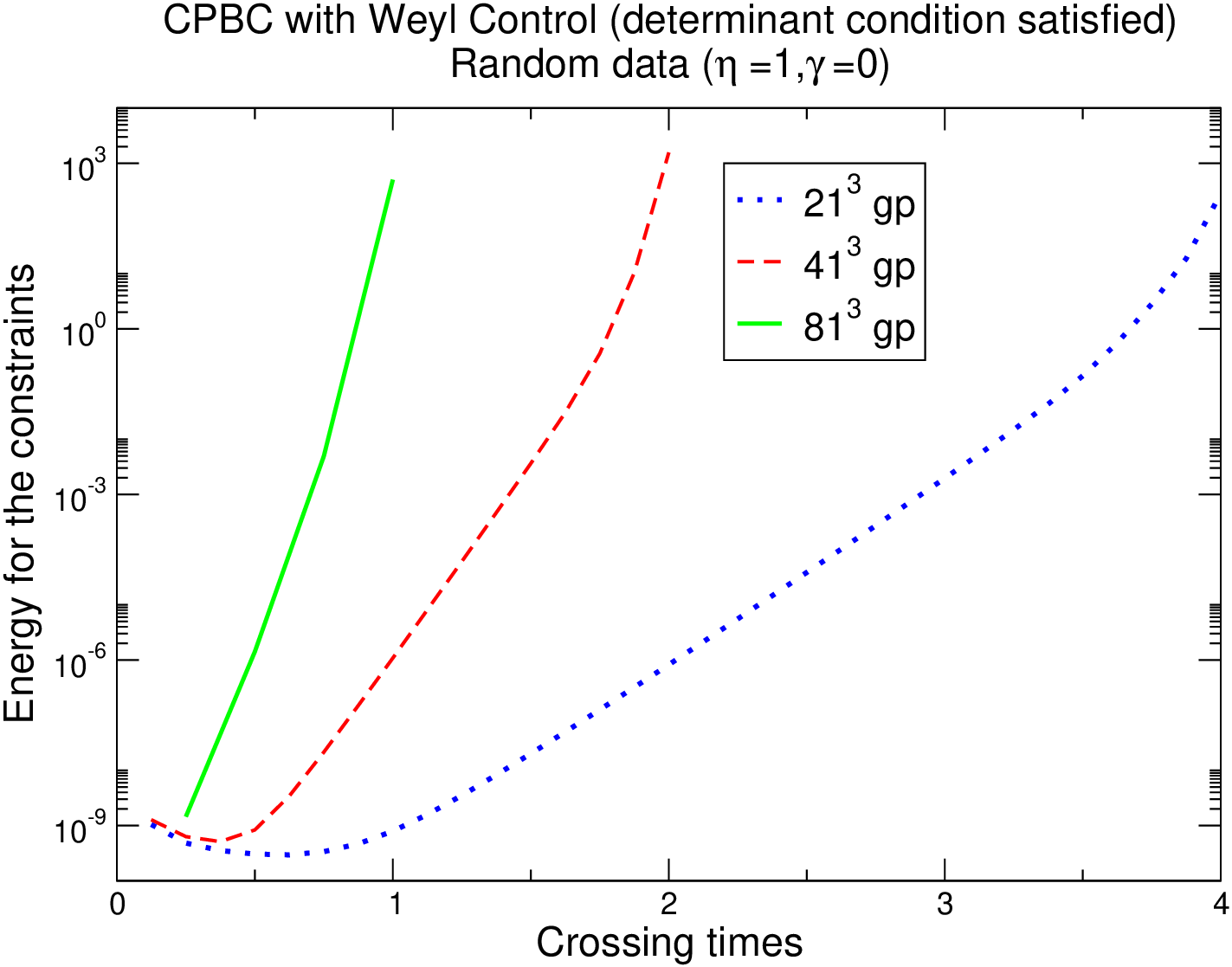}
  \includegraphics*[height=6cm]{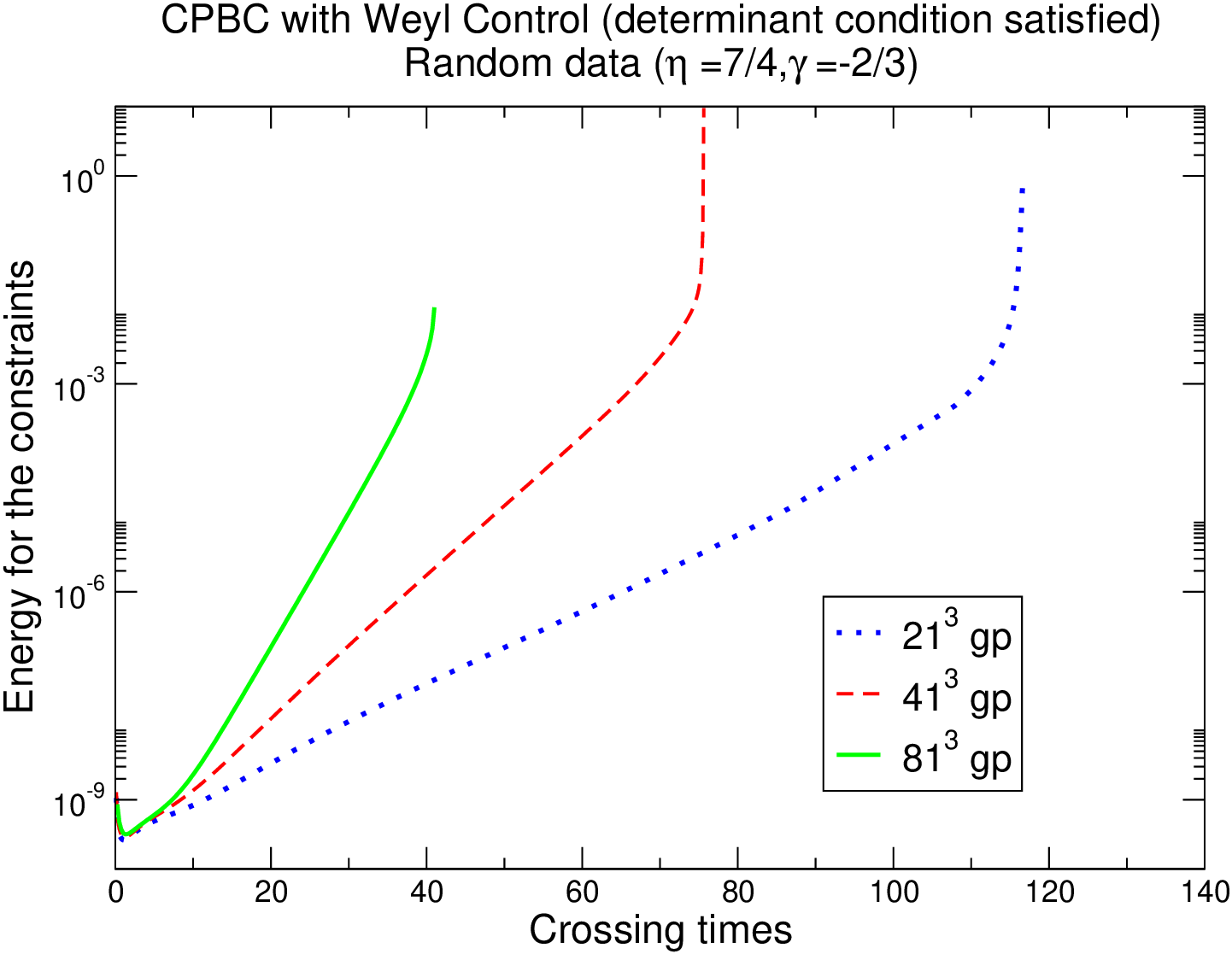}
  \includegraphics*[height=6cm]{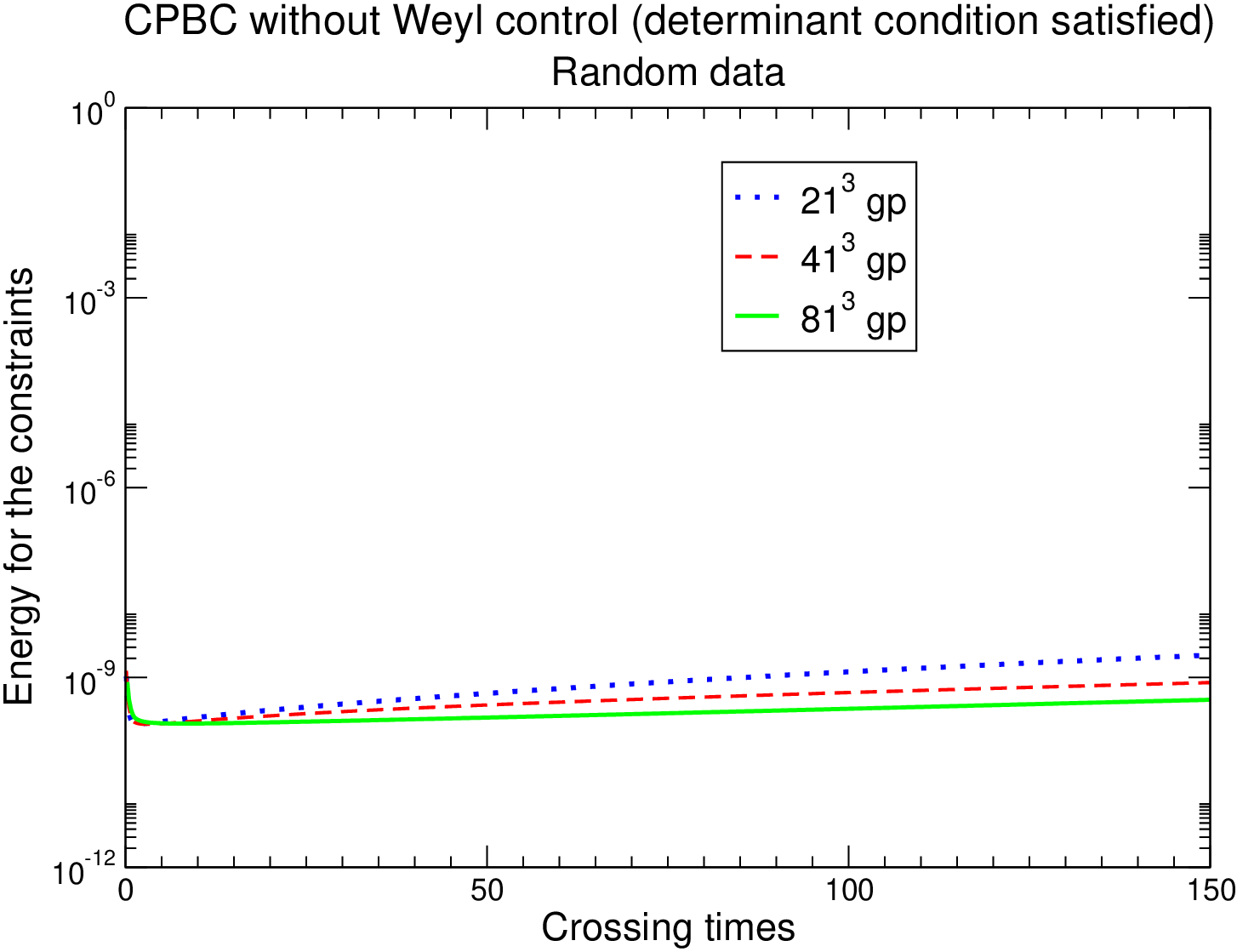}
  \includegraphics*[height=6cm]{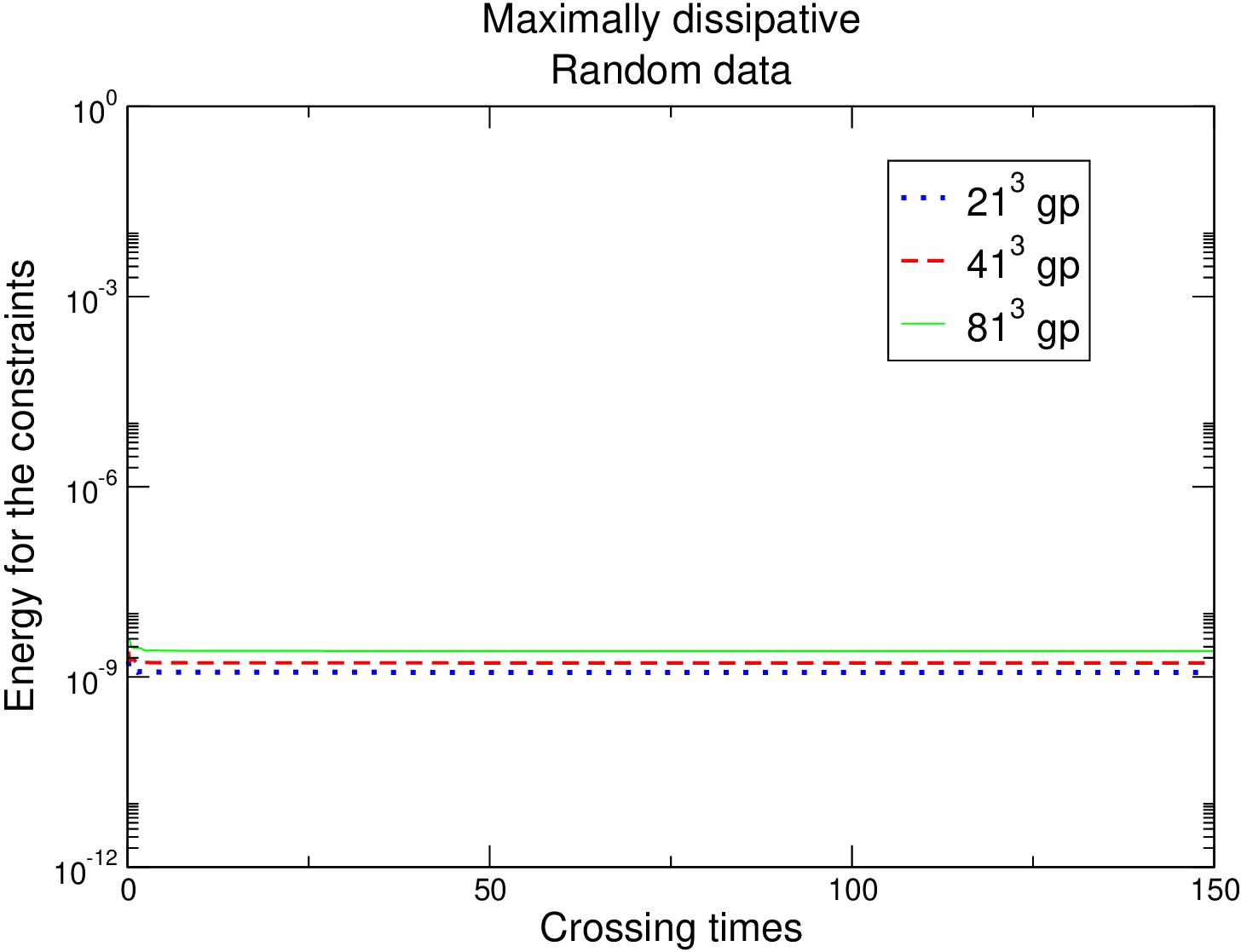}
\caption{Random data runs for the D2-1 derivative, with Courant factor
  $\lambda =0.5$ and dissipation $\sigma_{diss} = 0.05$. The energy of the
  constraints is defined to be the sum over all gridpoints of the sum of the
  square of the components of the constraint variables divided by the number
  of gridpoints.}
\label{random} 
\end{center} 
\end{figure}

For these runs, the Courant factor is $\lambda = 0.5$ and the
resolution varies from $21^3$, $41^3$ to $81^3$ gridpoints. The
spatial derivatives are discretized using the D2-1 operator, and a
dissipation parameter of $\sigma_{diss}=0.05$ is chosen. Recall that
random initial {\em and} boundary data is given here. In Figure
\ref{random} we show the energy of the constraints versus light
crossing time. The energy of the constraints is defined to be the sum
over all gridpoints of the sum of the square of the components of the
constraint variables divided by the number of gridpoints.

As expected, in the completely ill posed case, the energy of the main
variables grows at fixed time as resolution is increased, which makes the code
crash in the timescale of less than a crossing time (there is also an increase
in the energy at fixed resolution as a function of time, though this does not
necessarily represent a numerical instability). Figure \ref{random} shows that
there is also a similar kind of growth in the energy for the constraints. The
timescale and explosive kind of the numerical instability are similar to those
found in initial value problems that are completely ill posed due to the the
presence of complex eigenvalues in the principal part \cite{CPST-Convergence,
  CPST-Stab}. This similarity is, indeed, expected, as at the analytical level
the problem admits exponentially in time growing modes, where the exponential
factor increases with the frequency as shown in section \ref{Sect:FL}. {\em
  Recall, however, that the main evolution system is symmetric hyperbolic;
  thus the boundary conditions are responsible for the instabilities.}

In the CPBC case with Weyl control the runs also show evidence of a numerical
instability both in the main and constraint variables, though the growth rate
is somehow smaller than before. What is somehow unexpected is the resolution
dependent growth in the constraint variables. After all, at the continuum, the
constraints' evolution is governed by a symmetric hyperbolic system with
maximally dissipative boundary conditions which constitutes a well posed
problem by itself.  However, it seems that the main system is unstable and
that at the discrete level, the instabilities do have an effect on the
constraint variables. In order to gain a better insight into this problem we
have analyzed the semi-discrete problem, where only the space derivatives are
discretized. Even in the simpler case of linearizations about Minkowski
spacetime we found that while the discrete constraints do obey a symmetric
hyperbolic system it is far from obvious that the way in which we implement
the constraint-preserving boundary conditions allow for a semi-discrete energy
estimate. It also seems difficult to represent the boundary conditions in a
different way such that a semi-discrete energy estimate can be shown for the
constraint propagation system. We will not attempt to address this question
further in this article. Notice that even if we found such a discretization,
the system could still suffer from the presence of more weakly ill posed gauge
or physical modes that are undetected by the determinant condition. An
explicit example of such weakly ill posed modes is given in \ref{App:SFS}.

As opposed to the previous cases, in the CPBC case without Weyl control the
runs strongly suggest that the resulting systems is numerically stable and
that the continuum problem might be well posed.  At early times the
constraints' energy decrease while at around $100$ crossing times the
constraint variables start to slowly grow in time. However, the growth rate
does not become larger with increasing resolution, as opposed to the previous
cases.

Finally, the random runs with maximally dissipative boundary conditions also
strongly suggest that the system is numerically stable, though this is
expected because in this case we know that the IBVP is well posed.

%%%%%%%%%%%%%%%%%%%%%%%%%%%%%%%%%%%%%%%%%%%%%%%%%%%%%%%%%%%%%%%%%%%%
\subsection{Brill wave evolutions}
%%%%%%%%%%%%%%%%%%%%%%%%%%%%%%%%%%%%%%%%%%%%%%%%%%%%%%%%%%%%%%%%%%%%

\begin{figure}[ht] 
\begin{center} 
  \includegraphics*[height=6cm]{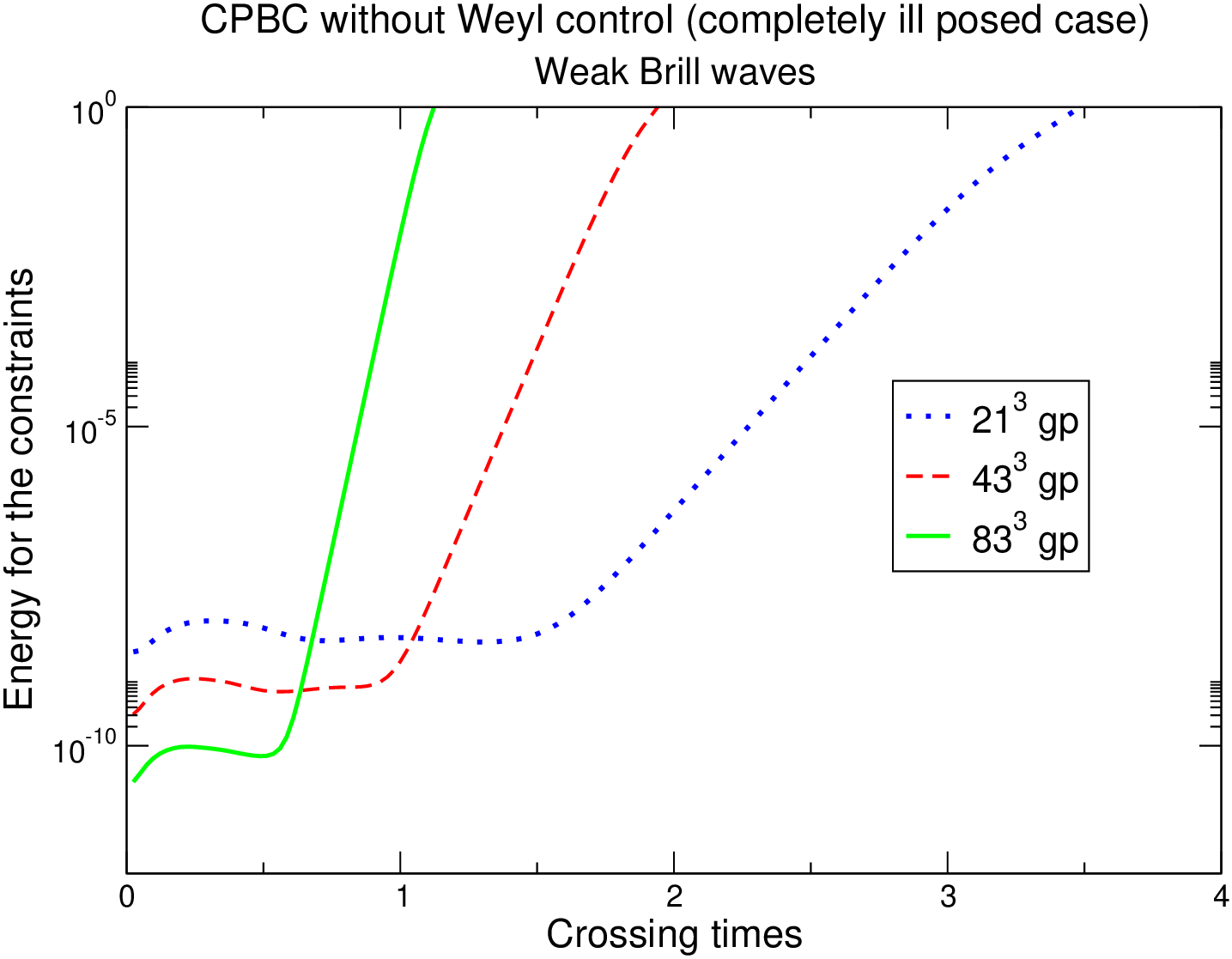}
  \includegraphics*[height=6cm]{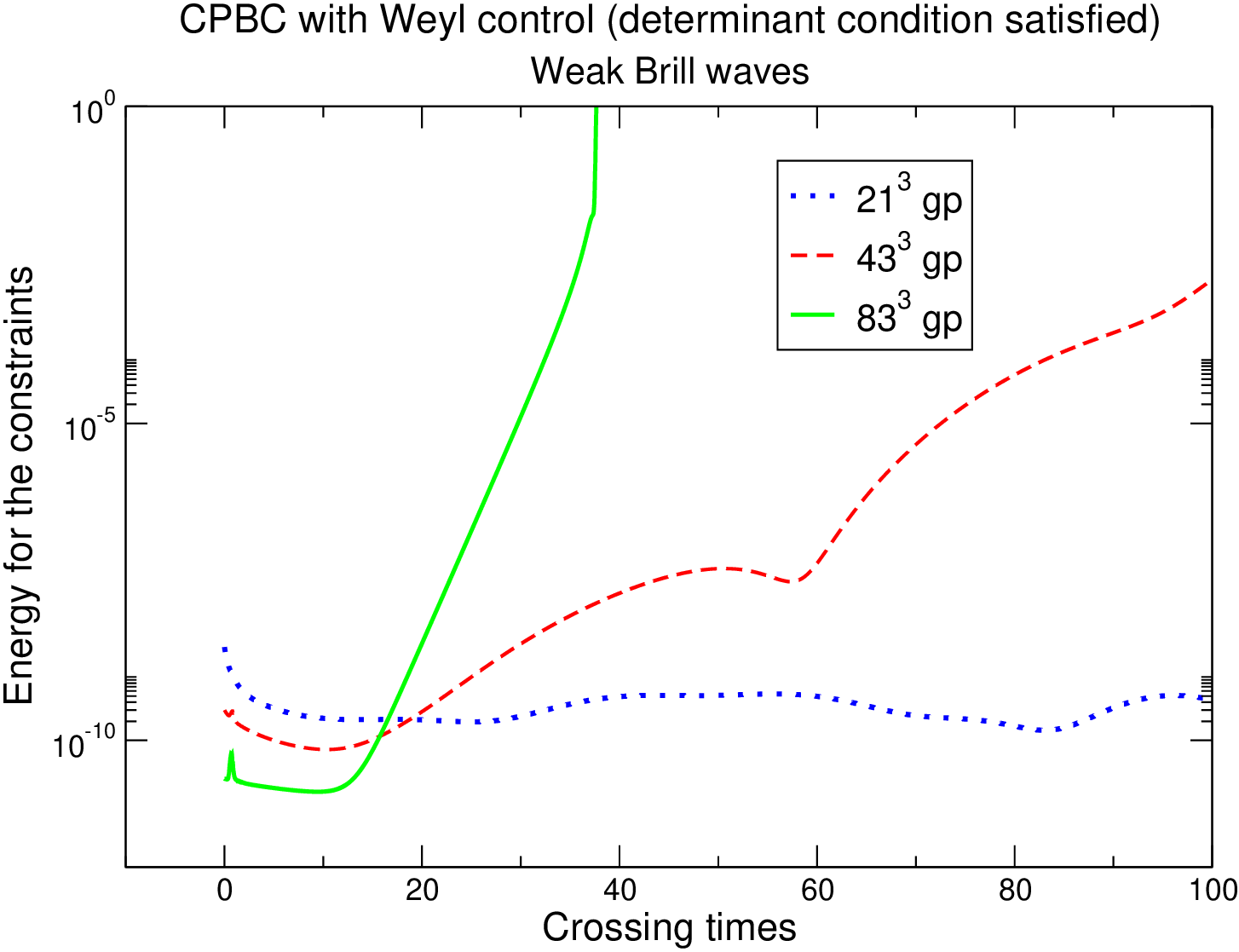}
  \includegraphics*[height=6cm]{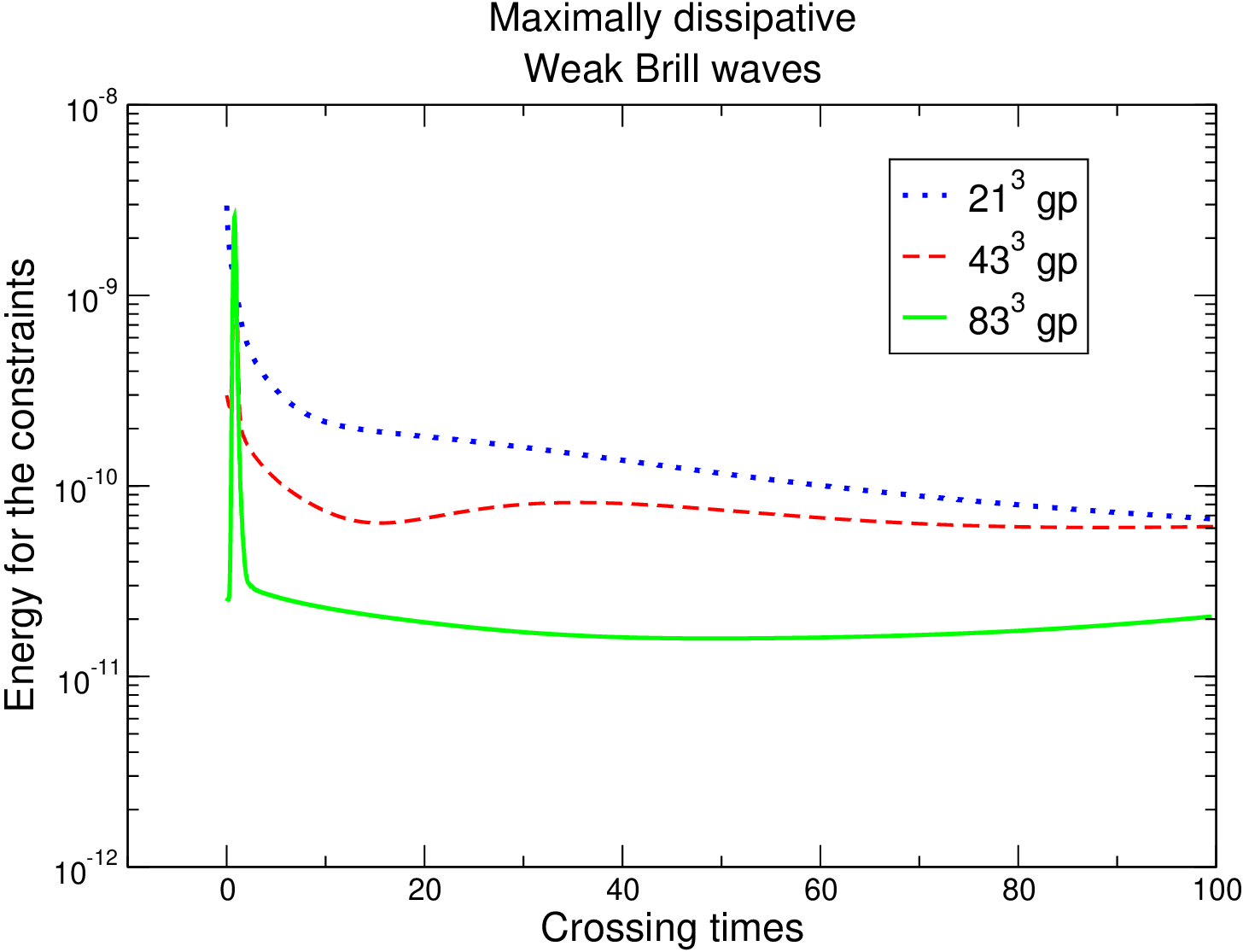}
  \includegraphics*[height=6cm]{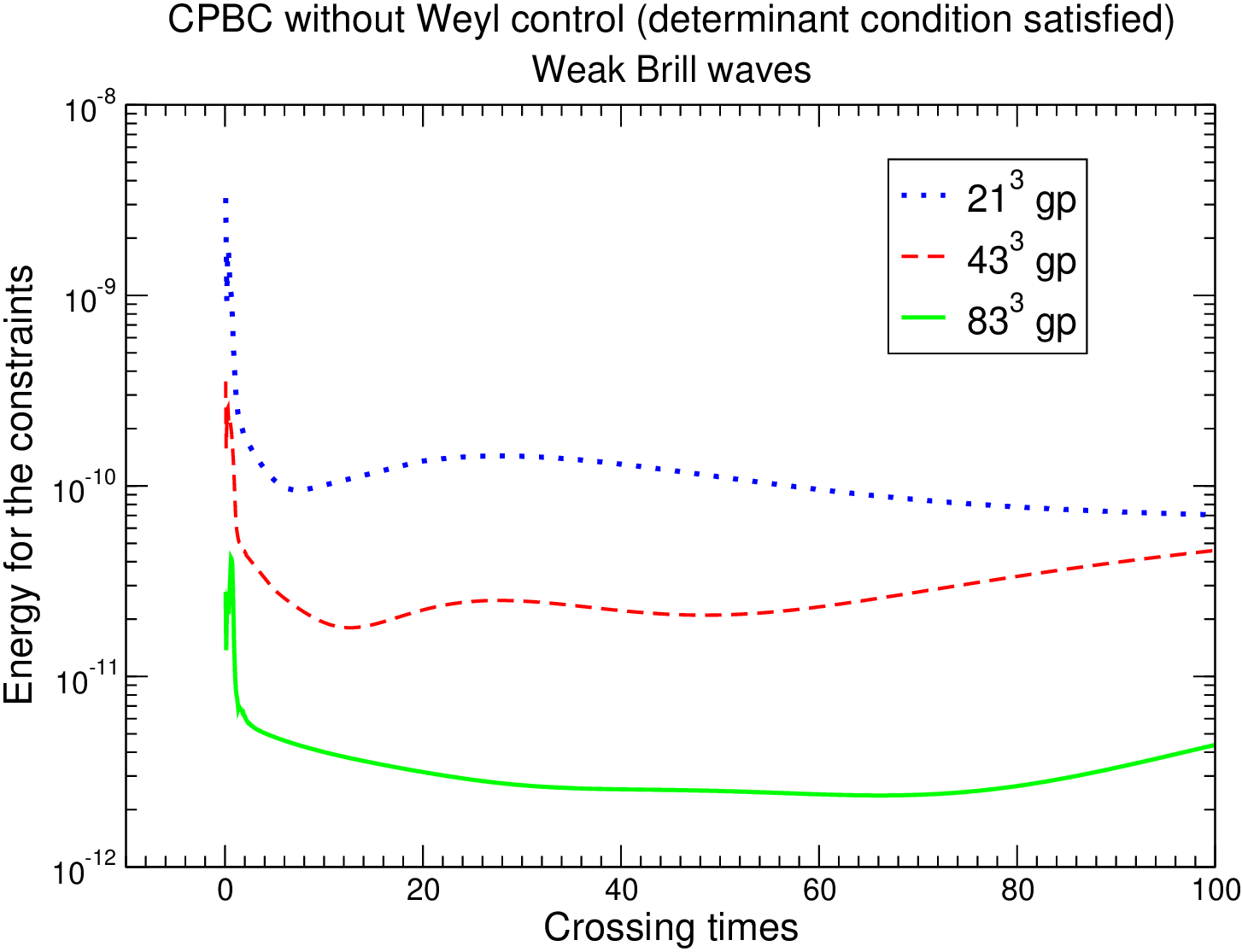}
\caption{Weak Brill runs for the four formulations of the IBVP used in the
  simulations of Fig.\ref{random}. The expectations based on that figure
  regarding numerical stability (or its lack) are here confirmed.}
\label{weak_brill}
\end{center} 
\end{figure}

In these simulations we use the smaller Courant factor of $\lambda =
0.25$.  Figure \ref{weak_brill} shows weak Brill wave runs for the
four formulations of the IBVP described in Section
\ref{formulations}. The expectations based on the random data runs of
the previous section are confirmed. Namely, the completely ill posed
CPBC case is manifestly numerically unstable. The CPBC with Weyl
control case that satisfies the determinant condition seems to be
unstable as well, though in a ``weaker'' sense while the two remaining
cases (the maximally dissipative one and the CPBC without Weyl control
case that satisfies the determinant condition) seem to be numerically
stable.

\begin{figure}[ht] 
\begin{center} 
  \includegraphics*[height=6cm]{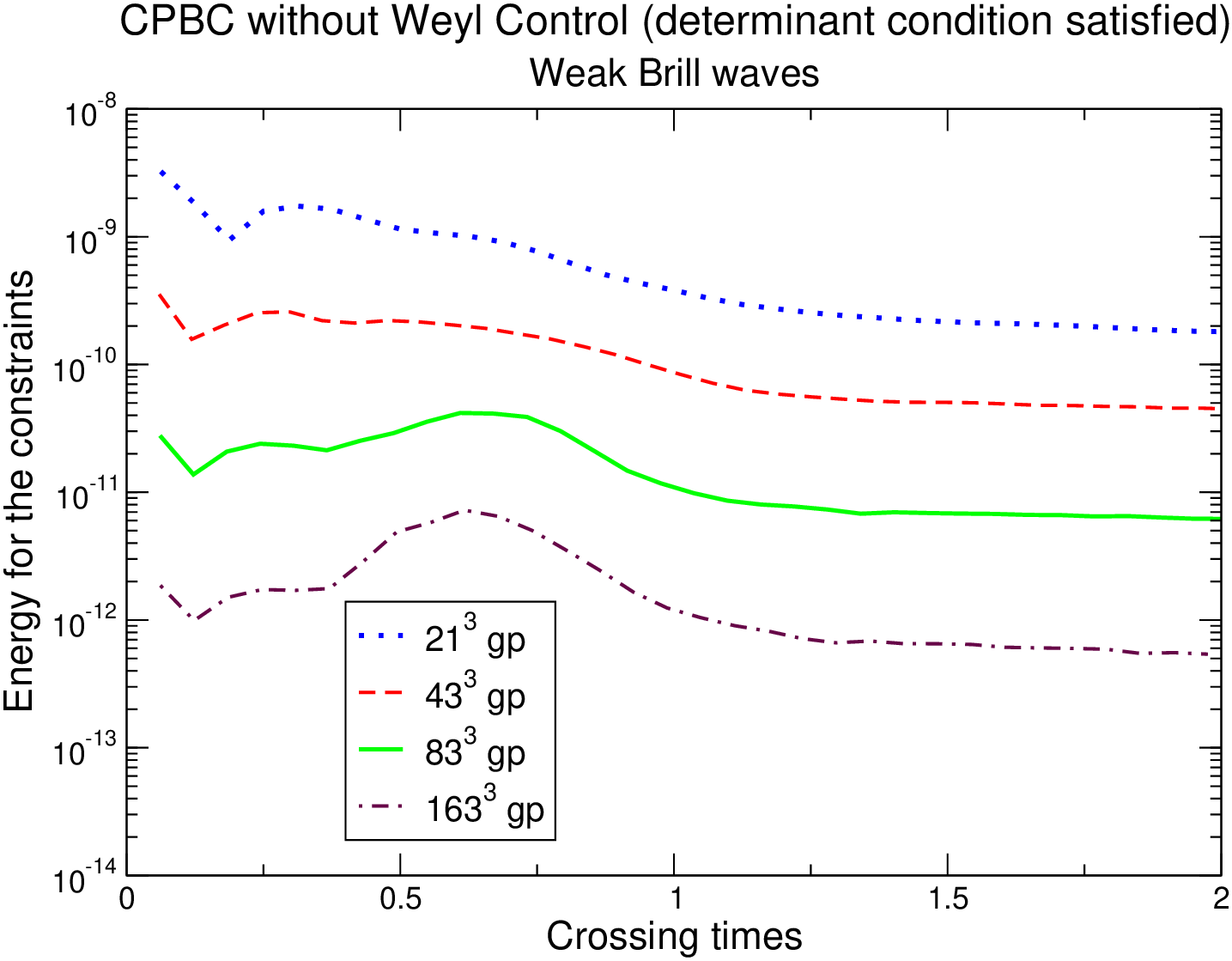}
  \includegraphics*[height=6cm]{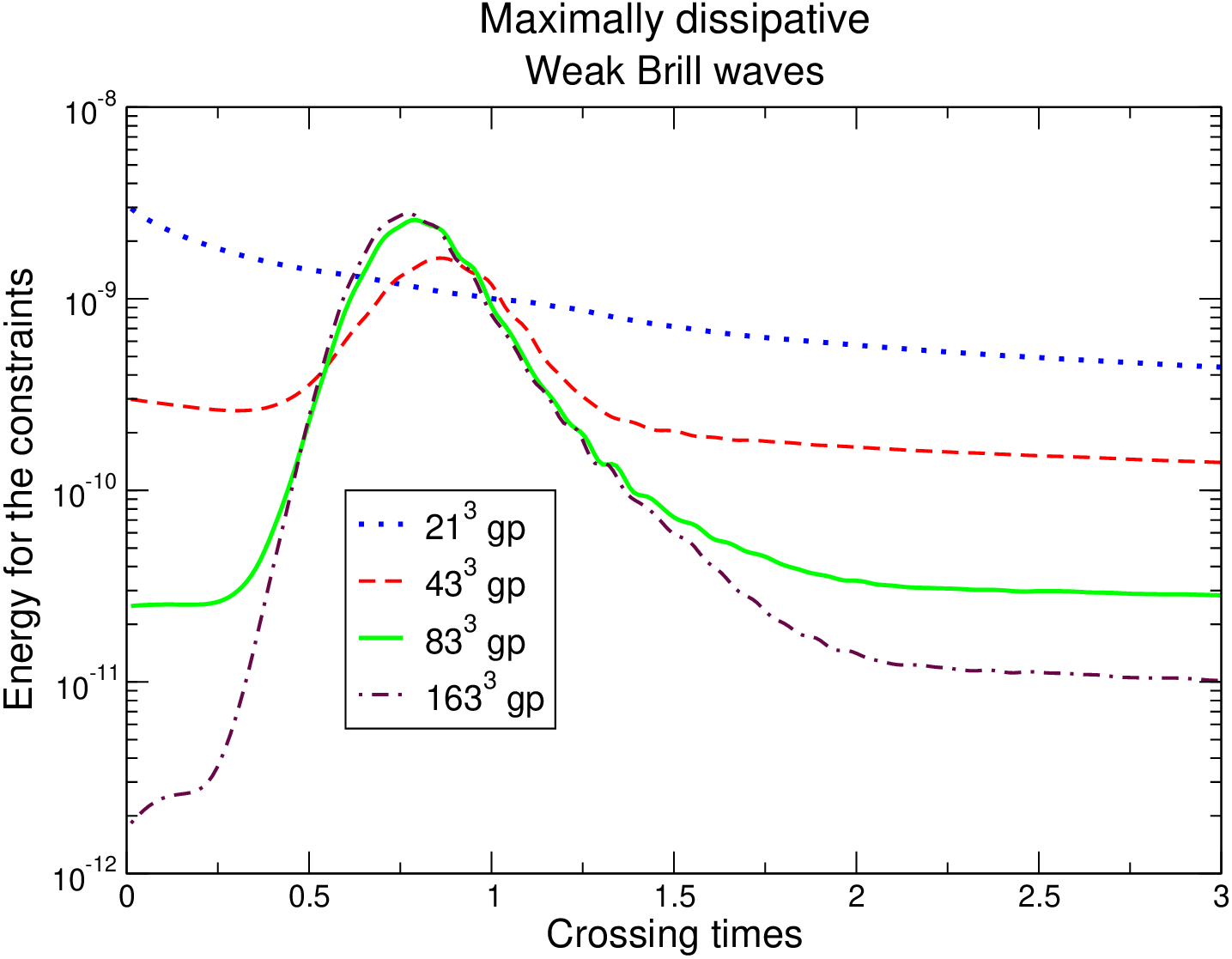}
\caption{Same runs as previous figure for the two numerically stable
  cases, but on a shorter timescale and with more resolution. Notice the lack
  of convergence to zero in the maximally dissipative case, as opposed to the
  constraint-preserving boundary conditions case. }
\label{weak_brill_short}
\end{center} 
\end{figure}

Figure \ref{weak_brill_short} shows the last two cases on a shorter timescale
and with a higher resolution run added. {\em Notice how in the CPBC case the
  constraint variables seem to converge to zero, while in the maximally
  dissipative case with the same initial data and resolution the lack of
  convergence to zero is evident}.

%%%%%%%%%%%%%%%%%%%%%%%%%%%%%%%%%%%%%%%%%%%%%%%%%%%%%%%%%%%%%%%%%%%%
\subsection{Gauge solutions}
%%%%%%%%%%%%%%%%%%%%%%%%%%%%%%%%%%%%%%%%%%%%%%%%%%%%%%%%%%%%%%%%%%%%

\begin{figure}[ht] 
\begin{center} 
  \includegraphics*[height=6cm]{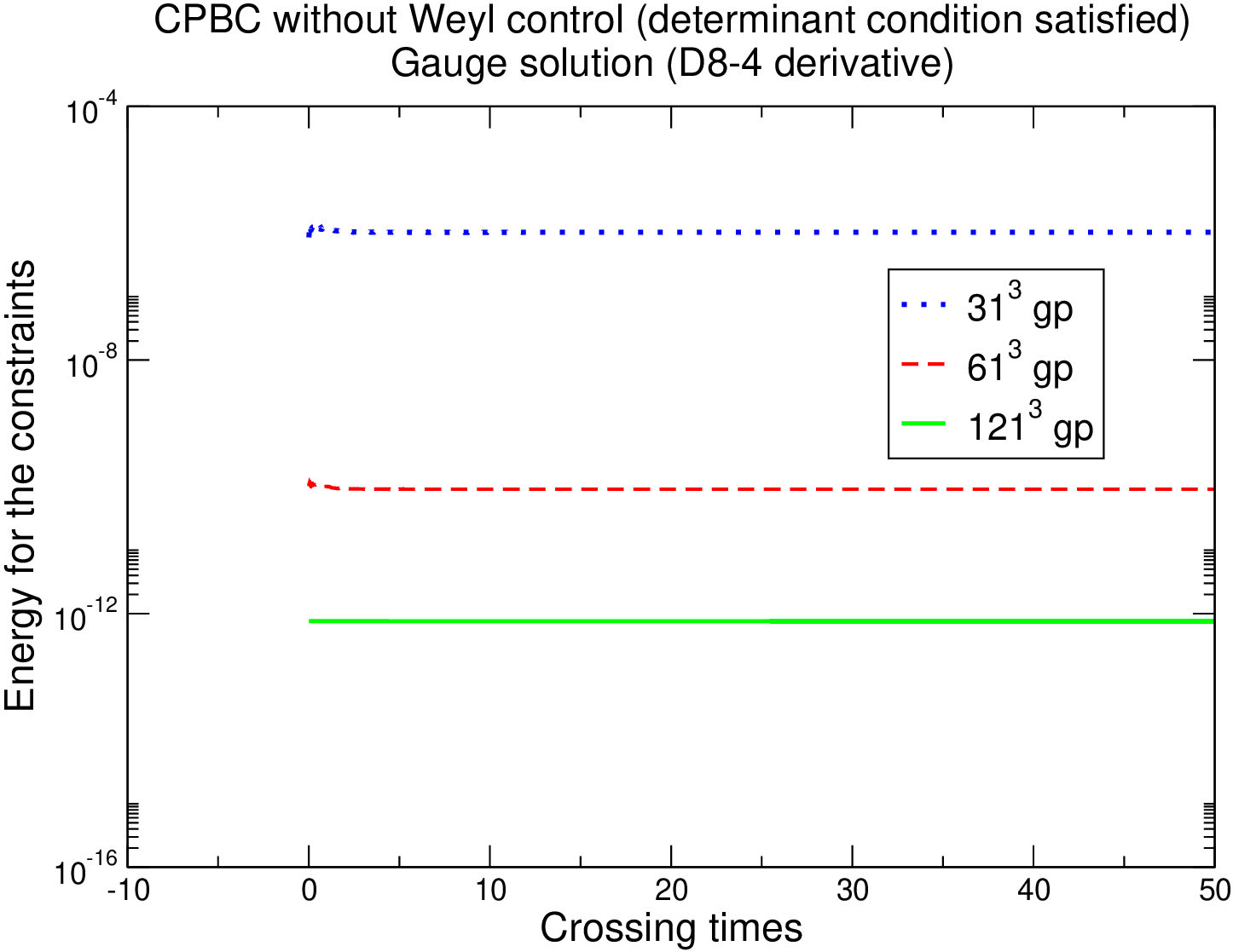}
\caption{Gauge solution simulations with a high order scheme, Courant factor
  $\lambda = 0.25$, and dissipation $\sigma_{diss}=10^{-4}$ }
\label{gauge}
\end{center} 
\end{figure}

Here, we concentrate on the CPBC without Weyl control case. Since the initial
data is given in analytic form we use the high order accurate finite
differencing operator D8-4. The results are shown in Figure \ref{gauge}.
Notice that the high order accurate scheme results in a much faster
convergence of the constraints to zero as resolution is increased. Here, the
Courant factor is $\lambda = 0.25$ and the dissipation parameter
$\sigma_{diss}=10^{-4}$.

%%%%%%%%%%%%%%%%%%%%%%%%%%%%%%%%%%%%%%%%%%%%%%%%%%%%%%%%%%%%%%%%%%%%
\subsection{Injecting pulses of gravitational radiation through the boundaries}
%%%%%%%%%%%%%%%%%%%%%%%%%%%%%%%%%%%%%%%%%%%%%%%%%%%%%%%%%%%%%%%%%%%%

\begin{figure}[ht] 
\begin{center} 
\includegraphics*[height=6cm]{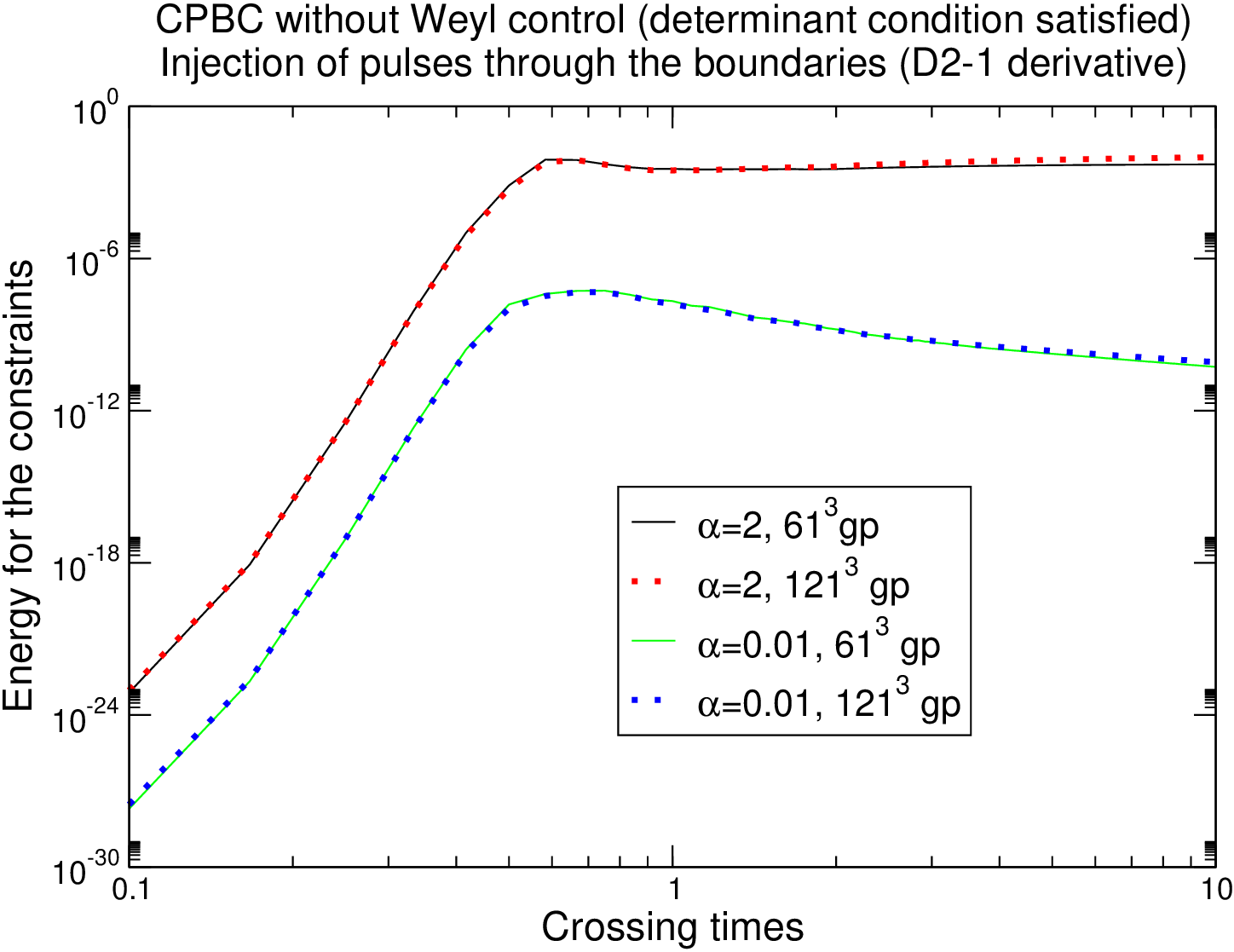} 
\caption{Injecting pulses through the boundaries. Here,the Courant
factor is $\lambda = 0.25$ and the dissipation parameter is
$\sigma_{diss}=10^{-4}$. The energy starts from roundoff as
the initial data corresponds to flat spacetime. The energy curves for
the highest resolutions have been scaled by a factor of four in a way
that overlapping with the curves for to the coarser resolution
corresponds to second order convergence. }
\label{col_energy}
\end{center} 
\end{figure}

As in the previous gauge simulations, we use the CPBC formulation
without Weyl control. However, we have found that in this case for the
same resolutions used in the previous simulations a numerical
instability shows up after some time. This is probably due to the fact
that the dissipative operator is zero near the boundary and is not
negative semi-definite with respect to the scalar product for which
SBP holds. Therefore for this case we have used the D2-1 operator,
with Courant factor $\lambda=0.25$, and dissipation parameter
$\sigma_{diss}=0.05$. In Figure \ref{col_energy} we show the energy of
the constraints as a function of time for two resolutions and the two
amplitudes $\alpha=0.01$ and $\alpha=2$. The energy starts from zero,
as the initial data consists of Minkowski spacetime. After a short
time the energy quickly grows as the pulses are injected into the
domain through the six boundaries, until the pulses have been
completely injected at roughly one crossing time, time at which the
energy stays approximately constant (at fixed resolution).

\begin{figure}[ht] 
\begin{center} 
  \includegraphics*[height=6cm]{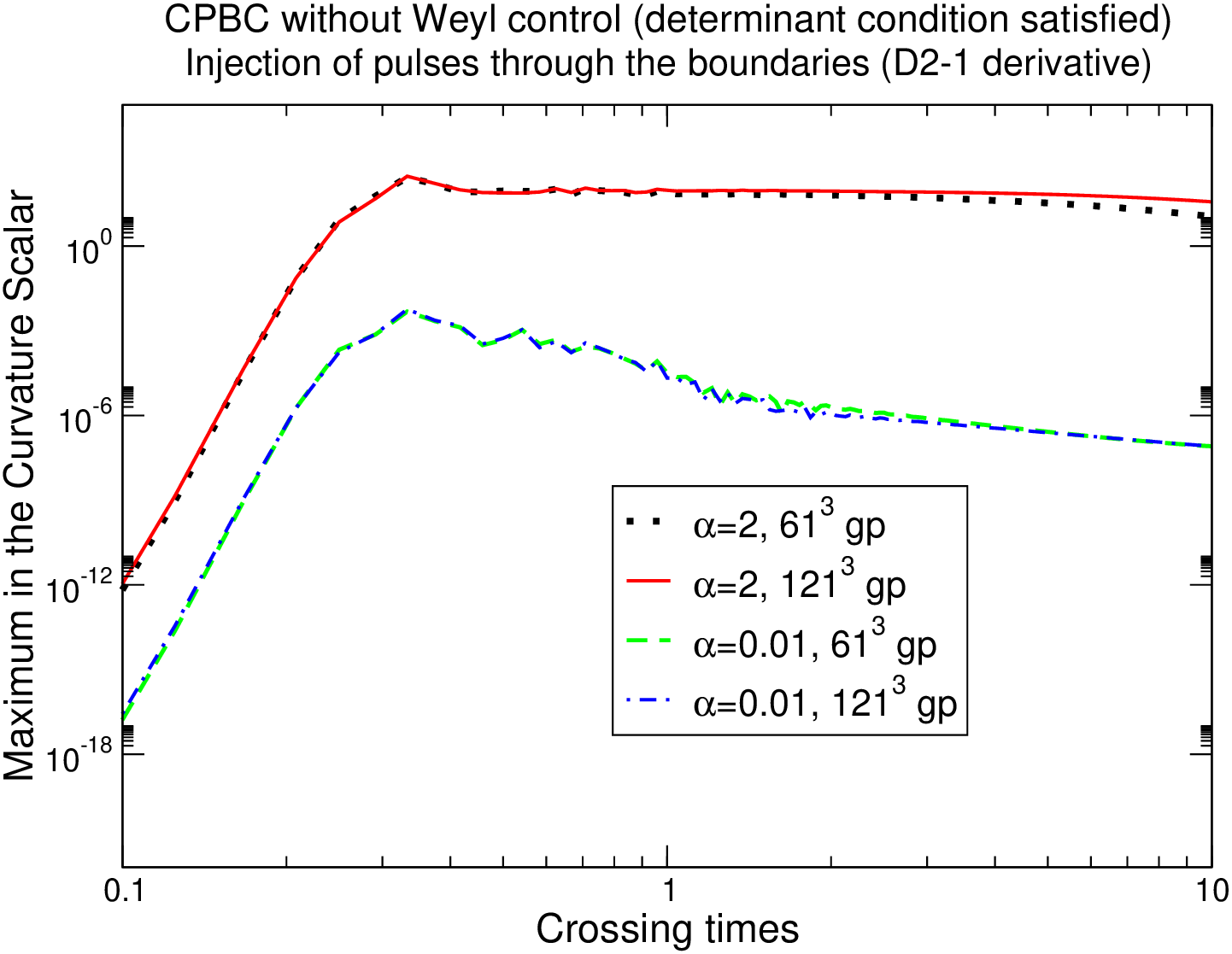}
\caption{Same simulations as those shown in the previous figure,
  showing here the maximum (in the computational domain) of the
  curvature invariant $R_{abcd}R^{abcd}$. The values achieved by this
  invariant in the simulations with amplitude $\alpha=2$ are
  comparable to values found inside a Schwarzschild black hole. This
  curvature is produced {\em solely} by the injection of pulses
  through the boundaries. }
\label{col_curvature}
\end{center} 
\end{figure}

Figure \ref{col_curvature} shows the maximum (in the computational
domain) of the curvature invariant $J:=R_{abcd}R^{abcd} =
8(E^2-B^2)$. This curvature invariant starts at zero as well and
remains small (compared to one) for amplitude $\alpha=0.01$ while for
amplitude $\alpha=2$ it increases to very large values as the pulses
are injected. For example, at $t=1.2$ we have $J \approx 3\times
10^2$. To have an idea of how strong the curvature is, recall that for
the Schwarzschild solution $J=48m^2/r^6$, where $m$ is the mass of the
black hole and $r$ is the area radial coordinate. Therefore, a
curvature of $J \approx 3\times 10^2$ would correspond to being inside
a black hole of mass $m=1$ at $r\approx 0.7$. This curvature is being
caused {\em solely} by the injection of pulses through the boundaries,
showing that the latter are able to handle very non-linear dynamics.

%%%%%%%%%%%%%%%%%%%%%%%%%%%%%%%%%%%%%%%%%%%%%%%%%%%%%%%%%%%%%%%%%%%%
\section{Conclusions}
\label{Sect:Conc}
%%%%%%%%%%%%%%%%%%%%%%%%%%%%%%%%%%%%%%%%%%%%%%%%%%%%%%%%%%%%%%%%%%%%

We derived a family of outer boundary conditions for first order hyperbolic
formulations of Einstein's field equations with live gauges. These boundary
conditions have the property of being constraint-preserving in the sense that
they guarantee that any smooth solution to the evolution equations subject to
these boundary conditions automatically satisfy the constraints if so
initially.  Furthermore, we have discussed different possibilities for using
constraint-preserving boundary conditions in order to control the gauge and
physical degrees of freedom at the boundary. One of these possibilities (which
we called CPBC with Weyl control) is attractive from a physical point of view
since it provides boundary data to the Weyl scalars $\Psi_0$ and $\Psi_4$,
which at null infinity represent the in- and outgoing radiation. Furthermore,
these scalars are gauge-invariant for linearizations about the Kerr metric, an
approximation that should be good in many simulations provided the boundaries
can be moved sufficiently far from the strong field region.  We also discussed
simpler ways (which we called CPBC without Weyl control) of controlling the
physical degrees of freedom; however, their physical interpretation is less
clear.

Next, we analyzed the stability of the resulting IBVP by analytical and
numerical means. By considering high-frequency perturbations we obtained the
corresponding frozen coefficient problem which is linear and can be analyzed
using Fourier-Laplace transformations. In this way, one obtains a determinant
condition which is a necessary condition for the well posedness of the
problem. The satisfaction of this condition restricts the freedom in the
choice of parameters in the formulation and the boundary conditions. In
particular, we found that the violation of the determinant condition leads to
the presence of ill posed constraint-violating or gauge modes. These modes are
ill posed in the sense that they grow exponentially in time with an
exponential factor that is unbounded. One could say that the boundary
conditions are responsible for the presence of these modes since, in the
absence of boundaries, the initial value problem is well posed because the
evolution equations are strongly hyperbolic. A further example of boundary
conditions leading to an instability of an otherwise stable initial-value
problem is given in \ref{App:SFS}.

Next, we performed three-dimensional numerical simulations. In order to do so,
we extended an earlier finite-differencing code and implemented the
constraint-preserving boundary conditions. We first performed a robust
stability test which consists in specifying random initial and boundary data
for different resolutions and checking that the time evolution of the fields
does not exhibit unbounded resolution-dependent growth. We found that the set
of CPBC without Weyl control considered in this article is robustly stable in
this sense. We also compared the results from these boundary conditions with
results obtained by simply freezing the ingoing fields of the main evolution
system to their initial values. While both systems are robustly stable, we
found very strong evidence for the freezing boundary condition to yield {\em
  constraint-violating} solutions: In contrast to the CPBC, the constraint
variables seem to converge to a nonzero value for freezing boundary
conditions. We further tested the CPBC without Weyl control by evolving a
gauge solution and weak Brill waves, and found that in both cases the
constraint variables seem to converge to zero. Finally, as an application, we
considered a situation in which one starts with flat initial data and injects
``pulses of gravitational radiation'' through the boundaries, of enough
amplitude to create very large curvature in the interior.

We expect the CPBC without Weyl control to be useful for many applications,
like improving the accuracy and stability of current binary black hole and
binary neutron star simulations or for a successful implementation of
characteristic or perturbative matching techniques \cite{ccm,cpm1,cpm2,cpm3}

Unfortunately, our numerical results for the CPBC with Weyl control do not
pass the robust stability test, although the instability in this case is much
weaker than for the case of CPBC that violate the determinant condition. There
are several possible causes for this instability. First, it is not clear if
this instability is caused by a ``bad'' discretization of the problem -- the
continuum IBVP being well posed -- or if the continuum problem is actually ill
posed and thus the cause for the instability observed. Another possible source
of problem is the cubical domain used in our simulations, which has edges and
corners. It might be that we violate some compatibility conditions at such
points and that this causes an instability\footnote{See, for instance,
  \cite{Gioel, CPRST} for a detailed study of compatibility conditions in
  domains with corners.}. Whatever the cause of the instability might be, it
would be nice to have a better analytic insight into the problem. Although the
determinant condition used here is necessary for the well posedness of the
IBVP it is not sufficient, not even for the frozen coefficient problem. When
making such a statement we have to specify in what sense we expect the IBVP to
be well posed. In the linear regime, we expect the problem to be well posed
with respect to a Hilbert space that controls the $L^2$ norm of the fields and
the $L^2$ norm of the constraint variables. In this case, there are explicit
examples \cite{RS} that show that the satisfaction of the determinant
condition is {\em not} sufficient for well posedness. A different examples
which applies to the IBVP considered in this article is given in
\ref{App:SFS}. Clearly, it would be much more satisfactory to derive CPBC with
and without Weyl control for which well posedness can be guaranteed.

%%%%%%%%%%%%%%%%%%%%%%%%%%%%%%%%%%%%%%%%%%%%%%%%%%%%%%%%%%%%%%%%%%%%
\section{Acknowledgments}
%%%%%%%%%%%%%%%%%%%%%%%%%%%%%%%%%%%%%%%%%%%%%%%%%%%%%%%%%%%%%%%%%%%%

This research was supported in part by the NSF under Grants No: PHY0244335,
PHY0326311, INT0204937 to Louisiana State University, the Horace Hearne Jr.
Institute for Theoretical Physics, NSF Grant No. PHY-0099568 to Caltech, and
NSF Grants No. PHY0354631 and PHY0312072 to Cornell University. This research
used the resources of the Center for Computation and Technology at Louisiana
State University, which is supported by funding from the Louisiana
legislature's Information Technology Initiative.

We are especially indebted to Mark Miller for many discussions, suggestions
and help throughout this project. We also thank L. Buchman, C. Gundlach, L.
Kidder, L. Lehner, L. Lindblom, G. Nagy C. Palenzuela, O. Reula, M. Scheel, S.
Teukolsky and J. York for many helpful discussions and comments.

\appendix
%%%%%%%%%%%%%%%%%%%%%%%%%%%%%%%%%%%%%%%%%%%%%%%%%%%%%%%%%%%%%%%%%%%%
\section{A derivation of the main evolution and constraint propagation system}
\label{App:MECPS}
%%%%%%%%%%%%%%%%%%%%%%%%%%%%%%%%%%%%%%%%%%%%%%%%%%%%%%%%%%%%%%%%%%%%

In this appendix we re-derive the main evolution system, carefully keeping
track of the constraints that are being used, and find the constraint
propagation system with the help of the (twice contracted) Bianchi identities.
We start with the $3+1$ split of the four-dimensional Ricci tensor,
\begin{equation}
^{(4)} R_{ij} = {^{(3)} R_{ij}} - \frac{1}{N} {^{(3)} \nabla_i}
{^{(3)} \nabla_j} N - 2 K_{il} K^l_{\; j} + K K_{ij} - \partial_0 K_{ij}\, ,
\label{Eq:fourRicci}
\end{equation}
where $^{(3)} R_{ij}$ and $^{(3)} \nabla_i$ are the Ricci tensor and
the covariant derivative, respectively, belonging to the three metric
$g_{ij}$, $N$ is the lapse, $K_{ij}$ the extrinsic curvature, and
$\partial_0$ denotes the operator $N^{-1}(\partial_t -
\pounds_\beta)$. In order to obtain first order equations we rewrite
the spatial derivatives of the three-metric as
\begin{equation}
\partial_k g_{ij} = d_{kij} - C_{kij}\; ,
\label{Eq:splitChris}
\end{equation}
where $d_{kij}$ are new fields and $C_{kij} = 0$ are constraints.
Correspondingly, we can split the Christoffel symbols, $^{(3)}\Gamma^k_{\;\;
  ij}$, belonging to the three metric as
\begin{equation}
^{(3)} \Gamma^k_{\;\; ij} = \Gamma^k_{\;\; ij} - \hat{\Gamma}^k_{\;\; ij}\; ,
\end{equation}
where
\begin{equation}
\Gamma^k_{\; ij} = \frac{1}{2} g^{kl} \left( 2d_{(ij)l} - d_{lij} \right),
\end{equation}
are Christoffel symbols belonging to a new connection $\nabla$ which is
torsion-free but not necessarily metric (we have $\nabla_k g_{ij} = -C_{kij}$,
so $\nabla$ is only metric if the constraints $C_{kij} = 0$ are satisfied),
and where
\begin{equation}
\hat{\Gamma}^k_{\; ij} = \frac{1}{2} g^{kl} \left( 2C_{(ij)l} - C_{lij} \right).
\end{equation}
Since the symbols $\hat{\Gamma}^k_{\; ij}$ correspond to the difference
between the connections $^{(3)} \nabla$ and $\nabla$, they are actually the
components of tensor field. Substituting Eq. (\ref{Eq:splitChris}) into the
expression for the Ricci tensor, we obtain
\begin{equation}
{^{(3)} R_{ij}} = R_{ij} - \hat{R}_{ij}\; ,
\label{Eq:Subst1}
\end{equation}
where $R_{ij}$ is given by Eq. (\ref{Eq:Rij}) and where
\begin{equation}
\hat{R}_{ij} = \nabla_k \hat{\Gamma}^k_{\; ij} - \nabla_{(i} \hat{\Gamma}^k_{\; j)k}
 - \hat{\Gamma}^k_{\; lk} \hat{\Gamma}^l_{\; ij} + \hat{\Gamma}^k_{\; lj} \hat{\Gamma}^l_{\; ik}
 - \Gamma^l_{\; ij} C^k_{\; kl} + \frac{1}{2} C_{(i}^{\;\; kl} d_{j)lk} - g^{kl} C_{k(ij)l}\; .
\end{equation}
Here, we have defined $C_{lkij} = \partial_{[l} C_{k]ij}$. Note that the
hatted quantities vanish if $C_{kij} = 0$.  Next, we replace the gradient of
the logarithm of the lapse by a new field $A_i$ minus a corresponding
constraint variable,
\begin{equation}
\frac{\partial_i N}{N} = A_i - C_i^{(A)},
\label{Eq:splitgradnabN}
\end{equation}
and rewrite
\begin{eqnarray}
\frac{1}{N}  {^{(3)} \nabla_i} {^{(3)} \nabla_j} N 
 &=& \nabla_{(i} A_{j)} + A_i A_j - \nabla_{(i} C_{j)}^{(A)}
\nonumber\\
 &-& 2 A_{(i} C_{j)}^{(A)} + C_i^{(A)} C_j^{(A)} 
  + \hat{\Gamma}^k_{\;\; ij}(A_k - C_k^{(A)}).
\label{Eq:Subst2}
\end{eqnarray}
Using Eqs. (\ref{Eq:Subst1}) and (\ref{Eq:Subst2}), we rewrite Eq.
(\ref{Eq:fourRicci}) as
\begin{equation}
\partial_0 K_{ij} = R_{ij} - \nabla_{(i} A_{j)} - A_i A_j 
  - 2 K_i^{\;\; l} K_{lj} + K K_{ij} - \Lambda_{ij}\; ,
\label{Eq:KijId}
\end{equation}
where
\begin{equation}
\Lambda_{ij} \equiv {^{(4)} R_{ij}} + \hat{R}_{ij} - \nabla_{(i} C_{j)}^{(A)} 
 - 2 A_{(i} C_{j)}^{(A)} + C_i^{(A)} C_j^{(A)} 
 + \hat{\Gamma}^k_{\;\; ij}(A_k - C_k^{(A)})
\label{Eq:Lambdaij}
\end{equation}
groups together the four Ricci tensor (which vanishes in vacuum) and the
constraint variables. An evolution equation for $K_{ij}$ in vacuum can be
obtained from the identity (\ref{Eq:KijId}) by setting $\Lambda_{ij}$ to zero.
However, in order to obtain a strongly hyperbolic evolution system, one needs
to set $\Lambda_{ij}$ equal to suitable combinations of the constraint
variables (see below). Next, in order to obtain evolution equations for the
new fields $d_{kij}$ and $A_i$, we apply the operator $\partial_0$ on Eqs.
(\ref{Eq:splitChris},\ref{Eq:splitgradnabN}) and use the commutation relation
\begin{eqnarray}
[ \partial_0\, , \partial_k ] T_{i_1 i_2 ... i_r} 
 &=& \frac{\partial_k N}{N} \partial_0 T_{i_1 i_2 ... i_r}
\nonumber\\
 &+& \frac{1}{N}\left( T_{s i_2 ... i_r} \partial_k\partial_{i_1} \beta^s + ... 
  + T_{i_1 i_2 ... i_{r-1} s} \partial_k\partial_{i_r} \beta^s \right),
\label{Eq:ComRel}
\end{eqnarray}
for any $r$-rank symbol $T_{i_1 i_2 ... i_r}$, where the Lie derivative
$\pounds_\beta$ of $T_{i_1 i_2 ... i_r}$ is formally defined by
\begin{displaymath}
\pounds_\beta T_{i_1 i_2 ... i_r} = \beta^k\partial_k T_{i_1 i_2 ... i_r} 
 + T_{k i_2 ... i_r} \partial_{i_1} \beta^k + ... + T_{i_1 i_2 ... i_{r-1} k} \partial_{i_r} \beta^k.
\end{displaymath}
Using the evolution equations $\partial_0 g_{ij} = -2K_{ij}$ and $\partial_0 N
= -F(N,K,x^\mu)$ for the three metric and the lapse, respectively, we obtain
\begin{eqnarray}
\partial_0 d_{kij} &=& -2\partial_k K_{ij} - 2 A_k K_{ij} 
 + \frac{2}{N} g_{l(i} \partial_{j)}\partial_k\beta^l + \Lambda_{kij}\; ,
\label{Eq:dkijId}\\
\partial_0 A_i &=& -\frac{\partial F}{\partial N}\, A_i 
 - \frac{1}{N}\,\frac{\partial F}{\partial K}\left( g^{kl}\partial_i K_{kl} - d_i^{\;\; kl} K_{kl} \right)
 - \frac{1}{N} \frac{\partial F}{\partial x^i} + \Lambda_i\; ,
\label{Eq:AiId}
\end{eqnarray}
where
\begin{eqnarray}
\Lambda_{kij} &\equiv& \partial_0 C_{kij} + 2 C_k^{(A)} K_{ij}\; ,
\label{Eq:Lambdakij}\\
\Lambda_i &\equiv& \partial_0 C_i^{(A)} + \frac{\partial F}{\partial N} C_i^{(A)} - \frac{1}{N}\,\frac{\partial F}{\partial K} C_i^{\;\; kl} K_{kl}\; .
\label{Eq:Lambdai}
\end{eqnarray}
Finally, we rewrite the Hamiltonian and momentum constraint.  Let $^{(4)}
G_{\mu\nu}$ be the Einstein tensor, and let $\mu=0$ denote the contraction
with the vector field $\partial_0$.  Then, we have
\begin{displaymath}
^{(4)} G_{00} = C - \hat{C}, \qquad
^{(4)} G_{0j} = -(C_j - \hat{C}_j) ,
\end{displaymath}
where the expressions for $C$ and $C_j$ are given by Eqs.
(\ref{Eq:C},\ref{Eq:Ci}) and where
\begin{equation}
\hat{C} = \frac{1}{2} g^{kl} \hat{R}_{kl}\; ,\qquad
\hat{C}_j = \frac{1}{2} g^{kl} (C_{ikl} - 2 C_{kli}) K^i_{\;\; j} + \frac{1}{2} K^{kl} C_{jkl}\; .
\end{equation}
The main evolution equations are $\partial_0 g_{ij} = -2K_{ij}$, $\partial_0 N
= -F(N,K,x^\mu)$, and Eqs. (\ref{Eq:KijId},\ref{Eq:dkijId},\ref{Eq:AiId})
where one sets the quantities
\begin{eqnarray}
\Lambda_{ij}(\gamma,\zeta) &\equiv& \Lambda_{ij} + \gamma\, g_{ij} C + \zeta\, g^{kl} C_{k(ij)l}\; ,\\
\Lambda_{kij}(\eta,\chi) &\equiv& \Lambda_{kij} - \eta g_{k(i} C_{j)} - \chi\, g_{ij} C_k\; ,\\
\Lambda_{i}(\xi) &\equiv& \Lambda_{i} - \xi C_i\; ,
\end{eqnarray}
to zero.

With this information it is not very difficult to find the constraint
propagation system using the commutation relation (\ref{Eq:ComRel}) and the
twice contracted Bianchi identities (written in 3+1 form)
\begin{eqnarray}
 \partial_0 {^{(4)} G_{00}} &=&
  + \frac{1}{N^2} {^{(3)}\nabla}^i \left( N^2 {^{(4)} G_{0i}} \right)
  + 2 K {^{(4)} G_{00}} + K^{ij} {^{(4)} R_{ij}} - K g^{ij} {^{(4)} R_{ij}}, 
\label{Eq:Bianchi1}\\
\partial_0 {^{(4)} G_{0j}} &=&
  + \frac{1}{N^2} {^{(3)}\nabla}_j \left( N^2 {^{(4)} G_{00}} \right) 
  + K {^{(4)} G_{0j}}
\nonumber\\
 &+& \frac{1}{N} \nabla^i\left( N {^{(4)} R_{ij}} \right)
  - \frac{1}{N} {^{(3)}\nabla}_j \left( N g^{rs}{^{(4)} R_{rs}} \right).
\label{Eq:Bianchi2}
\end{eqnarray}
Substituting $G_{00} = C - \hat{C}$, $G_{0j} = -(C_j - \hat{C}_j)$ and using
the equations $\Lambda_{ij}(\gamma,\zeta) = 0$, $\Lambda_{kij}(\eta,\chi) = 0$
and $\Lambda_{i}(\xi) = 0$ and Eqs.
(\ref{Eq:Lambdaij},\ref{Eq:Lambdakij},\ref{Eq:Lambdai}), a lengthy but
straightforward calculation yields
\begin{eqnarray}
\partial_0 C &=& -\left( 1 + \chi - \frac{\eta}{2} \right) g^{kl}\partial_l C_k + l.o.,\\
\partial_0 C_j &=& -(1 + 2\gamma)\partial_j C 
 - g^{kl} g^{is}\partial_s\left( C_{k[ij]l} + \frac{1}{2} C_{ijkl} - \zeta C_{k(ij)l} \right)
\nonumber\\
 &-& g^{is}\partial_s C_{ij}^{(A)} + l.o.,\\
\partial_0 C_{kij} &=& l.o.,\\
\partial_0 C_{lkij} &=& \frac{\eta}{2}\left( g_{i[k} \partial_{l]} C_j 
  + g_{j[k} \partial_{l]} C_i \right) + \chi\, g_{ij} \partial_{[l} C_{k]} + l.o.,\\
\partial_0 C_k^{(A)} &=& l.o.,\\
\partial_0 C_{ij}^{(A)} &=& \xi\,\partial_{[l} C_{k]} + l.o.,
\end{eqnarray}
where we have defined $C_{ij}^{(A)} = N^{-1}\partial_{[i} (N C_{j]}^{(A)})$
and where the lower order terms, $l.o.$, are linear algebraic expressions in
the constraint variables, with coefficients that depend on the main variables
and their spatial derivatives. The hard part of the calculation is to show
that no spatial derivatives of $C_{kij}$ and $C_i^{(A)}$ other than the
antisymmetric ones (which can be re-expressed in terms of the constraint
variables $C_{lkij}$ and $C_{ij}^{(A)}$, respectively) enter the lower order
terms.

%%%%%%%%%%%%%%%%%%%%%%%%%%%%%%%%%%%%%%%%%%%%%%%%%%%%%%%%%%%%%%%%%%%%
\section{Characteristic fields and strong hyperbolicity}
\label{App:CharFields&SH}
%%%%%%%%%%%%%%%%%%%%%%%%%%%%%%%%%%%%%%%%%%%%%%%%%%%%%%%%%%%%%%%%%%%%

Given a first order evolution system with principal symbol ${\cal A}(n)$, the
characteristic fields with respect to a fixed direction $n_i$ are defined to
be the projections of the main fields onto the eigenspaces of ${\cal A}(n)$.
In order to find the characteristic fields for the symbol defined by Eq.
(\ref{Eq:PrincipalSymb}), it is convenient to choose an orthonormal basis of
three-vectors $e_1$, $e_2$, $e_3$, such that $e_1^i = n^i = g^{ij} n_j$ and
express the main variables $K_{ij}$, $d_{kij}$ and $A_i$ with respect to this
basis:
\begin{equation}
K_{ab} = K_{ij} e_a^i e_b^j\, , \qquad
d_{cab} = d_{kij} e_c^k  e_a^i e_b^j\, , \qquad
A_a = A_i e_a^i\, .
\end{equation}
Here and in the following, we assume that $n_i$ is normalized such that
$g^{ij} n_i n_j = 1$. Assuming that $\lambda_1$, $\lambda_2$ and $\lambda_3$
are positive (which is a necessary condition for strong hyperbolicity) and
defining
\begin{equation}
\Omega = \frac{1 + 2\lambda_1 + \lambda_2 - 4\lambda_3}{2(\lambda_1-\lambda_2)}\, ,
\quad \hbox{if $\lambda_1\neq\lambda_2$ and $\Omega$ arbitrary otherwise,}
\end{equation}
we can express the characteristic fields as ($n$ refers to the the index $a=1$
and $A,B,C,...$ to the indices $a=2,3$)
\begin{eqnarray}
v_{nn}^{(\pm)} &=& K_{nn} + \Omega K^A_A \pm \frac{1}{\sqrt{\lambda_1}}
(D_{nn} + \Omega D^A_A), \\
v_{AA}^{(\pm)} &=& K^A_A \pm \frac{1}{\sqrt{\lambda_2}}\, D^A_A\; ,\\
v_{nA}^{(\pm)} &=& K_{nA} \pm \frac{1}{\sqrt{\lambda_3}}\, D_{nA}\; ,\\
\hat{v}_{AB}^{(\pm)} &=& \hat{K}_{AB} \pm \hat{D}_{AB}\; ,
\end{eqnarray}
where
\begin{eqnarray*}
D_{nn} &=& \frac{\zeta}{2}\, d_{nnn} + \frac{1}{2}(1-\zeta)b_n-\frac{1}{2}\, d_n - A_n +\frac{\gamma}{2}(b_n-d_n),
\\
D^A_A &=& \frac{1}{2}\delta^{AB}\left[ -d_{nAB} + (1+\zeta) d_{ABn} \right] + \gamma(b_n-d_n),
\\
D_{nA} &=& -\frac{1}{4}(1-\zeta)(d_{nnA}-b_A) + \frac{1}{4}(1+\zeta) d_{Ann} - \frac{1}{4}\, d_A - \frac{1}{2}\, A_A\; ,
\\
\hat{D}_{AB} &=& \frac{1}{2}\left[ -d_{nAB} + (1+\zeta) d_{(AB)n} \right]
 - \frac{1}{4}\,\delta_{AB}\delta^{CD}\left[ -d_{nCD} + (1+\zeta) d_{CDn} \right].
\end{eqnarray*}
Here, $d_k = g^{ij} d_{kij}$, $b_j = g^{ki} d_{kij}$ and $\hat{K}_{AB} =
K_{AB} - \delta_{AB}\delta^{CD} K_{CD}/2$.  $v_{nn}^{(\pm)}$,
$v_{AA}^{(\pm)}$, $v_{AA}^{(\pm)}$, $\hat{v}_{AB}^{(\pm)}$ have the speeds
$\mu = \pm\sqrt{\lambda_1}$, $\mu = \pm\sqrt{\lambda_2}$, $\mu =
\pm\sqrt{\lambda_3}$, and $\mu = \pm 1$, respectively. (These are the speeds
with respect to the normal derivative operator. The speeds with respect to the
time evolution vector field are obtained from these after the transformation
$\mu \mapsto N\mu + \beta^i n_i$.) The remaining characteristic fields have
speeds $\mu = 0$. For the following representation, we assume that the
conditions (\ref{Eq:ConstrSym1},\ref{Eq:ConstrSym2}), which are necessary for
the constraint propagation system to be symmetrizable, hold. Defining $\omega
= (\eta - 2\chi - 2)^{-1}$, the zero speed characteristic fields are
\begin{eqnarray}
&& N,\\
&& g_{ij}\; ,\\
v_{Ann}^{(0)} &=& d_{Ann} - \chi\omega (b_A - d_A), \\
v_{AnB}^{(0)} &=& d_{AnB} - \frac{1}{2}\, \eta\omega \delta_{AB} (b_n - d_n), \\
v_{ABC}^{(0)} &=& d_{ABC} - \eta\omega \delta_{A(B} (b_{C)} - d_{C)}) 
  - \chi\omega \delta_{BC} (b_A - d_A), \\
v_n^{(0)} &=& A_n + \sigma\omega\left[ (2+\eta+3\chi) b_n - (2\eta+\chi)d_n \right]
 - (2\sigma + \xi)\omega (b_n - d_n), \\
v_A^{(0)} &=& A_A - \xi\omega(b_A - d_A).
\end{eqnarray}
Having obtained the characteristic fields explicitly, it is not difficult to
verify the additional smoothness requirements for the system to be strongly
hyperbolic: Namely, we have to construct a symmetric positive definite matrix
${\bf H} = {\bf H}(p,n,u)$ which symmetrizes the principal symbol ${\cal
  A}(p,n,u)$ defined in Eq. (\ref{Eq:PrincipalSymb}). The system is called
strongly hyperbolic if ${\bf H}$ depends smoothly on $p$, $n$, $u$ and is such
that the matrix ${\bf H} {\cal A}$ is symmetric for all $p$, $n$, $u$. The
smoothness requirements is needed for the pseudo-differential calculus while
the symmetry condition allows for appropriate energy estimates. The matrix
${\bf H}$ can be obtained from the quadratic form which is built by summing
over the square of the eigenfields:
\begin{eqnarray}
u^T {\bf H} u &=& N^2 + \delta^{ij}\delta^{kl} g_{ik} g_{jl} 
 + \delta^{AB} v_{Ann}^{(0)} v_{Bnn}^{(0)} 
 + \delta^{AB}\delta^{CD} v_{AnC}^{(0)} v_{BnD}^{(0)}
\nonumber\\
 &+& \delta^{AB}\delta^{CD}\delta^{EF} v_{ABC}^{(0)} v_{DEF}^{(0)}
  + (v_n^{(0)})^2 + \delta^{AB} v_{A}^{(0)} v_{B}^{(0)} 
\nonumber\\
 &+& \sum\limits_{\pm} \left\{ (v^{(\pm)}_{nn})^2 + (v^{(\pm)}_{AA})^2 + \delta^{AB} v_{nA}^{(\pm)} v_{nB}^{(\pm)}
  + \delta^{AB}\delta^{CD} \hat{v}_{AC}^{(\pm)} \hat{v}_{BD}^{(\pm)} \right\}.
\nonumber
\end{eqnarray}
Clearly, this quadratic form depends smoothly on $n_i$ since it can be written
in terms of contractions with $g^{ij}$ and $n^i$. ${\bf H}$ is also smooth in
the other variables provided that the parameters are smooth functions, and
provided that the function $\Omega$ can be chosen smoothly. The latter
condition is nontrivial if there are crossing points in the values swept by
$\lambda_1$ and $\lambda_2$. In all our simulations, we choose smooth
parameters such that $\lambda_1 = \lambda_2 = \lambda_3 = 1$ and set
$\Omega=0$, so the system is strongly hyperbolic.

Finally, we give the inverse transformation which allows to recover the main
variables from the characteristic ones. First, compute
\begin{eqnarray}
\hat{K}_{AB} &=& \frac{1}{2}( \hat{v}_{AB}^{(+)} + \hat{v}_{AB}^{(-)}),\\
K_{nA} &=& \frac{1}{2}( v_{nA}^{(+)} + v_{nA}^{(-)}),\\
K^A_A &=& \frac{1}{2}( v_{AA}^{(+)} + v_{AA}^{(-)}),\\
K_{nn} &=& \frac{1}{2}( v_{nn}^{(+)} + v_{nn}^{(-)}) - \Omega K^A_A\; ,
\end{eqnarray}
from which $K_{ab}$ is obtained. Next, compute
\begin{eqnarray}
\hat{D}_{AB} &=& \frac{1}{2}\left( \hat{v}_{AB}^{(+)} - \hat{v}_{AB}^{(-)} \right),\\
D_{nA} &=& \frac{\sqrt{\lambda_3}}{2}( v_{nA}^{(+)} - v_{nA}^{(-)}),\\
D^A_A &=& \frac{\sqrt{\lambda_2}}{2}( v_{AA}^{(+)} - v_{AA}^{(-)}),\\
D_{nn} &=& \frac{\sqrt{\lambda_1}}{2}( v_{nn}^{(+)} - v_{nn}^{(-)}) - \Omega D^A_A\; .
\end{eqnarray}
Next, set
\begin{eqnarray}
\left( \begin{array}{c} b_n \\ d_n \end{array} \right) &=&
M_1 \left( \begin{array}{c} 2 D_{AA} - \zeta\, \delta^{AB} v_{AnB}^{(0)} \\ 
  D_{nn} + D_{AA} + v_n^{(0)} \end{array} \right),\\
%%%%%%%%%%%%%%%%%%%%%%%%
\left( \begin{array}{c} b_A \\ d_A \end{array} \right) &=&
M_2 \left( \begin{array}{c} v_{Ann}^{(0)} + \delta^{BC} v_{ABC}^{(0)} \\ 
 2 D_{nA} + v_A^{(0)} - \frac{1}{2}(1-\zeta)\delta^{BC}v_{BCA}^{(0)} 
 + \frac{1}{2}\delta^{BC} V_{ABC}^{(0)} - \frac{\zeta}{2} v_{Ann}^{(0)} 
\end{array} \right),
\end{eqnarray}
where
\begin{eqnarray}
M_1 &=& \frac{1}{\sigma}\left( \begin{array}{cc} 
 \frac{ 2+3\gamma-2\omega\xi + 2\omega\sigma(2\eta+\chi-2) }{ 2(1 + 2\gamma + \eta\zeta\omega) } & -1 \\
 \frac{ 2+3\gamma-2\omega\xi + 2\omega\sigma(\eta+3\chi) }{ 2(1 + 2\gamma + \eta\zeta\omega) } & -1
\end{array} \right),\\
%%%%%%%%%%%%%%%%%%%
M_2 &=& \left( \begin{array}{cc} 1 & \frac{\omega(\eta+3\chi) + 1}{2\omega\lambda_3} \\ 
  1 & \frac{\eta+3\chi}{2\lambda_3} \end{array} \right).
\end{eqnarray}
Notice that as a consequence of the conditions
(\ref{Eq:ConstrSym1},\ref{Eq:ConstrSym2}), $1 + 2\gamma + \eta\zeta\omega =
-\omega[ (1+2\gamma)(2+2\chi-\eta) - \eta\zeta] < 0$ and the matrix $M_1$ is
well defined. Using this, compute
\begin{eqnarray}
d_{Ann} &=& v_{Ann}^{(0)} + \chi \omega (b_A - d_A), \\
d_{AnB} &=& v_{AnB}^{(0)} + \frac{1}{2}\, \eta\omega \delta_{AB} (b_n - d_n), \\
d_{ABC} &=& v_{ABC}^{(0)} + \eta\omega \delta_{A(B} (b_{C)} - d_{C)}) 
  + \chi\omega \delta_{BC} (b_A - d_A), \\
A_n &=& v_n^{(0)} -\sigma\omega\left[ (2+\eta+3\chi) b_n - (2\eta+\chi)d_n \right]
 + (2\sigma + \xi)\omega (b_n - d_n), \\
A_A &=& v_A^{(0)} + \xi\omega(b_A - d_A), \\
%%%%%%%%%%%%%%%%%%%%%%%%%%%%%%%%%%%%%%%%%%%%%%%
d_{nAB} &=& -2\hat{D}_{AB} + (1+\zeta) d_{(AB)n} + \delta_{AB}\left[ \gamma(b_n-d_n) - D^C_C \right], \\
d_{nnA} &=& \zeta^{-1}\left[ 2 D_{nA} - \frac{1}{2}(1+\zeta)\delta^{BC}(d_{BCA} - d_{ABC}) + A_A \right]
 + b_A - \frac{1}{2}\, d_A\; ,\\
d_{nnn} &=& 2\zeta^{-1}\left[ D_{nn} -\frac{1}{2}(1-\zeta) b_n + \frac{1}{2}\, d_n + A_n - \frac{\gamma}{2}\, (b_n-d_n) \right].
\end{eqnarray}
from which one can compute the components of $d_{kij}$ and $A_i$ with respect
to the orthonormal basis.

Finally, one obtains the coordinate components of the main variables by
contracting with the dual basis $\theta^a_i$, which is defined by
$\theta^a_i(e_b^i) = \delta^a_b$, $a,b=1,2,3$:
\begin{equation}
K_{ij} = K_{ab} \theta^a_i\theta^b_j\, , \qquad
d_{kij} = d_{cab} \theta^c_k\theta^a_i\theta^b_j\, , \qquad
A_i = A_a \theta^a_i\, .
\end{equation}

%%%%%%%%%%%%%%%%%%%%%%%%%%%%%%%%%%%%%%%%%%%%%%%%%%%%%%%%%%%%%%%%%%%%
\section{A special family of solutions}
\label{App:SFS}
%%%%%%%%%%%%%%%%%%%%%%%%%%%%%%%%%%%%%%%%%%%%%%%%%%%%%%%%%%%%%%%%%%%%

In this appendix we show explicitly that the linearized IBVP with Weyl
control cannot be well posed in $L^2$ if the parameter $\Omega$
defined in Eq.  (\ref{Eq:Omega}) is one; independent on whether or not
the determinant condition is satisfied. In order to see this, we
consider the following family of solutions:

Let $w_j$ be a one-form that satisfies $\partial^j w_j = 0$,
$\partial^i\partial_i w_j = 0$, and $\partial_t w_j = 0$, and let
\begin{eqnarray}
K_{ij} &=& \partial_{(i} w_{j)}\; ,\\
d_{kij} &=& -2t \partial_k\partial_{(i} w_{j)}\; ,
\label{Eq:Sol}\\
A_i &=& 0.
\end{eqnarray}
It is not difficult to check that these expressions satisfy the
evolution equations (\ref{Eq:LinKij},\ref{Eq:Lindkij},\ref{Eq:LinAi})
and the constraints $\partial^k(d_k - b_k) = 0$, $\partial_{[l}
d_{k]ij} = 0$, $\partial^i K_{ij} - \partial_j K = 0$. For systems
without boundaries, these solutions are trivial if appropriate fall
off conditions on the fields are demanded since then the harmonic
condition on $w_j$ implies that it must vanish. However, if boundaries
are present, $w_j$ may be nontrivial. The electric and magnetic
components of the linearized Weyl tensor corresponding to the
solutions (\ref{Eq:Sol}) are
\begin{displaymath}
E_{ij} = 0, \qquad
B_{ij} = \frac{1}{2} \partial_{(i } \varepsilon_{j)rs} \partial^r w^s,
\end{displaymath}
and $B_{ij}$ vanishes if the one-form $w_j$ is closed. In particular, this is
true if $w_j$ is exact, i.e. if $w_j = \partial_j f$ for some time-independent
harmonic function $f$. In this case, we also have
\begin{equation}
v^{(+)}_{nn} = (1-\Omega)\left( \partial_n^2 f - \frac{\zeta t}{\sqrt{\lambda_1}}\, \partial_n^3 f \right).
\end{equation}
Therefore, if $\Omega=1$, the family of solutions (\ref{Eq:Sol}) with
$w_j = \partial_j f$ and $f$ harmonic and time-independent shows that
the linearized IBVP with the boundary conditions
(\ref{Eq:CPBC},\ref{Eq:Gauge},\ref{Eq:MaxDissWeyl}) is not well posed
in $L^2$ since then the boundary conditions are satisfied with
homogeneous data and since the initial data depends only on second
derivatives of $f$ whereas $d_{kij}$ depends on third derivatives of
$f$ for $t > 0$. This results in frequency dependent growth of the
solution of the form $|k| t$, where $k$ is a characteristic wave
number of the initial data.

%%%%%%%%%%%%%%%%%%%%%%%%%%%%%%%%%%%%%%%%%%%%%%%%%%%%%%%%%%%%%%%%%%%%


\begin{thebibliography}{10}
%%%%%%%%%%%%%%%%%%%%%%%%%%%%%%%%%%%%%%%%%%%%%%%%%%%%%%%%%%%%%%%%%%%%
  
\bibitem{miller} M.~Miller, P.~Gressman and W.~M.~Suen, Towards a Realistic
  Neutron Star Binary Inspiral: Initial Data and Multiple Orbit Evolution in
  Full General Relativity, Phys.\ Rev.\ D {\bf 69} (2004) 064026.
  
\bibitem{max_slicing} A. Lichnerowicz, L\'\,Int\'{e}gration des \'{e}quations
  de la gravitation relativiste et le probl\`{e}me des n corps, J. Math. Pure
  Appl. {\bf 23} (1944) 37--63.
  
\bibitem{min_distortion} L.~Smarr and J.~W.~.~York, Kinematical Conditions In
  The Construction Of Space-Time, Phys.\ Rev.\ D {\bf 17} (1978) 2529--2551.
  
\bibitem{cp_AM} M.~Anderson and R.~A.~Matzner, Extended Lifetime in
  Computational Evolution of Isolated Black Holes, arXiv:gr-qc/0307055.
  
\bibitem{cp_S} E. Schnetter, Ph.D. thesis, Universit\"at T\"ubingen (2003),
  http://w210.ub.uni-tuebingen.de/dbt/volltexte/ 2003/819/.
  
\bibitem{cp_Caltech} M.~Holst, L.~Lindblom, R.~Owen, H.~P.~Pfeiffer,
  M.~A.~Scheel and L.~E.~Kidder, Optimal Constraint Projection for Hyperbolic
  Evolution Systems, Phys.\ Rev.\ D {\bf 70} (2004) 084017.
  
\bibitem{ccm} J.~Winicour, Characteristic Evolution and Matching, Living Rev.
  Relativity 4, (2001), 3.
  
\bibitem{cpm1} L.~Rezzolla, A.~M.~Abrahams, R.~A.~Matzner, M.~E.~Rupright and
  S.~L.~Shapiro, Cauchy-perturbative matching and outer boundary conditions:
  Computational studies, Phys.\ Rev.\ D {\bf 59} (1999) 064001.
  
\bibitem{cpm2} M.~E.~Rupright, A.~M.~Abrahams and L.~Rezzolla,
  Cauchy-perturbative matching and outer boundary conditions. I: Methods and
  tests, Phys.\ Rev.\ D {\bf 58} (1998) 044005.
  
\bibitem{cpm3} A.~M.~Abrahams {\it et al.}  [Binary Black Hole Grand Challenge
  Alliance Collaboration], Gravitational wave extraction and outer boundary
  conditions by perturbative matching, Phys.\ Rev.\ Lett.\ {\bf 80} (1998)
  1812--1815.
  
\bibitem{LaxPh} P.D. Lax, and R.S. Phillips, Local Boundary Conditions for
  Dissipative Symmetric Linear Differential Operators,, Commun. Pure Appl.
  Math. {\bf 13} (1960) 427--455.
  
\bibitem{Friedrichs} K.O. Friedrichs, Symmetric Positive Linear Differential
  Equations, Commun. Pure Appl. Math. {\bf 11} (1958) 333--418.
  
\bibitem{Stewart-Book} J.~M.~Stewart, {\em Advanced general relativity},
  Cambridge University Press (1996).
  
\bibitem{Teukolsky} S.~A.~Teukolsky, Perturbations Of A Rotating Black Hole.
  1. Fundamental Equations For Gravitational Electromagnetic, And Neutrino
  Field Perturbations, Astrophys.\ J.\ {\bf 185} (1973) 635--647.
  
\bibitem{FN} H.~Friedrich and G.~Nagy, The initial boundary value problem for
  Einstein's vacuum field equations, {\it Comm. Math. Phys.}  {\bf 201} (1999)
  619--655.
  
\bibitem{BB} J.~M.~Bardeen and L.~T.~Buchman, Numerical tests of evolution
  systems, gauge conditions, and boundary conditions for 1D colliding
  gravitational plane waves, {\it Phys.\ Rev.\ D} {\bf 65} (2002) 064037.
  
\bibitem{Rauch} J. Rauch, Symmetric positive systems with boundary
  characteristics of constant multiplicity, Trans. Am. Math. Soc. {\bf 291}
  (1985) 167--187.
  
\bibitem{Secchi1} P. Secchi, The initial boundary value problem for linear
  symmetric hyperbolic systems with characteristic boundary of constant
  multiplicity, Diff. Int. Eq. {\bf 9} (1996) 671--700.
  
\bibitem{Secchi2} P. Secchi, Well-Posedness of Characteristic Symmetric
  Hyperbolic Systems, Arch. Rat. Mech. Anal. {\bf 134} (1996) 155--197.
  
\bibitem{IR} M.~S.~Iriondo and O.~A.~Reula, On free evolution of
  selfgravitating, spherically symmetric waves, Phys.\ Rev.\ D {\bf 65} (2002)
  044024.
  
\bibitem{Stewart} J.~M.~Stewart, The Cauchy problem and the initial boundary
  value problem in numerical relativity {\it Class. Quantum Grav.} {\bf 15}
  (1998) 2865--2889.
  
\bibitem{SSW} B.~Szilagyi, B.~G.~Schmidt and J.~Winicour, Boundary conditions
  in linearized harmonic gravity, {\it Phys.\ Rev.\ D} {\bf 65} (2002) 064015.
  
\bibitem{SW} B.~Szilagyi and J.~Winicour, Well-Posed Initial-Boundary
  Evolution in General Relativity, {\it Phys.\ Rev.\ D} {\bf 68} (2003)
  041501.
  
\bibitem{CPRST} G.~Calabrese, J.~Pullin, O.~Reula, O.~Sarbach and M.~Tiglio,
  Well posed constraint-preserving boundary conditions for the linearized
  Einstein equations, {\it Commun.\ Math.\ Phys.}  {\bf 240} (2003) 377--395.
  
\bibitem{CS} G.~Calabrese and O.~Sarbach, Detecting ill posed boundary
  conditions in General Relativity, {\it J.\ Math.\ Phys.}  {\bf 44} (2003)
  3888--3899.
  
\bibitem{Gioel} G. Calabrese, Ph.D. thesis (unpublished).
  
\bibitem{Frittelli} S.~Frittelli and R.~Gomez, Boundary conditions for
  hyperbolic formulations of the Einstein equations, {\it Class.\ Quant.\ 
    Grav.}  {\bf 20} (2003) 2379--2392, Einstein boundary conditions for the
  3+1 Einstein equations, {\it Phys.\ Rev.\ D} {\bf 68} (2003) 044014,
  Einstein boundary conditions in relation to constraint propagation for the
  initial-boundary value problem of the Einstein equations, {\it Phys.\ Rev.\ 
    D} {\bf 69} (2004) 124020, Einstein boundary conditions for the Einstein
  equations in the conformal-traceless decomposition, {\it Phys.\ Rev.\ D}
  {\bf 70} (2004) 064008.
  
\bibitem{GMG1} C.~Gundlach and J.~M.~Mart\'{\i}n-Garc\'{\i}a, Symmetric
  hyperbolic form of systems of second-order evolution equations subject to
  constraints, {\it Phys.\ Rev.\ D} {\bf 70} (2004) 044031.
  
\bibitem{GMG2} C.~Gundlach and J.~M.~Mart\'{\i}n-Garc\'{\i}a, Symmetric
  hyperbolicity and consistent boundary conditions for second-order Einstein
  equations, {\it Phys.\ Rev.\ D} {\bf 70} (2004) 044032.
  
\bibitem{RS} O.~Reula and O.~Sarbach, A model problem for the
  initial-boundary value formulation of Einstein's field equations,
  {\it Journal of Hyperbolic Differential Equations}, {\bf 2} (2005)
  397--435.
  
\bibitem{CLT} G.~Calabrese, L.~Lehner and M.~Tiglio, Constraint-preserving
  boundary conditions in numerical relativity, {\it Phys.\ Rev.\ D} {\bf 65}
  (2002) 104031.
  
\bibitem{LS-FatMax} L.~Lindblom, M.~A.~Scheel, L.~E.~Kidder, H.~P.~Pfeiffer,
  D.~Shoemaker and S.~A.~Teukolsky, Controlling the Growth of Constraints in
  Hyperbolic Evolution Systems, {\it Phys.\ Rev.\ D} {\bf 69} (2004) 124025.
  
\bibitem{CPBC-Bona} C.~Bona, T.~Ledvinka, C.~Palenzuela-Luque and M.~Zacek,
  Constraint-preserving boundary conditions in the Z4 Numerical Relativity
  formalism, arXiv:gr-qc/0411110.
  
\bibitem{ST} O.~Sarbach and M.~Tiglio, Exploiting gauge and constraint freedom
  in hyperbolic formulations of Einstein's equations, Phys.\ Rev.\ D {\bf 66}
  (2002) 064023.
  
\bibitem{EC-FR} S.~Frittelli and O.~A.~Reula, First-order symmetric-hyperbolic
  Einstein equations with arbitrary fixed gauge, {\it Phys.\ Rev.\ Lett.} {\bf
    76}, (1996) 4667--4670.
  
\bibitem{EC-AY} A.~Anderson and J.~W.~.~York, Fixing Einstein's equations,
  {\it Phys.\ Rev.\ Lett.} {\bf 82} (1999) 4384--4387.
  
\bibitem{EC-Hern} S.~D.~Hern, Ph.D. thesis, University of Cambridge, 1999,
  arXiv:gr-qc/0004036.
  
\bibitem{EC-KST} L.~E.~Kidder, M.~A.~Scheel and S.~A.~Teukolsky, Extending the
  lifetime of 3D black hole computations with a new hyperbolic system of
  evolution equations, {\it Phys.\ Rev.\ D} {\bf 64} (2001) 064017.
  
\bibitem{BM-Live1} C.~Bona, J.~Masso, E.~Seidel and J.~Stela, A New formalism
  for numerical relativity, Phys.\ Rev.\ Lett.\ {\bf 75} (1995) 600--603.
  
\bibitem{BM-Live2} C.~Bona, J.~Masso, E.~Seidel and J.~Stela, First order
  hyperbolic formalism for numerical relativity, Phys.\ Rev.\ D {\bf 56}
  (1997) 3405--3415.
  
\bibitem{Kreiss} H.~O.~Kreiss, Initial boundary value problems for hyperbolic
  systems, {\it Commun. Pure Appl. Math.} {\bf 23} (1970) 277--298.
  
\bibitem{KL-Book} H.O. Kreiss, J. Lorenz, {\em ``Initial-Boundary Value
    Problems and the Navier-Stokes Equations,''} Academic Press, (1989).
  
\bibitem{GKO-Book} B. Gustafsson, H. Kreiss, and J. Oliger, {\em ``Time
    dependent problems and difference methods,''} John Wiley \& Sons, New York
  (1995).
  
\bibitem{MO} A. Majda and S. Osher, Initial-Boundary Value Problems for
  Hyperbolic Equations with Uniformly Characteristic Boundary, Commun. Pure
  Appl. Math. {\bf 28} (1975) 607--675.
  
\bibitem{robust1} B.~Szilagyi, R.~Gomez, N.~T.~Bishop and J.~Winicour, Cauchy
  boundaries in linearized gravitational theory, Phys.\ Rev.\ D {\bf 62}
  (2000) 104006.
  
\bibitem{robust2} M.~Alcubierre {\it et al.}, Toward standard testbeds for
  numerical relativity, Class.\ Quant.\ Grav.\ {\bf 21} (2004) 589--613.
  
\bibitem{CPST-Convergence} G.~Calabrese, J.~Pullin, O.~Sarbach and M.~Tiglio,
  Convergence and stability in numerical relativity, Phys.\ Rev.\ D {\bf 66}
  (2002) 041501.
  
\bibitem{BrillWaves} D. Brill, On the Positive Definite Mass of the
  Bondi-Weber-Wheeler Time-Symmetric Gravitational Waves, Ann. Phys. {\bf 7}
  (1959) 466--483.
  
\bibitem{DynControl} M.~Tiglio, Dynamical control of the constraints growth in
  free evolutions of Einstein's equations, arXiv:gr-qc/0304062.
  
\bibitem{TLN-BlackHole} M.~Tiglio, L.~Lehner and D.~Neilsen, 3D simulations of
  Einstein's equations: symmetric hyperbolicity, live gauges and dynamic
  control of the constraints, Phys.\ Rev.\ D {\bf 70} (2004) 104018.
  
\bibitem{kreiss_scherer} H. Kreiss and G. Scherer, ``On the existence of
  energy estimates for difference approximations for hyperbolic systems'',
  Tech.~Report, Dept.~of Scientific Computing, Uppsala University, 1977.
  
\bibitem{kreiss_oliger} H. Kreiss and J. Oliger, ``Methods for the approximate
  solution of time dependent problems'' (Geneva: GARP Publication Series,
  1973).
  
\bibitem{strand} B. Strand, Summation by Parts for Finite Difference
  Approximations for $d/dx$, Journal of Computational Physics {\bf 110} (1994)
  47--67.
  
\bibitem{cactus} http://www.cactuscode.org
  
\bibitem{critical} M.~Alcubierre, G.~Allen, B.~Brugmann, G.~Lanfermann,
  E.~Seidel, W.~M.~Suen and M.~Tobias, Gravitational collapse of gravitational
  waves in 3D numerical relativity, Phys.\ Rev.\ D {\bf 61} (2000) 041501.
  
\bibitem{bssn} M.~Alcubierre, G.~Allen, B.~Brugmann, E.~Seidel and W.~M.~Suen,
  Towards an understanding of the stability properties of the 3+1 evolution
  equations in general relativity, Phys.\ Rev.\ D {\bf 62} (2000) 124011.
  
\bibitem{numerics-Let} G.~Calabrese, L.~Lehner, D.~Neilsen, J.~Pullin,
  O.~Reula, O.~Sarbach and M.~Tiglio, Novel finite-differencing techniques for
  numerical relativity: application to black hole excision, Class.\ Quant.\ 
  Grav.\ {\bf 20} (2003) L245--L252.
  
\bibitem{numerics1} G.~Calabrese, L.~Lehner, O.~Reula, O.~Sarbach and
  M.~Tiglio, Summation by parts and dissipation for domains with excised
  regions, Class.\ Quant.\ Grav.\ {\bf 21} (2004) 5735--5758.
  
\bibitem{numerics2} L.~Lehner, D.~Neilsen, O.~Reula and M.~Tiglio, The
  discrete energy method in numerical relativity: Towards long-term stability,
  Class.\ Quant.\ Grav.\ {\bf 21} (2004) 5819--5848.
  
\bibitem{multipatch} L. Lehner, O. Reula and M. Tiglio, in preparation.
  
\bibitem{CPST-Stab} G.~Calabrese, J.~Pullin, O.~Sarbach and M.~Tiglio,
  Stability properties of a formulation of Einstein's equations, Phys.\ Rev.\ 
  D {\bf 66} (2002) 064011.


\end{thebibliography}
\end{document}